\documentclass[11pt,preprintnumbers,superscriptaddress,amsmath,amssymb,nofootinbib]{revtex4-1}
\usepackage{graphicx}  
\usepackage{dcolumn}
\usepackage{bm}
\usepackage{amssymb}
\usepackage{amsmath}
\usepackage{epsfig}    
\usepackage{color}
\usepackage{slashed}
\usepackage[pdftex]{hyperref} 

\graphicspath{{./png/}}
\allowdisplaybreaks[1]

\begin{document} 

\title{Full next-to-leading-order calculations of Higgs boson decay rates in models with non-minimal scalar sectors}


\preprint{OU-HET 1006}
\preprint{KA-TP-12-2019}

\author{Shinya Kanemura}
\email{kanemu@het.phys.sci.osaka-u.ac.jp}
\affiliation{Department of Physics, Osaka University, Toyonaka, Osaka 560-0043, Japan}
\author{Mariko Kikuchi}
\email{kikuchi@kct.ac.jp}
\affiliation{National Institute of Technology, Kitakyushu College, Kitakyushu, Fukuoka 802-0985, Japan}
\author{Kentarou~Mawatari}
\email{kentarou.mawatari@het.phys.sci.osaka-u.ac.jp}
\affiliation{Department of Physics, Osaka University, Toyonaka, Osaka 560-0043, Japan}

\author{Kodai Sakurai}
\email{kodai.sakurai@kit.edu}
\affiliation{Department of Physics, Osaka University, Toyonaka, Osaka 560-0043, Japan}
\affiliation{Institute for Theoretical Physics, Karlsruhe Institute of Technology, 76131 Karlsruhe, Germany}

\author{Kei Yagyu}
\email{yagyu@het.phys.sci.osaka-u.ac.jp}
\affiliation{Department of Physics, Osaka University, Toyonaka, Osaka 560-0043, Japan}

\begin{abstract}

We present a complete set of decay rates of the Higgs boson with the mass of 125 GeV at the full next-to-leading order 
in a variety of extended Higgs models; i.e., a model with an additional real singlet scalar field, 
four types of two Higgs doublet models and the inert doublet model. 
All the one-loop contributions due to QCD and electroweak interactions as well as scalar interactions
are taken into account, and the calculations are systematically performed. 
Branching ratios for all the decay modes are evaluated in these models, and patterns of 
deviations in each decay mode from the standard model predictions are comprehensively analyzed. 
We show how these models with extended Higgs sectors can be distinguished 
by using our calculation of the branching ratios and future precision measurements of the Higgs boson decays. 

\end{abstract}
\maketitle

\newpage
\section{Introduction}

After the discovery of a Higgs boson, the Standard Model (SM) has been completed in a sense that
the existence of all the predicted particles was confirmed experimentally. 
In the SM, the minimal form with an isospin doublet scalar field is introduced as the Higgs sector.
Although the discovered Higgs boson shows similar properties to that of the SM under the current experimental and theoretical 
uncertainties, the possibility that the Higgs sector takes a non-minimal form is not excluded at all, and 
its exploration is one of the central interests of current and future high-energy physics. 
If the Higgs sector is extended from the minimal form, it has a different structure which can be classified 
by the number of scalar fields, their representations, symmetries and strength of coupling constants. 
There should be strong connection between these properties of extended Higgs sectors and the physics behind 
the electroweak (EW) symmetry breaking. 
Furthermore, a non-minimal structure of Higgs sectors could solve the problems which cannot be explained in the SM 
such as neutrino masses, dark matter and baryon asymmetry of the Universe.
Therefore, the Higgs sector is one of the most important probes of new physics beyond the SM. 

The most important property of a non-minimal Higgs sector is the prediction of multiple scalar bosons. 
Thus, discovery of additional scalar bosons will be a clear evidence of extended Higgs sectors.  
At the LHC, direct searches for a new particle are being performed continuously. 
On the other hand, existence of such additional scalar fields generally affects the couplings of the SM-like Higgs boson
to the particles in the SM 
by the effect of mixing and the quantum correction, yielding deviations from the predictions of the SM. 
Therefore, detecting such deviations by precision measurements is also a strong signature for models with extended Higgs sectors. 
Moreover, from the pattern of deviations in various Higgs boson couplings we can indirectly 
distinguish the shape of the Higgs sector and further determine new physics models~\cite{Kanemura:2014bqa}. 

At the LHC, direct searches for additional Higgs bosons have been performed 
via bosonic channels~\cite{Khachatryan:2016yec,Aaboud:2017yyg,Sirunyan:2017nrt,Aaboud:2017itg,Aaboud:2017fgj,Khachatryan:2016cfx,Aaboud:2017cxo} and 
fermionic channels~\cite{Khachatryan:2015tra,Khachatryan:2016qkc,Aaboud:2017sjh}. 
From the non-observation of the signature, parameters of each extended Higgs sector such as masses and couplings are constrained.  
In addition, some of the Higgs boson couplings have been measured at the LHC~Run-I~\cite{Khachatryan:2016vau} 
and Run-II~\cite{Sirunyan:2018koj,ATLAS-CONF-2018-031} experiments. 
Although the current data from these measurements are consistent with the SM predictions, the experimental and theoretical 
uncertainties are not small yet; e.g., about 20\% for the Higgs boson couplings to weak bosons and 
typically 20--50\% for the Yukawa couplings of the third generation at the 95\% confidence level. 
Above experimental uncertainties can be much reduced at future colliders; 
e.g., at the High-Luminosity LHC (HL-LHC)~\cite{CMS:2013xfa,ATLAS:2013hta}, 
the International Linear Collider (ILC)~\cite{Baer:2013cma,Fujii:2017vwa,Asai:2017pwp}, 
the Future Circular Collider (FCC)~\cite{Gomez-Ceballos:2013zzn}, 
the Circular Electron Positron Collider (CEPC)~\cite{CEPC-SPPCStudyGroup:2015csa}, 
the Compact LInear Collider (CLIC)~\cite{CLIC:2016zwp} and so on. 
For example, at the ILC with the collision energy of 250 GeV and the luminosity of 2 ab$^{-1}$, 
some of the Higgs boson couplings are expected to be measured with ${\cal O}(1\%)$ level or better~\cite{Fujii:2017vwa}. 

In order to extract information on new physics from 
these precision measurements at future experiments, accurate calculations with higher-order corrections
are required in models with various extended Higgs sectors. 
Radiative corrections to the SM-like Higgs boson  vertices have been studied in various Higgs sectors such as,  
for example, a model with a real isospin singlet Higgs field (HSM)~\cite{Bojarski:2015kra,Kanemura:2015fra,Kanemura:2016lkz,He:2016sqr}, 
two Higgs doublet models (THDMs)~\cite{Arhrib:2003ph,Kanemura:2004mg,LopezVal:2010vk,Kanemura:2014dja,Kanemura:2015mxa,Krause:2016oke},  
the inert doublet model (IDM)~\cite{Arhrib:2015hoa,Kanemura:2016sos}, the Higgs triplet model~\cite{Aoki:2012yt,Aoki:2012jj}
and the Georgi-Machacek model~\cite{Chiang:2017vvo,Chiang:2018xpl}. 
In order to see differences of the prediction among these models, 
it is quite important to calculate the renormalized Higgs boson vertices with a consistent and systematic way. 
Recently, we have published a numerical program {\tt H-COUP} (version 1.0)~\cite{Kanemura:2017gbi}
to compute a set of SM-like Higgs boson vertices at one-loop level 
in various extended Higgs models; i.e., the HSM, four types of THDMs and the IDM.  
Other numerical tools are also available 
to calculate Higgs boson decays with radiative corrections in models with extended Higgs sectors; e.g.,  
{\tt Prophecy4f}~\cite{Altenkamp:2017kxk,Altenkamp:2018bcs} and {\tt 2HDECAY}~\cite{Krause:2018wmo,Krause:2019oar}. 

In this paper, we present a complete set of the decay rates of the SM-like Higgs boson ($h$) 
including the $h \to WW^*$ mode at the full next-to-leading order (NLO) in QCD 
and EW as well as scalar interactions in the HSM, four types of THDMs and the IDM\footnote{The decay rates of the 
one-loop induced processes, i.e, $h \to \gamma\gamma$, $Z\gamma$ and $gg$, are calculated at NLO in QCD and at leading order in EW. }. 
Some important results have already been highlighted in our letter paper~\cite{Kanemura:2018yai}, in which 
the calculation of the partial decay rate of the $h \to WW^*$ mode was not yet included. 
We then calculate the branching ratios of the SM-like Higgs boson at NLO in these models. 
One-loop calculations are consistently performed based on the on-shell renormalization scheme 
for EW parameters~\cite{Hollik:1988ii,Kniehl:1993ay,Kanemura:2017wtm} and the modified minimal subtraction ($\overline{\text{MS}}$) scheme 
for QCD corrections~\cite{tHooft:1973mfk} in these models with the extended Higgs sectors. 
We discuss the amount of the NLO corrections of the Higgs boson decay rates in each model with detailed descriptions of the computation. 
We show various correlations of the deviation in the branching ratios from the SM predictions 
under constraints of the perturbative unitarity~\cite{Lee:1977eg}, vacuum stability~\cite{Deshpande:1977rw},  
conditions to avoid wrong vacua~\cite{Frere:1983ag} and experimental constraints. 
Finally, we investigate the possibility to discriminate the extended Higgs sectors from the difference of the prediction among the models. 
 
This paper is organized as follows. 
In Sec.~\ref{sec:model}, we introduce the HSM, the THDMs and the IDM. 
In Sec.~\ref{sec:decay}, we present analytic formulae for the decay rates of the Higgs boson at NLO. 
EW corrections in each decay mode are discussed in detail. 
In Sec.~\ref{sec:numerical}, we show numerical results of the total width, the branching ratios and 
correlations of the branching ratios. 
Conclusions are given in Sec.~\ref{sec:conc}. 
In Appendix, explicit formulae for the NLO calculations are presented. 

\section{Models with non-minimal Higgs sectors\label{sec:model}}

In this section, we define the HSM, the THDM and the IDM in order. 
Before moving on to the discussion on each extended Higgs sector, 
let us briefly explain constraints on a parameter space, as their basic notion are common to models with the extended Higgs sectors. 

First of all, the size of dimensionless parameters in the potential can be constrained 
by imposing the perturbative unitarity bound which has originally 
been introduced in Ref.~\cite{Lee:1977eg} to obtain the upper limit on the Higgs boson mass in the SM. 
Using the equivalence theorem~\cite{Cornwall:1974km}, 
this bound requires that the magnitude of partial wave amplitudes for the elastic scatterings 
of 2 body to 2 body scalar boson processes, including the Nambu-Goldstone (NG) bosons, does not exceed a certain value. 
Each eigenvalue of the $s$-wave amplitude denoted as $a_0^i$ is required to satisfy: 
\begin{align}
|a_0^i| \leq \xi, 
\end{align}
where $\xi=1$~\cite{Lee:1977eg} or 1/2~\cite{Gunion:1989we}.  
We here take $\xi = 1/2$. 
We note that each $a_0^i$ only depends on the scalar quartic couplings as 
only scalar contact interactions contribute to the scattering process at the high-energy limit. 

Next, the vacuum stability bound provides an independent constraint on scalar quartic couplings. 
It requires that the Higgs potential is bounded from below in any direction with large field values. 
This condition is schematically written by 
\begin{align}
V^{(4)} >0, \label{eq:vs}
\end{align}
where $V^{(4)}$ represents quartic terms of the Higgs potential. 
Although in the SM this condition is trivially satisfied by taking the scalar quartic coupling to be positive, 
in models with non-minimal Higgs sectors it is given by a set of inequalities in terms of scalar quartic couplings~\cite{Deshpande:1977rw}. 

Furthermore, in extended Higgs sectors, wrong local 
vacua can generally appear in addition to the true vacuum giving the correct value of the Fermi constant $G_F$. 
Thus, we have to avoid parameter regions which realize the depth of such wrong vacua to be deeper than that of the true one \footnote{There is still a possibility of a meta-stable electroweak vacuum even if such situations are realized. {By taking into account such a possibility a constraint from wrong vacua is more moderate. For example, see Ref.~\cite{Branchina:2018qlf}.} }. 
The condition to avoid the wrong vacua can be written by combinations of dimensionful and dimensionless parameters in the potential~\cite{Frere:1983ag}, 
so that it can provide an independent constraint from the above two constraints. 

Apart from these theoretical constraints, we need to take into account bounds from experimental data. 
At the LEP/SLC experiments, various EW observables have been precisely measured such as the masses and widths of the weak gauge bosons. 
These precise measurements can be used to constrain new physics effects which can indirectly be entered into the self-energy diagrams for weak gauge bosons. 
Such indirect effect, so called oblique corrections, is conveniently parameterized by the $S$, $T$ and $U$ parameters introduced by Peskin and Takeuchi~\cite{Peskin:1990zt,Peskin:1991sw}, which 
are expressed in terms of two point functions of the weak bosons. 
From the global fit of EW parameters~\cite{Baak:2012kk}, new physics effects on the $S$ and $T$ parameters under $U = 0$ are constrained by
\begin{align}
S = 0.05\pm 0.09, \quad T = 0.08\pm 0.07, 
\end{align}
with the correlation factor of $+0.91$ and the reference values of the masses of SM Higgs boson and top quark being $m_h^{\text{ref}} = 126$ GeV and $m_t^{\text{ref}} = 173$ GeV, respectively.
Flavor experiments also provide important constraints on a parameter space of extended Higgs models, particularly models with a multi-doublet structure. 
We will discuss these constraints in more detail in Sec.~\ref{sec:2hdm} about THDMs.   
As mentioned in Introduction, additional scalars have been directly searched at the LHC~\cite{Khachatryan:2016yec,Aaboud:2017yyg,Sirunyan:2017nrt,Aaboud:2017itg,Aaboud:2017fgj,Khachatryan:2016cfx,Aaboud:2017cxo,Khachatryan:2015tra,Khachatryan:2016qkc,Aaboud:2017sjh},
and some of the constraints are interpreted in THDMs.
Moreover, the Higgs coupling measurements also constrain the mixing parameters in THDMs~\cite{Khachatryan:2016vau,Sirunyan:2018koj,ATLAS-CONF-2018-031}
by using the so-called $\kappa$ framework~\cite{Heinemeyer:2013tqa}.
We note that the $\kappa$ framework is constructed by the leading order (LO) relation,
and hence the interpretation of such constraints at higher-order level might not be straightforward.
The application of these constraints to each extended Higgs sector will be discussed in the following subsections. 

\subsection{Higgs singlet model \label{sec:hsm}}

The Higgs sector of the HSM is composed of an isospin doublet scalar field $\Phi$ with the hypercharge $Y = 1/2$
and a real singlet field $S$ with $Y = 0$. 
These scalar fields are parameterized as 
\begin{align}
\Phi=\left(\begin{array}{c}
G^+\\
\frac{1}{\sqrt{2}}(v+\phi+iG^0)
\end{array}\right),\quad
S=v_S^{} + s,  \label{hsm_f}
\end{align}
where $v$ is the vacuum expectation value (VEV) of the doublet filed which is related to the Fermi constant $G_F$ by 
$v = (\sqrt{2}G_F)^{-1/2}\simeq 246$ GeV, while $v_S^{}$ is the VEV 
of the singlet field.
Because the singlet field does not contribute to the EW symmetry breaking, 
the component fields $G^\pm$ and $G^0$ in the doublet field correspond to the NG bosons which are absorbed into the weak bosons.  

The most general Higgs potential is written as
 \begin{align}
V_{\text{HSM}} =&\, m_\Phi^2|\Phi|^2+\lambda_\Phi |\Phi|^4  
+\mu_{\Phi S}^{}|\Phi|^2 S+ \lambda_{\Phi S} |\Phi|^2 S^2 
+t_S^{}S +m^2_SS^2+ \mu_SS^3+ \lambda_SS^4,\label{Eq:HSM_pot}
\end{align}
where all the parameters are real. By the reparameterization of the Higgs potential, 
we can take any value of $v_S^{}$ without changing physical results~\cite{Chen:2014ask}. 
Hence, we take $v_S^{} = 0$ throughout the paper. 

In the HSM, there are two physical neutral Higgs bosons. Their mass eigenstates are defined as 
\begin{align}
\begin{pmatrix}
s \\
\phi
\end{pmatrix} = R(\alpha)
\begin{pmatrix}
H \\
h
\end{pmatrix}~~\text{with}~~R(\theta) = 
\begin{pmatrix}
c_\theta & -s_ \theta \\
s_\theta & c_\theta
\end{pmatrix},    \label{mat_r}
\end{align}
where $\alpha$ is the mixing angle, and we define the domain of $\alpha$ by 
$-\pi/2 \leq \alpha \leq \pi/2$.
Throughout the paper, we use the shorthand notation for the trigonometric function as 
$s_\theta \equiv \sin\theta$ and $c_\theta \equiv \cos\theta$. 
We identify $h$ as the discovered Higgs boson with a mass of 125 GeV. 
After solving the tadpole conditions, the squared masses of these Higgs bosons are expressed as  
\begin{align}
 &m_H^2=M_{11}^2c^2_\alpha +M_{22}^2s^2_\alpha +M_{12}^2s_{2\alpha} , \label{mbh}\\
 &m_h^2=M_{11}^2s^2_\alpha +M_{22}^2c^2_\alpha -M_{12}^2s_{2\alpha} , \label{mh}\\
 &\tan 2\alpha=\frac{2M_{12}^2}{M_{11}^2-M_{22}^2}, \label{tan2a}
 \end{align}  
where $M^2_{ij}$ ($i,j=1,2$) are the squared mass matrix elements in the basis of ($s,\phi$). 
Each element is given by 
 \begin{align}
M^2_{11}= M^2+ v^2\lambda_{\Phi S}  ,\quad M^2_{22}=2v^2\lambda_\Phi ,\quad M^2_{12}=v\mu_{\Phi S}^{}, \label{mij}
\end{align}
with $M^2 = 2m_S^2$. 
The seven parameters in the potential are then expressed by the following five input parameters 
\begin{align}
&m_H^{},~~M^2,~~\mu_S,~~\lambda_S,~~c_\alpha, \label{input:hsm}
\end{align}
{and the two parameters $m_h$ and $v$ that are fixed by experiments. }

Let us briefly discuss the other relevant parts of the Lagrangian in the HSM. 
The kinetic term is given by 
\begin{align}
{\cal L}_{\text{kin}}^{\text{HSM}} = |D_\mu \Phi|^2 +\frac{1}{2}(\partial_\mu S)^2, 
\end{align}
where $D_\mu$ is the covariant derivative for the Higgs doublet. Because the singlet field $S$ does not have the gauge interaction, 
the additional Higgs boson $H$ couples to weak bosons only through the $\phi$ component of $H$. 
Thus, the gauge-gauge-scalar type interactions are given as 
\begin{align}
{\cal L}_{\text{kin}}^{\text{HSM}} \supset gm_W^{}(c_\alpha W_\mu^+ W^{-\mu}h  + s_\alpha W_\mu^+ W^{-\mu}H) + \frac{g_Z^{}m_Z^{}}{2}(c_\alpha Z_\mu Z^\mu h + s_\alpha Z_\mu Z^\mu H), \label{eq:hsm-v}
\end{align}
where $g$ is the weak gauge coupling and $g_Z^{} = g/\cos\theta_W$ with $\theta_W$ being the weak mixing angle. 
The Yukawa interactions are written by the same form as those in the SM:
\begin{align}
{\cal L}_Y^{\text{HSM}} = -Y_u\bar{Q}_L \tilde{\Phi}\,u_R^{} - Y_d\bar{Q}_L \Phi \, d_R^{} - Y_e\bar{L}_L \Phi \, e_R^{} + \text{h.c.}, 
\end{align}
where $\tilde{\Phi} = i\sigma_2 \Phi^*$, and we do not show the flavor indices here.  
In the above equation, $Q_L$, $L_L$, $u_R$, $d_R$ and $e_R$ are respectively the left-handed quark doublet, lepton doublet, right-handed up-type quark singlet, down-type quark singlet and charged lepton singlet.  
The singlet field does not couple to fermions, so that the interaction terms for $h$ and $H$ are extracted as 
\begin{align}
{\cal L}_Y^{\text{HSM}} \supset -\frac{m_f}{v}(c_\alpha \bar{f}f h + s_\alpha \bar{f}f H). \label{eq:hsm-f}
\end{align}
As it is seen in Eqs.~(\ref{eq:hsm-v}) and (\ref{eq:hsm-f}), the SM-like Higgs boson $h$ couplings are universally suppressed by the factor of $c_\alpha$ as compared to the corresponding SM values.  

As we already mentioned at the beginning of this section, 
the parameters in the potential can be constrained by the unitarity, the vacuum stability and the condition to avoid wrong vacua. 
For the unitarity bound, there are four independent eigenvalues given in Refs.~\cite{Cynolter:2004cq,Kanemura:2016lkz}. 
In this paper, we use the expression for the eigenvalues given in Ref.~\cite{Kanemura:2016lkz}, 
where the same notation of the potential parameters as that in this paper is applied. 
The necessary and sufficient condition to satisfy the vacuum stability is given by~\cite{Pruna:2013bma}
\begin{align}
\lambda_\Phi > 0, \quad \lambda_S > 0, \quad 2\sqrt{\lambda_\Phi \lambda_S} + \lambda_{\Phi S} > 0. 
\end{align}
For the condition to avoid these wrong vacua is found in {Ref.~\cite{Espinosa:2011ax,Chen:2014ask,Lewis:2017dme}}. 
We use the expression given in Ref.~\cite{Kanemura:2016lkz}. 
In the HSM, one-loop corrected two point functions for weak bosons are found in Ref.~\cite{Lopez-Val:2014jva}. 
Imposing the bound from the $S$ and $T$ parameters, we can obtain the upper limit on $m_H^{}$ depending on the value of $c_\alpha$. 
Constraints on the mass of the additional Higgs boson and the mixing angle from the LHC data have been studied in Refs.~\cite{Robens:2015gla,Robens:2016xkb,Blasi:2017zel,Gu:2017ckc}.

\subsection{Two Higgs doublet model \label{sec:2hdm}}

The Higgs sector of the THDM is composed of two isospin doublet scalar fields $\Phi_1$ and $\Phi_2$ with $Y = 1/2$. 
These doublets are parameterized as 
\begin{align}
\Phi_i = \left(\begin{array}{c}
w_i^+\\
\frac{1}{\sqrt{2}}(v_i + h_i + iz_i)
\end{array}\right),\hspace{3mm}(i=1,2),  \label{eq:doublets}
\end{align} 
where $v_1$ and $v_2$ are the VEVs of two doublets with $v = \sqrt{v_1^2 + v_2^2}$, and their ratio is expressed by $\tan\beta = v_2/v_1$. 

Having two doublet fields with the same quantum charges causes 
dangerous flavor changing neutral currents (FCNCs) at tree level, because both doublets couple to each type of fermions. 
In order to avoid such FCNCs, we impose a discrete $Z_2$ symmetry, where two doublets transform as 
$\Phi_1 \to +\Phi_1$ and  $\Phi_2 \to -\Phi_2$. 
One can introduce the soft breaking term of the $Z_2$ symmetry in the potential, retaining the 
good property of the flavor sector. In the following, we discuss the THDM with the softly-broken $Z_2$
symmetry and the CP-conservation. 

The most general Higgs potential is given by
\begin{align}
V_{\text{THDM}}=& m_1^2|\Phi_1|^2+m_2^2|\Phi_2|^2-m_3^2(\Phi_1^\dagger \Phi_2 +\text{h.c.})\notag\\
& +\frac{\lambda_1}{2}|\Phi_1|^4+\frac{\lambda_2}{2}|\Phi_2|^4+\lambda_3|\Phi_1|^2|\Phi_2|^2+\lambda_4|\Phi_1^\dagger\Phi_2|^2
+\frac{\lambda_5}{2}\left[(\Phi_1^\dagger\Phi_2)^2+\text{h.c.}\right],  \label{eq:pot-thdm}
\end{align}
where the $m_3^2$ term softly breaks the $Z_2$ symmetry. 
The $m_3^2$ and $\lambda_5$ parameters are taken to be real as we consider the CP-conserving case. 
The scalar mass eigenstates can then be defined as follows: 
\begin{align}
\left(\begin{array}{c}
w_1^\pm\\
w_2^\pm
\end{array}\right)&=R(\beta)
\left(\begin{array}{c}
G^\pm\\
H^\pm
\end{array}\right),\quad 
\left(\begin{array}{c}
z_1\\
z_2
\end{array}\right)
=R(\beta)\left(\begin{array}{c}
G^0\\
A
\end{array}\right),\quad
\left(\begin{array}{c}
h_1\\
h_2
\end{array}\right)=R(\alpha)
\left(\begin{array}{c}
H\\
h
\end{array}\right), \label{mixing}
\end{align}
where $H^\pm$ and $A$ are the charged and CP-odd Higgs bosons, respectively, 
while $H$ and $h$ are the 
CP-even Higgs bosons. Similar to the HSM case, we identify $h$ {as} the discovered Higgs boson {with a} mass of 125 GeV.  
{We define the domain of $\alpha$ and $\beta$ to be $-\pi/2 \leq \alpha \leq 0$ and $0 \leq \beta \leq \pi/2$, respectively, 
so that the viable range for  $\beta-\alpha$ is expressed as $0\leq \beta-\alpha \leq \pi$.}

After solving two tadpole conditions for $h_1$ and $h_2$, squared masses of the charged and CP-odd 
Higgs bosons are given by 
\begin{align}
m_{H^\pm}^2 = M^2-\frac{v^2}{2}(\lambda_4+\lambda_5),\quad m_A^2&=M^2-v^2\lambda_5,  \label{mass1}
\end{align}
where $M^2=m_3^2/(s_\beta c_\beta)$ which describes the soft-breaking scale of the $Z_2$ symmetry. 
For {the} two CP-even Higgs bosons, {the} squared mass matrix elements $M_{ij}^2$ 
in the Higgs basis~\cite{Davidson:2005cw} are given by 
\begin{align}
M_{11}^2&=v^2(\lambda_1c^4_\beta+\lambda_2s^4_\beta +2\lambda_{345}s^2_{\beta}c^2_{\beta}) ,  \notag \\
M_{22}^2&=M^2 + \frac{v^2}{4}s^2_{2\beta}(\lambda_1+\lambda_2-2\lambda_{345}), \label{m22}  \\
M_{12}^2&=\frac{v^2}{2} s_{2\beta}( -\lambda_1c^2_\beta +  \lambda_2s^2_\beta  + \lambda_{345}c_{2\beta}),  \notag
\end{align}
with $\lambda_{345}\equiv \lambda_3+\lambda_4+\lambda_5$. 
The squared masses of {the} two CP-even Higgs bosons and the mixing angle $\beta-\alpha$ are  
expressed in terms of the matrix elements $M_{ij}^2$ as  
\begin{align}
&m_H^2= M_{11}^2c^2_{\beta-\alpha} +  M_{22}^2s^2_{\beta-\alpha} -  M_{12}^2s_{2(\beta-\alpha)} , \label{111}  \\
&m_h^2 =  M_{11}^2s^2_{\beta-\alpha} +  M_{22}^2c^2_{\beta-\alpha} + M_{12}^2s_{2(\beta-\alpha)},  \label{222} \\
&\tan 2(\beta-\alpha)= -\frac{2M_{12}^2}{M_{11}^2-M_{22}^2}.  \label{333}
\end{align} 
The eight parameters in the potential are then expressed by the following six input parameters 
\begin{align}
m_H^{},~~m_A^{},~~m_{H^\pm}^{},~~M^2,~~\tan\beta,~~s_{\beta-\alpha},  \label{input:thdm}
\end{align} 
and {the} two parameters $m_h$ and $v$ {that are fixed} by experiments. 
In addition to these parameters, there is a degree of freedom of the sign of $c_{\beta-\alpha}$.

Let us discuss the other relevant parts of the Lagrangian. 
The kinetic term for the Higgs doublets are written as
\begin{align}
{\cal L}_{\text{kin}}^{\text{THDM}} = |D_\mu \Phi_1|^2 + |D_\mu \Phi_2|^2. 
\end{align} 
In the mass eigenbasis of the Higgs bosons, the gauge-gauge-scalar type interaction terms are extracted as     
\begin{align}
{\cal L}_{\text{kin}}^{\text{THDM}} \supset gm_W^{}(s_{\beta-\alpha} W_\mu^+ W^{-\mu}h  + c_{\beta-\alpha} W_\mu^+ W^{-\mu}H) + \frac{g_Z^{}m_Z^{}}{2}(s_{\beta-\alpha} Z_\mu Z^\mu h  + c_{\beta-\alpha} Z_\mu Z^\mu H). 
\end{align}

\begin{table}[t]
\begin{center}
{\renewcommand\arraystretch{1.2}
\begin{tabular}{l|ccccccc|ccc}\hline\hline
&$\Phi_1$&$\Phi_2$&$Q_L$&$L_L$&$u_R$&$d_R$&$e_R$&$\zeta_u$ &$\zeta_d$&$\zeta_e$ \\\hline
Type-I &$+$&
$-$&$+$&$+$&
$-$&$-$&$-$&$\cot\beta$&$\cot\beta$&$\cot\beta$ \\\hline
Type-II&$+$&
$-$&$+$&$+$&
$-$
&$+$&$+$& $\cot\beta$&$-\tan\beta$&$-\tan\beta$ \\\hline
Type-X (lepton specific)&$+$&
$-$&$+$&$+$&
$-$
&$-$&$+$&$\cot\beta$&$\cot\beta$&$-\tan\beta$ \\\hline
Type-Y (flipped) &$+$&
$-$&$+$&$+$&
$-$
&$+$&$-$& $\cot\beta$&$-\tan\beta$&$\cot\beta$ \\\hline\hline
\end{tabular}}
\caption{$Z_2$ charge assignments in four types of Yukawa interactions, 
and the $\zeta_f$ ($f=u,d,e$) factors appearing in Eq.~(\ref{eq:yuk-thdm}). 
}
\label{tab:z2}
\end{center}
\end{table}
The most general Yukawa interactions under the $Z_2$ symmetry are written as 
\begin{align}
{\cal L}_Y^{\text{THDM}} 
&= -Y_u\bar{Q}_L \tilde{\Phi}_i \,u_R^{} - Y_d\bar{Q}_L \Phi_j \, d_R^{} - Y_e\bar{L}_L \Phi_k \, e_R^{} + \text{h.c.}, 
\end{align}
where the subscripts $i$, $j$ and $k$ are 1 or 2. 
These indices are fixed when we determine the $Z_2$ charge for right-handed fermions. 
As in Table~\ref{tab:z2}, there are four independent types of Yukawa interactions depending on the assignment of the 
$Z_2$ charge~\cite{Barger:1989fj,Grossman:1994jb,Aoki:2009ha}. 
The interaction terms for the physical Higgs bosons are then extracted as 
\begin{align}
{\cal L}_Y^{\text{THDM}} \supset
&-\sum_{f=u,d,e}\frac{m_f}{v}\left[ (s_{\beta-\alpha} + \zeta_f c_{\beta-\alpha})\bar{f}fh
+ (c_{\beta-\alpha} - \zeta_f s_{\beta-\alpha}){\overline f}fH-2iI_f \zeta_f\bar{f}\gamma_5fA\right)\notag\\
&-\frac{\sqrt{2}}{v}\left[V_{ud}\bar{u}
\left(m_d\zeta_d\,P_R-m_u\zeta_uP_L\right)d\,H^+
+m_e\zeta_e\bar{\nu}P_Re^{}H^+ +\text{h.c.}\right],  \label{eq:yuk-thdm}
\end{align}
with $I_f = 1/2\,(-1/2)$ for $f = u\,(d,e)$ and $V_{ud}$ is the Cabibbo-Kobayashi-Maskawa matrix element. 

Similar to the HSM, parameters in the THDMs can be constrained by both the theoretical and experimental constraints. 
For the unitarity bound, there are 12 independent eigenvalues of the 
$s$-wave amplitude matrix~\cite{Kanemura:1993hm,Akeroyd:2000wc,Ginzburg:2005dt,Kanemura:2015ska}. 
We use the expression for the eigenvalues given in Ref.~\cite{Kanemura:2015mxa}. 
The vacuum stability bound is sufficiently and necessarily satisfied by imposing the following {conditions}~\cite{Deshpande:1977rw, Sher:1988mj, Nie:1998yn, Kanemura:1999xf} 
\begin{align}
\lambda_1 > 0,\quad 
\lambda_2 > 0,\quad 
\sqrt{\lambda_1\lambda_2} + \lambda_3 + \text{MIN}(0,\lambda_4+\lambda_5,\lambda_4 - \lambda_5) > 0. 
\end{align}
In addition, the wrong vacua can be avoided by taking $M^2 \geq 0$~\cite{Branchina:2018qlf}. We thus only take the positive value of $M^2$ in the following discussion. 
The expressions of the two point functions for the weak bosons in the THDM are found in Refs.~\cite{Toussaint:1978zm,Bertolini:1985ia,Peskin:2001rw,Grimus:2008nb,Kanemura:2011sj}. 
Constraints on the parameters in THDMs from the LHC data have been discussed in Refs.~\cite{Bernon:2015qea,Dorsch:2016tab,Han:2017pfo,Arbey:2017gmh,Chang:2015goa,Blasi:2017zel,Gu:2017ckc}.

Differently from the HSM, constraints from flavor experiments are important to be taken into account in the THDM. 
These bounds particularly  provide the lower limit on the mass of {the} charged Higgs boson $m_{H^\pm}$ depending on the type of Yukawa interaction and $\tan\beta$. 
For example, from the $B_s \to X_s\gamma$ data, $m_{H^\pm}$ hs to be greater than about 600 GeV at 95\% confidence level in the Type-II and Type-Y THDMs with $\tan\beta \gtrsim 2$, while 
${\cal O}(100)$ GeV of $m_{H^\pm}$ is allowed in the Type-I and Type-X THDMs with $\tan\beta \gtrsim 2$~\cite{Misiak:2017bgg}.
Constraints on $m_{H^\pm}$ and $\tan\beta$ from various flavor observables are also shown in Ref.~\cite{Haller:2018nnx} 
in four types of the THDMs. 

\subsection{Inert doublet model}

The contents of the scalar sector in the IDM are the same as those in the THDM. 
In the THDM, the $Z_2$ symmetry can be softly-broken by introducing the $m_3^2$ term in the potential, 
while in the IDM it is assumed to be unbroken even after the EW symmetry breaking. 
Thus, the potential is obtained from Eq.~(\ref{eq:pot-thdm}) with $m_3^2 = 0$. 
In addition, the second Higgs doublet $\Phi_2$ is supposed not 
to develop the nonzero VEV, otherwise the $Z_2$ symmetry is spontaneously broken.

In the IDM, {the} component scalar fields of $\Phi_1$ and $\Phi_2$ do not mix with each other due to the unbroken $Z_2$ symmetry. 
Therefore, we can identify these scalar bosons $(w_2^\pm,z_2,h_2,h_1)$ defined in Eq.~(\ref{eq:doublets}) 
with the mass eigenstates $(H^\pm,A,H,h)$. 
The lightest neutral inert scalar boson can be a candidate of dark matter as it cannot decay into a pair of SM particles. 

The mass formulae for the scalar bosons are changed from those of the THDMs, not just because of the absence of the $m_3^2$ term, but
also the absence of the tadpole condition for $h_2$. 
These are given as follows: 
\begin{align}
m_h^2 &= \lambda_1 v^2,\\
m_{H}^2 &= M^2 + \frac{v^2}{2}(\lambda_3+\lambda_4+\lambda_5),\\
m_{A}^2 &= M^2 + \frac{v^2}{2}(\lambda_3+\lambda_4-\lambda_5), \\
m_{H^\pm}^2 &= M^2 + \frac{v^2}{2}\lambda_3,
\end{align}
where $M^2 = m_2^2$. 
We then choose the following five {parameters to be free input parameters} of the IDM
\begin{align}
m_H^{},~~m_A^{},~~m_{H^\pm}^{},~~M^2,~~\lambda_2,   \label{input:idm}
\end{align}
{and the two fixed} parameters $m_h$ and $v$. 

The same conditions for the perturbative unitarity and vacuum stability in the THDM can be applied to 
the IDM, because these bounds are given in terms of the scalar quartic couplings. 
The condition to guarantee the inert vacuum with $(\langle \Phi_1^0 \rangle, \langle \Phi_2^0 \rangle = (v/\sqrt{2},0))$ is given by~\cite{Ginzburg:2010wa}, 
\begin{align}
\frac{m_1^2}{\sqrt{\lambda_1}} < \frac{M^2}{\sqrt{\lambda_2}} . \label{eq:wv_idm}
\end{align}
Since the tadpole condition makes $m_1^2$ negative, and the vacuum stability condition constraints $\lambda_1$ and $\lambda_2$ to be positive, 
the condition given in Eq.~(\ref{eq:wv_idm}) is satisfied by taking $M^2 > 0$. 
We refer {to} this condition as the one to avoid wrong vacua, according to the other two models discussed above. 
For the constraints of the $S$ and $T$ parameters, we can use the same {expressions} as those in the THDM with $s_{\beta-\alpha} = 1$. 

In the IDM, constraints on the masses of the additional Higgs bosons from collider experiments are relatively weak
since the additional scalars do not couple to SM fermions.
The constraints from the LEP and the LHC have been studied in Refs.~\cite{Lundstrom:2008ai,Gustafsson:2010zz} and Refs.~\cite{Belanger:2015kga,Belyaev:2016lok}, respectively.
Dark matter constraints from relic density and direct detection also limit the parameter space;
see, e.g., Refs.~\cite{Belyaev:2016lok,Ilnicka:2018def} for details.

\section{Decay rates of the SM-like Higgs boson at one loop\label{sec:decay}}

In this section, we discuss the decay rates of the SM-like Higgs boson; i.e., 
$h \to f\bar{f}$, 
$h \to ZZ^* \to Zf\bar{f}$ and  $h \to WW^* \to Wf\bar{f}'$ at NLO in EW and QCD. 
The loop induced decay rates $h \to \gamma\gamma$, $h \to Z\gamma$ and $h \to gg$ are also discussed at NLO in QCD. 

We outline our one-loop calculations.  
For the computation of the EW corrections, we adopt the modified on-shell renormalization scheme which has been defined in Ref.~\cite{Kanemura:2017wtm}, while 
for the QCD corrections we apply the $\overline{\text{MS}}$ scheme. 
In the on-shell scheme, all the counterterms appearing in the decay rates of the $h \to f\bar{f}$ and $h \to VV^* \to Vf\bar{f}$ modes are determined in terms of the 
one particle irreducible (1PI) diagrams for one- and two-point functions of Higgs bosons, gauge bosons and fermions by imposing a set of the renormalization conditions. 
Adding these counterterms, one can obtain the ultra-violet (UV) finite one-loop corrected vertices. 

The on-shell scheme is a physically quite natural renormalization scheme, and it is suitable to apply to EW observables as they include well defined scales such as the weak boson masses. 
However, it has been known that the on-shell scheme introduces gauge dependent counterterms, particularly in some mixing parameters~\cite{Yamada:2001px}. 
In extended Higgs sectors, a mixing between Higgs bosons can generally appear. 
We thus apply the so-called pinch technique to remove the gauge dependence in the renormalized vertex functions to our computations~\cite{Bojarski:2015kra,Krause:2016oke,Kanemura:2017wtm}.

Apart from the UV divergences, there appear infrared (IR) divergences when we calculate virtual photon loop contributions.
Such IR divergences can exactly be cancelled by adding contributions of real photon emissions, where 
the finite QED corrections are common to those in the SM.
The analytic expressions of QED corrections are known for $h\to f\bar f$~\cite{Kniehl:1991ze,Dabelstein:1991ky,Bardin:1990zj} and $h\to Zf\bar{f}$~\cite{Kniehl:1993ay}. 
Thus, we simply switch off the photon-loop contributions, and use these analytic formulae in our computation as the QED correction part.
For $h\to Wf\bar{f}'$, on the other hand, 
we cannot separate EW corrections into QED and weak corrections.
Therefore, by using the phase-space slicing method~\cite{Harris:2001sx},
we numerically evaluate both the virtual and real corrections to $h\to Wf\bar{f}'$, see Appendix~\ref{sec:real_photon} for details. 
For QCD corrections, we use the similar technique to remove IR divergences coming from virtual gluon loop contributions. 

In our renormalization calculation, we choose 
the fine structure constant $\alpha_{\rm em}$, the Fermi constant $G_F$ and the $Z$ boson mass $m_Z$ as the input parameters for the EW parameters.
In this scheme, all the other EW parameters such as $v$, $m_W^{}$ and $s_W^{}$ are outputs. 
Using the on-shell definition of the weak mixing angle; i.e., $s_W^2 = 1 - m_W^2/m_Z^2$~\cite{Sirlin:1980nh} and 
the modified relation among the EW parameters:
\begin{align}
G_F =\frac{\pi \alpha_{\text{em}}}{\sqrt{2}s_W^2m_W^2}\frac{1}{1-\Delta r}=\frac{1}{\sqrt{2}v^2}\frac{1}{1-\Delta r}, \label{mod-gf}
\end{align} 
we can calculate the renormalized squared W boson mass as
\begin{align}
(m_W^2)_{\rm reno} = \frac{m_Z^2}{2}\left[1+\sqrt{1-\frac{4\pi\alpha_{\rm em}}{\sqrt{2}G_Fm_Z^2(1-\Delta r)}} \right]. \label{rmw}
\end{align} 
In Eq.~(\ref{mod-gf}), $\Delta r$ is calculated by~{\cite{Sirlin:1980nh}}
\begin{align}
\Delta r = \frac{\text{Re}\hat{\Pi}_{WW}(0)}{m_W^2} + \frac{\alpha_{\text{em}}}{4\pi s_W^2}\left(6 + \frac{7-4s_W^2}{2s_W^2}\log c_W^2 \right), \label{delr}
\end{align} 
{where $\hat{\Pi}_{WW}$ is the renormalized $W$ boson two-point function and the second term corresponds to vertex corrections and box diagram contributions to the muon decay rate.} 
Including these three EW parameters, we choose the following parameters as the SM inputs: 
\begin{align}
\alpha_{\rm em},~~m_Z,~~G_F,~~\Delta\alpha_{\rm em},~~\alpha_s,~~m_t,~~m_b,~~m_c,~~m_\tau,~~m_h, \label{inputs}
\end{align}
where $\Delta \alpha_{\rm em}$ is the shift of the fine structure constant $\alpha_{\text{em}}$ from zero energy to $m_Z$. 
We also input the parameters in the potential given in Eqs.~(\ref{input:hsm}), (\ref{input:thdm}) and (\ref{input:idm}) in the HSM, THDMs and IDM, respectively. 
We note that the parameters $c_{\alpha}$ and $s_{\beta-\alpha}$ in the HSM and THDMs, respectively, do not physically mean the mixing angles for the CP-even Higgs bosons after the renormalization.

\subsection{Renormalized vertices\label{sec:ren-vert}}

Important ingredients for calculations of decay rates of the Higgs boson are renormalized Higgs boson vertices. 
In our computations, the $hf\bar{f}$, $hV^\mu V^{\nu}$ $(V=W,Z)$ and $h {\cal V}^\mu {\cal V}^{\prime\nu}$ $({\cal V V'} = \gamma\gamma,~Z\gamma,~gg)$ vertices are relevant, 
where the $h {\cal V}^\mu {\cal V}^{\prime\nu}$ vertices are one-loop induced. 
Each of these vertices can be decomposed into several form factors depending on their Lorentz structure as written below. 
The {\tt H-COUP} program (ver. 1.0)~\cite{Kanemura:2017gbi} provides numerical values of these renormalized form factors in the extended Higgs sectors.\footnote{
We are now preparing the next version of {\tt H-COUP} program (ver. 2.0)~\cite{future} providing numerical values of decay rates of $h$ at NLO which are calculated in this paper. } 
We fully use {\tt H-COUP} in our numerical evaluation of the form factors.

The renormalized $hf\bar{f}$ vertices can be decomposed into the following form factors: 
\begin{align}
\hat{\Gamma}_{h ff}(p_1^2,p_2^2,q^2)&=
\hat{\Gamma}_{h ff}^S+\gamma_5 \hat{\Gamma}_{h ff}^P+p_1\hspace{-3.5mm}/\hspace{2mm}\hat{\Gamma}_{h ff}^{V_1}
+p_2\hspace{-3.5mm}/\hspace{2mm}\hat{\Gamma}_{h ff}^{V_2}\notag\\
&\quad +p_1\hspace{-3.5mm}/\hspace{2mm}\gamma_5 \hat{\Gamma}_{h ff}^{A_1}
+p_2\hspace{-3.5mm}/\hspace{2mm}\gamma_5\hat{\Gamma}_{h ff}^{A_2}
+p_1\hspace{-3.5mm}/\hspace{2mm}p_2\hspace{-3.5mm}/\hspace{2mm}\hat{\Gamma}_{h ff}^{T}
+p_1\hspace{-3.5mm}/\hspace{2mm}p_2\hspace{-3.5mm}/\hspace{2mm}\gamma_5\hat{\Gamma}_{h ff}^{PT}, \label{eq:hff-form}
\end{align}
where $p_1^\mu$ $(p_2^\mu)$ is the incoming four-momentum of the fermion (anti-fermion), 
and $q^\mu  (= p_1^\mu + p_2^\mu)$ is the outgoing four-momentum of the Higgs boson. 
For the case with on-shell fermions; i.e., $p_1^2 = p_2^2 = m_f^2$, the following relations hold: 
\begin{align}
\hat{\Gamma}_{h ff}^{P} &= \hat{\Gamma}_{h ff}^{PT} = 0, \quad 
\hat{\Gamma}_{h ff}^{V_1} = -\hat{\Gamma}_{h ff}^{V_2},\quad 
\hat{\Gamma}_{h ff}^{A_1} = -\hat{\Gamma}_{h ff}^{A_2}. 
\end{align}
These relations are used for the calculation of the Higgs boson decay into fermions discussed in Sec.~\ref{sec:hff}. 

Next, the renormalized $hV^\mu V^\nu$ vertices are defined in terms of three renormalized form factors:
\begin{align}
\hat{\Gamma}_{hVV}^{\mu\nu}(p_1^2,p_2^2,q^2)&=g^{\mu\nu}\hat{\Gamma}_{h VV}^1 + \frac{p_1^\nu p_2^\mu}{m_V^2}\hat{\Gamma}_{h VV}^2 + i\epsilon^{\mu\nu\rho\sigma}\frac{p_{1\rho} p_{2\sigma}}{m_V^2}\hat{\Gamma}_{h VV}^3,  
\end{align}
where $p_1^\mu$ and $p_2^\mu$ are incoming four-momenta of the weak bosons, and $q^\mu$ is the outgoing four-momentum of the Higgs boson. 
Similarly, we can define the loop induced vertices as 
\begin{align}
\hat{\Gamma}_{h{\cal V V'}}^{\mu\nu}(p_1^2,p_2^2,q^2)&=g^{\mu\nu}\hat{\Gamma}_{h {\cal V V'}}^1
+\frac{p_1^\nu p_2^\mu}{q^2}\hat{\Gamma}_{h {\cal V V'}}^2
+i\epsilon^{\mu\nu\rho\sigma}\frac{p_{1\rho} p_{2\sigma}}{q^2}\hat{\Gamma}_{h {\cal V V'}}^3.   
\end{align}
For the on-shell photon and gluon with a four-momentum $p_i^\mu$, the Ward identity holds, i.e, $p_{i\mu}\hat{\Gamma}_{h{\cal V V}'}^{\mu\nu} = 0$. 
This gives the following relation 
\begin{align}
\hat{\Gamma}_{h {\cal V V'}}^2 = -\frac{q^2}{p_1\cdot p_2}\hat{\Gamma}_{h {\cal V V'}}^1. \label{ward}
\end{align}
This relation can be applied to the computation of the loop induced decays of the Higgs boson. 
For an off-shell photon appearing in the $h \to Z\gamma^* \to Zf\bar{f}$ decay mode at NLO, 
Eq.~(\ref{ward}) cannot be used, so that $\hat{\Gamma}_{h {\cal V V'}}^1$ and $\hat{\Gamma}_{h {\cal V V'}}^2$ separately appear, 
as it will be discussed in Sec.~\ref{sec:hzz}. 
We note that the form factor $\hat{\Gamma}_{h {\cal V V}'}^3$ is non-zero only when the Higgs boson is a CP-mixed state. 
In this paper, we consider the case with CP-conservation in the Higgs sector, so that this form factor becomes zero. 

All the renormalized form factors for the $hXX$ vertices defined above
are further decomposed into the tree level and one-loop level parts as follows:
\begin{align}
\hat{\Gamma}^i_{hXX}(p_1^2,p_2^2,q^2)&=\Gamma^{i,{\rm tree}}_{hXX} + \Gamma^{i,{\rm loop}}_{hXX}(p_1^2,p_2^2,q^2) . \label{eq:form-loop}
\end{align}
The tree level contribution to each form factor{,} denoted as $\Gamma^{i,{\rm tree}}_{hXX}${,} is given as 
\begin{align}
\Gamma^{S,{\rm tree}}_{hff} = \kappa_f m_f(\sqrt{2}G_F)^{1/2},\quad
\Gamma^{1,{\rm tree}}_{hVV} = 2\kappa_V m_V^2(\sqrt{2}G_F)^{1/2}, \label{form_tree}
\end{align}
where the scaling factors $\kappa_f^{}$ and $\kappa_V^{}$ are given in Table~\ref{tab:kappa} for each extended Higgs model.  
All the other form factors are zero at tree level. 
The one-loop contributions ($\Gamma^{i,{\rm loop}}_{hXX}$) are decomposed by 
\begin{align}
\Gamma^{i,{\rm loop}}_{hXX}(p_1^2,p_2^2,q^2) &= \Gamma^{i,{\rm 1PI}}_{hXX}(p_1^2,p_2^2,q^2) + \delta \Gamma^i_{hXX}. \label{gam_hff_loop}
\end{align}
The first and second terms of the right-hand side are the contribution from 1PI diagrams and counterterms, respectively. 
As we mentioned at the beginning of this section, counterterms are determined by a set of on-shell conditions by which 
these are written in terms of 1PI diagrams for one- and two-point functions with a fixed value of squared momenta. 
Analytic expressions for the contributions from 1PI diagrams and counterterms to these renormalized Higgs boson vertices 
are presented in {Ref.~\cite{Kanemura:2015fra} for the HSM, Refs.~\cite{Kanemura:2015mxa, Kanemura:2018esc} for the THDMs, and Refs.~\cite{Kanemura:2016sos,Kanemura:2018esc} for the IDM.} 

For the computation of the partial decay rates of $h \to VV^* \to Vf\bar{f}$, 
we also need to calculate the one-loop corrected $Vf\bar{f}$ vertices and box diagrams in addition to the above Higgs boson vertices. 
In the massless limit for the external fermions, the renormalized $Vf\bar{f}$ vertices are the same as those in the SM, while 
the contribution from the box diagrams is simply given by the SM expression multiplied by the scaling factor $\kappa_V^{}$. 
For the completeness, we present the analytic expressions for one-loop corrections to the $Vf\bar{f}$ vertices in Appendix~\ref{sec:vff} and those for the box corrections in Appendix~\ref{sec:box}. 

\begin{table}[t]
\begin{center}
\begin{tabular}{c|ccccccc|}\hline\hline
& HSM & THDMs & IDM \\\hline
$\kappa_f$ & $c_\alpha$ &$s_{\beta-\alpha} + \zeta_f c_{\beta-\alpha}$ & 1\\\hline
$\kappa_V$ & $c_\alpha$ &$s_{\beta-\alpha}$ & 1 \\\hline\hline
\end{tabular}
\caption{Scaling factors for the Higgs boson couplings to fermions ($\kappa_f^{}$) and weak bosons ($\kappa_V^{}$) in the extended Higgs models at tree level. 
The $\zeta_f$ factor in the THDMs is given in Table~\ref{tab:z2}. }
\label{tab:kappa}
\end{center}
\end{table}

\subsection{$h \to f\bar{f}$\label{sec:hff}}

At NLO, the partial decay rate of the $h\to f\bar{f}$ $(f \neq t)$ process can be written in terms of the EW correction part $\Delta_{\text{EW}}^f$ 
and the QCD correction part $\Delta_{\text{QCD}}^f$ as 
\begin{align}
\Gamma(h\to f\bar{f})=\Gamma_0(h\to f\bar{f}) \left[1 + \Delta_{\text{EW}}^f +  \Delta_{\text{QCD}}^f \right], 
\end{align} 
where $\Gamma_0$ is the decay rate at LO expressed as 
\begin{align}
\Gamma_0(h\to f\bar{f}) = \frac{N_c^f}{8\pi}m_h(\Gamma_{hff}^{S,\text{tree}})^2\left(1-\frac{4m_f^2}{m_h^2}\right)^{3/2}, \label{eq:hff-lo}
\end{align}
with $N_c^f$ being the color factor{,} i.e.,  $N_c^f = 3\,(1)$ for $f$ to be quarks (leptons). 
The expression for the tree level form factor $\Gamma_{hff}^{S,\text{tree}}$ is given in Eq.~(\ref{form_tree}). 
The EW corrections $\Delta_{\text{EW}}^f$ can be further decomposed into weak corrections $\Delta_{\text{weak}}^f$ and QED corrections $\Delta_{\text{QED}}^f$ as:  
\begin{align}
\Delta_{\text{EW}}^f = \Delta_{\text{weak}}^f + \Delta_{\text{QED}}^f. 
\end{align}
Here, the weak correction means contributions from $W$, $Z$ and scalar boson loops, namely all the loop contributions except for the photon and gluon loops. 
We also use this terminology in the later discussion. 
The expression for $\Delta_{\text{weak}}^f$ is given in terms of the form factors defined in Eqs.~(\ref{eq:hff-form}) and (\ref{eq:form-loop}) as 
\begin{align}
\Delta_{\text{weak}}^f =  \frac{2}{\Gamma_{hff}^{S,{\rm tree}}}{\rm Re}\left\{\left[\Gamma_{hff}^{S,{\rm loop}}
+2m_f\Gamma^{V_1,{\rm loop}}_{hff} 
+m_h^2\left(1-\frac{m_f^2}{m_h^2}\right)\Gamma_{hff}^{T,{\rm loop}}\right](m_f^2,m_f^2,m_h^2)\right\} - \Delta r , \label{eq:delew}
\end{align} 
where $\Delta r$ is given in Eq.~(\ref{delr}).

The QED (QCD) corrections are obtained by taking into account contributions from the virtual photon (gluon) loops and
those from the real photon (gluon) emissions.  
We can obtain simple expressions for these corrections by neglecting the term proportional to $m_f^2/m_h^2$.\footnote{In the numerical computation, 
we use the exact formula for the QED correction with the $m_f^2/m_h^2$ term, which is given in Ref.~\cite{Dabelstein:1991ky}. } 
For the leptonic decays, $f = \ell$, {the QED correction in the on-shell scheme is given by}~\cite{Kniehl:1991ze,Dabelstein:1991ky,Bardin:1990zj}
\begin{align}
\Delta_{\text{QED}}^\ell =  \frac{\alpha_{\text{em}}}{\pi}Q_\ell^2\left(\frac{9}{4} + \frac{3}{2}\log\frac{m_\ell^2}{m_h^2} \right). \label{eq:del_qed_f}
\end{align} 
%
For the hadronic decays, $f = q$, {the QED and QCD corrections in the $\overline{\text{MS}}$ scheme~\cite{Mihaila:2015lwa} with the renormalization scale $\mu$ are given by} 
\begin{align}
\Delta_{\text{QED}}^q =  \frac{\alpha_{\text{em}}}{\pi}Q_q^2\left(\frac{17}{4}  +\frac{3}{2}\log\frac{\mu^2}{m_h^2} \right),\quad 
\Delta_{\text{QCD}}^q = \frac{\alpha_s(\mu)}{\pi}C_F\left(\frac{17}{4} +\frac{3}{2}\log\frac{\mu^2}{m_h^2}\right), \label{eq:del_qcd_f}
\end{align}
where $C_F = 4/3$. In the numerical evaluation, we set $\mu = m_h$, and replace the quark mass in the tree level form factor $\Gamma_{hff}^{S,{\rm tree}}$ in Eq.~(\ref{eq:hff-lo}) by 
the running mass $\bar{m}_q (\mu = m_h)$. 
From Eqs.~(\ref{eq:del_qed_f}) and (\ref{eq:del_qcd_f}), we can see that there is no additional Higgs boson mass dependence in their expression, so that 
these corrections do not provide deviations in the Higgs boson decay rate from the SM prediction at NLO. 

\subsection{$h \to ZZ^* \to Zf\bar{f}$ \label{sec:hzz}}

\begin{figure}[!t]
        \centering
        \includegraphics[scale=0.2]{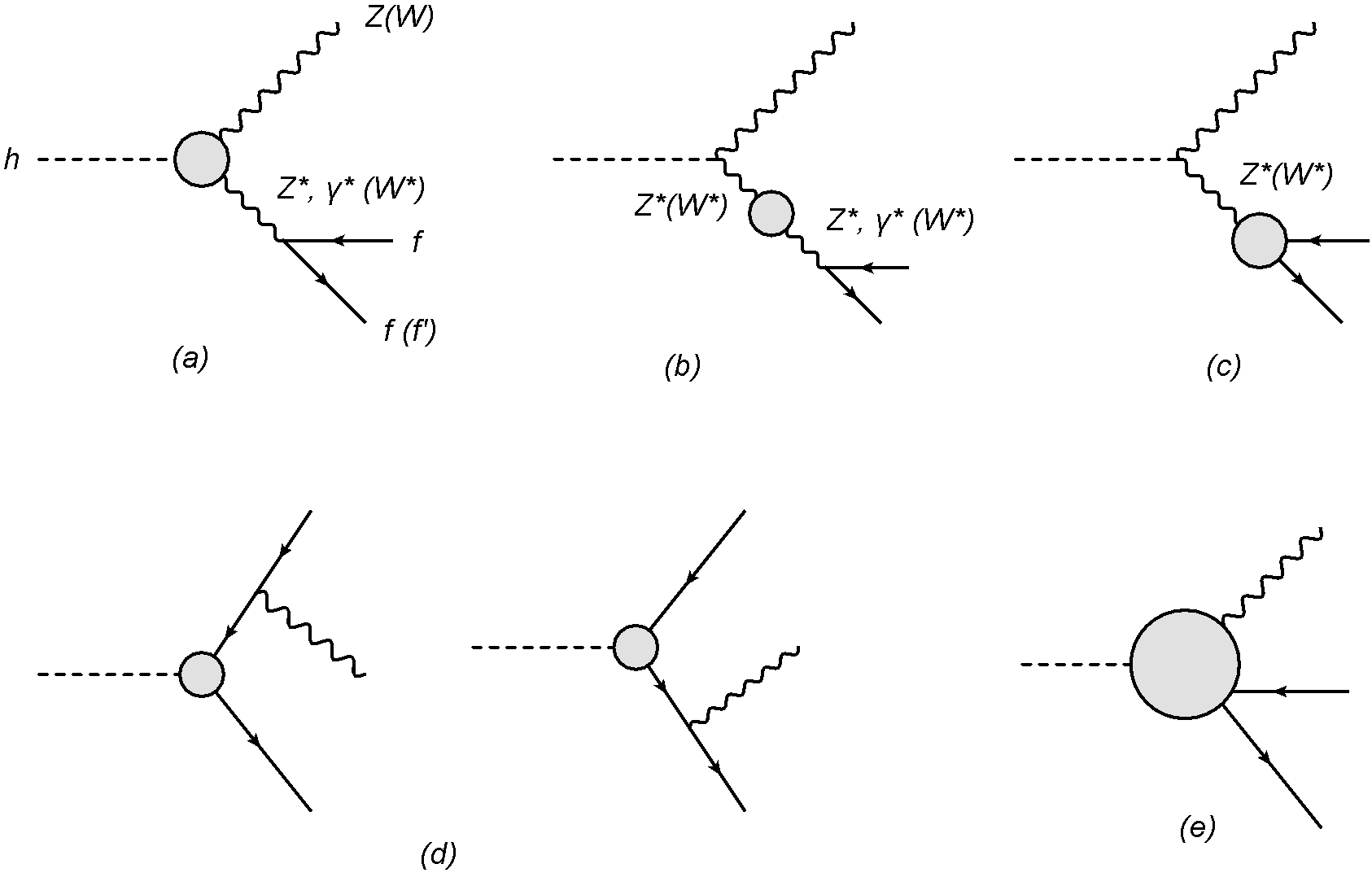}
     \vspace{1.0cm}
      \caption{Diagrams contributing to the $h\to ZZ^\ast\to Zf\bar{f}$ ($h\to WW^\ast\to Wf\bar{f}'$) mode at NLO. Each diagram denotes
the contributions from $hVV$ vertex corrections (a),  oblique corrections (b), $Vf\bar{f}$ vertex corrections  (c),  $hf\bar{f}$ vertex corrections 
(d) and box corrections (e). }
\label{FIG:hzff}
\end{figure}

We calculate the partial decay rate of the Higgs boson into a pair of weak bosons at NLO in this and next subsections. 
Because the mass of the discovered Higgs boson is about 125 GeV, at least one of the weak bosons must be off-shell. We thus calculate the process with 3-body final states{,} i.e., 
$h \to VV^* \to Vf\bar{f}$. 
Throughout this paper, we neglect the masses of external fermions in the $h \to VV^* \to Vf\bar{f}$ processes. 
In Fig.~\ref{FIG:hzff}, all the diagrams contributing to the process are shown. 

Similar to the $h \to f \bar{f}$ mode, the decay rate for the $h\to ZZ^\ast \to Z f\bar{f}$ mode at NLO is expressed as  
\begin{align}
\Gamma(h\to Z f\bar{f}) =\Gamma_0(h\to Z f\bar{f})\left[1+\Delta_{\rm EW}^Z  +\Delta_{\rm QCD}^Z  \right], 
\end{align}
and the EW correction can separately be expressed by the weak corrections and the QED corrections: 
\begin{align}
\Delta_{\rm EW}^Z = \Delta_{\rm weak}^Z + \Delta_{\rm QED}^Z. 
\end{align}
The LO contribution to the decay rate of $h\to Z f\bar{f}$ is expressed by
\begin{align}
 \Gamma_0(h\to Z f\bar{f}) &=  \int_0^{(m_h-m_Z)^2} |\overline{{\cal M}_0^Z}|^2\, ds,  \label{eq:gam0}
\end{align}
where $s$ is the Mandelstam variable defined by $(p_f^\mu + p_{\bar{f}}^\mu)^2$, and another variable $u$ defined by $(p_Z^\mu + p_{\bar{f}}^\mu)^2$ is already
integrated out in the squared tree level amplitude $|\overline{{\cal M}_0^Z}|^2$ expressed as 
\begin{align}
|\overline{{\cal M}_0^Z}|^2 =  \frac{g_Z^2(\Gamma_{hZZ}^{1,\text{tree}})^2}{256\pi^3m_h^3}\frac{v_f^2 + a_f^2}{(x_s - x_Z)^2}\frac{\lambda(x_Z,x_s)+12 x_Z x_s}{3x_Z}\lambda^{1/2}(x_Z,x_s), 
\label{eq:m0sq}
\end{align}
with $x_Z^{} = m_Z^2/m_h^2$, $x_s = s/m_h^2$ and $\lambda(x,y) = (1-x-y)^2-4xy$. 
The tree level form factor $\Gamma_{hZZ}^{1,\text{tree}}$ is given in Eq.~(\ref{form_tree}), and that for the $Zf\bar{f}$ vertex 
$v_f$ and $a_f$ is given in Eq.~(\ref{eq:vff_tree}). 

The QED ($\Delta_{\rm QED}^Z$) and QCD ($\Delta_{\rm QCD}^Z$) corrections only enter the $Zf\bar{f}$ vertex correction depicted {in} the diagram (c) in Fig.~\ref{FIG:hzff}. 
Their expressions are common to the SM given as~\cite{Kniehl:1993ay} 
\begin{align}
\Delta_{\rm QED}^Z = Q_f^2\frac{3\alpha_{\text{em}}}{4\pi},\quad 
\Delta_{\rm QCD}^Z = C_F\frac{3\alpha_s(\mu)}{4\pi}. \label{eq:qcd}
\end{align}
Although the diagrams (d) and (e) can also receive QED and QCD corrections, they vanish in the massless limit for the external fermions. 

The weak corrections $\Delta_{\rm weak}^Z$ are given by
\begin{align}
\Delta_{\rm weak}^Z & = \frac{2}{\Gamma_0}\int_0^{(m_h-m_Z)^2}ds\, |\overline{{\cal M}_0^Z}|^2 \Bigg\{
\text{Re}\left[ \frac{\Gamma_{hZZ}^{1,{\rm loop}}}{\Gamma_{hZZ}^{1,{\rm tree}}} + \frac{\bar{\lambda}(x_Z,x_s)}{x_Z^{}} \frac{\Gamma_{hZZ}^{2,{\rm loop}} }{\Gamma_{hZZ}^{1,{\rm tree}}}\right](m_Z^2,s,m_h^2) \notag \\ 
& +\frac{v_fQ_fc_Ws_W}{v_f^2+a_f^2}\frac{s - m_Z^2}{s} \text{Re}\left[\frac{\hat{\Gamma}_{hZ\gamma}^1}{\Gamma_{hZZ}^{1,{\rm tree}}} 
+ \bar{\lambda}(x_Z,x_s)\frac{\hat{\Gamma}_{hZ\gamma}^{2}}{\Gamma_{hZZ}^{1,{\rm tree}}} \right](m_Z^2,s,m_h^2) \notag\\
&+\frac{\text{Re}[v_f\Gamma^{V,{\rm loop}}_{Zff} + a_f\Gamma_{Zff}^{A,{\rm loop}}](0,0,s)}{v_f^2 + a_f^2} 
-\frac{\text{Re}\,\hat{\Pi}_{ZZ}(s)}{s-m_Z^2} -   \frac{v_fQ_fs_W^{}c_W^{}}{v_f^2 + a_f^2}\frac{\text{Re}\,\hat{\Pi}_{Z\gamma}(s)}{s} \Bigg\}\notag \\
& + \frac{1}{\Gamma_0}\int_0^{(m_h-m_Z)^2}ds \int_{u_{\text{min}}}^{u_{\text{max}}} du\, \text{Re}\left(T_{hff}^Z + B_Z\right) \notag\\
& - 2\Delta r - {\rm Re}\hat{\Pi}_{ZZ}^\prime(m_Z^2),  \label{eq:del_ew_z}
\end{align}
where
\begin{align}
 \bar{\lambda}(x,y) = \frac{1- x - y}{2}\frac{\lambda(x,y)}{\lambda(x,y) +12 x y}. 
\end{align}
The similar expression in the SM is found in Ref.~\cite{Kniehl:1993ay}. 
The first and second lines correspond to the contribution from the diagram (a) in Fig.~\ref{FIG:hzff}, where 
$\hat{\Gamma}_{hZ\gamma}^{1,2}$ are the renormalized form factors for the loop induced $hZ\gamma$ vertex. 
The analytic expressions for $\hat{\Gamma}_{hZ\gamma}^{1,2}$ are presented in Appendix~\ref{sec:hgg}. 
The third line corresponds to the contribution from the diagrams (b) and (c). 
The diagram (c) contains the $Vf\bar{f}$ vertex corrections, so that we need to prepare the calculation of the renormalized $Vf\bar{f}$ vertex denoted as $\hat{\Gamma}_{Vff}$ which 
will be implemented in the {\tt H-COUP} ver. 2.0~\cite{future}. 
In the massless limit of the external fermions, this correction becomes the same as the SM prediction. 
Details of the calculation of $\hat{\Gamma}_{Vff}$ are given in Appendix~\ref{sec:vff}. 
In the fourth line, the $T_{hff}^Z$ and $B_Z$ terms represent the contribution from the $hf\bar{f}$ vertex corrections and the box diagrams shown as the diagrams (d) and (e) in Fig.~\ref{FIG:hzff}, respectively. 
Both $T_{hff}^Z$ and $B_Z$ depend on the Mandelstam variable $u$ in loop functions, which has to be integrated out. 
The integration range of $u$ is given by  
\begin{align}
u_{\text{max},\text{min}} = \frac{m_h^2}{2}[1+x_Z^{} - x_s \pm \lambda^{1/2}(x_Z,x_s)]. 
\end{align}
The explicit formulae for $T_{hff}^Z$ and $B_Z$ are given in Appendix~\ref{sec:box}. 
Although the $hf\bar{f}$ vertex corrections can be calculated by using {\tt H-COUP} ver. 1.0~\cite{Kanemura:2017gbi}, we present the {explicit} analytic formulae 
for the contribution from the diagram (d) in Eq.~(\ref{eq:thff1}) in the massless limit of the external fermions. 
In this limit, both contributions from (d) and (e) become the SM predictions times the scaling factor $\kappa_V^{}$. 
We note that we need to add the contribution from the wave function renormalization of the external $Z$ boson{,} i.e., $\hat{\Pi}'_{ZZ}$, because 
in our on-shell scheme, the counterterm for the wave function renormalization (normally denoted as $\delta Z_Z$) is not fixed by the condition 
{which requires a vanishing  derivative} of the renormalized $Z$ boson two-point function, but it is determined by the other conditions, see Ref.~\cite{Kanemura:2017wtm}.

\subsection{$h \to WW^* \to Wf\bar{f}'$}

We compute the partial decay rate of $h \to WW^* \to Wf\bar{f}'$ mode at NLO. 
Feynman diagrams are shown in Fig.~\ref{FIG:hzff}. 
The decay rate is expressed as 
\begin{align}
\Gamma(h\to Wf\bar{f^\prime})&=\Gamma_0(h\to Wf\bar{f^\prime})\left[ 1+\Delta_{\rm EW}^W +\Delta_{\rm QCD}^W \right],
\end{align}
where $\Gamma_0$, $\Delta_{\text{EW}}^W$ and $\Delta_{\text{QCD}}^W$ are  
the contributions from LO, EW corrections and QCD corrections, respectively. 
The expression for the LO contribution is obtained from Eqs.~(\ref{eq:gam0}) and (\ref{eq:m0sq}) by replacing $Z \to W$ with $g_W^{} \equiv g/\sqrt{2}$, and that 
for the QCD corrections is the same as that given in Eq.~(\ref{eq:qcd}), because 
the gluon loop corrections only appear in the $W\bar{f}f'$ vertex similar to the $Z\bar{f}f$ vertex in the $h \to Zf\bar{f}$ decay~\cite{Denner:1991kt}.  

The EW corrections $\Delta_{\text{EW}}^W$ are given in a similar way to Eq.~(\ref{eq:del_ew_z}) as follows
\begin{align}
\Delta_{\rm EW}^W & = \frac{1}{\Gamma_0}\int_0^{(m_h-m_W)^2}ds\, |\overline{{\cal M}_0^W}|^2\Bigg\{
\text{Re}\left[ \frac{2\Gamma_{hWW}^{1,{\rm loop}}}{\Gamma_{hWW}^{1,{\rm tree}}} + \frac{\bar{\lambda}(x_W,x_s)}{x_W^{}} 
\frac{\Gamma_{hWW}^{2,{\rm loop}} }{\Gamma_{hWW}^{1,{\rm tree}}}\right](m_W^2,s,m_h^2) \notag \\ 
&+2\text{Re}[\Gamma^{V,{\rm loop}}_{Wff} + \Gamma_{Wff}^{A,{\rm loop}}](0,0,s) 
-\frac{2\text{Re}\,\hat{\Pi}_{WW}(s)}{s-m_W^2}\Bigg\}\notag \\
&+\frac{1}{\Gamma_0} \left[\int_0^{(m_h-m_W)^2}ds  \int_{u_{\text{min}}}^{u_{\text{max}}} du\, \left(T_{hff}^W + B_W\right) +\Gamma({h\to Wf\bar{f}^\prime \gamma}) \right]\notag\\
& - 2\Delta r - {\rm Re}\hat{\Pi}_{WW}^\prime(m_W^2),   \label{eq:del_ew_w}
\end{align}
where the expressions for the $Wf\bar{f}'$ vertex corrections $\Gamma^{V,{\rm loop}}_{Wff}$ and $\Gamma_{Wff}^{A,{\rm loop}}$
are given in Appendix~\ref{sec:vff}. 
The $hf\bar{f}$ vertex corrections $T_{hff}^W$ and the box diagrams $B_W$ are given in Appendix~\ref{sec:box}. 
In the {third} line, $\Gamma({h\to Wf\bar{f}^\prime \gamma})$ denotes the contribution from real photon emissions. 
Differently from the $h \to Zf\bar{f}$ mode, we cannot separate the QED corrections from the EW one, 
because the virtual photon loop can appear together with the $W$ boson loop in vertex corrections. 
Thus, we cannot obtain a simple expression for the QED correction such as Eq.~(\ref{eq:qcd}) in this decay mode. 
However, such IR divergence can be cancelled by adding the real photon emissions $\Gamma({h\to Wf\bar{f}^\prime \gamma})$. 
Detailed discussions for the treatment of the IR divergence are given in Appendix~\ref{sec:real_photon}. 
Similar to the case of $h \to Zf\bar{f}$, in the massless limit of the external fermions 
the contributions from {the} $Wf\bar{f}$ vertex become the same as those in the SM, while $T_{hff}^W$ and 
$B_W$ are given by the SM prediction multiplied by $\kappa_V^{}$.  

\subsection{$h \to \gamma\gamma$, $Z\gamma$, $gg$}

In addition to the $h \to f\bar{f}$ and $h \to VV^* \to Vf\bar{f}$ decays, the Higgs boson can also decay into $\gamma\gamma$, $Z\gamma$ and $gg$. 
The LO contributions to the decay rates arise from one-loop diagrams, and they can be expressed in terms of the renormalized 
$h{\cal VV}'$ (${\cal VV}' = \gamma\gamma,~Z\gamma,~gg$) vertices defined in Sec.~\ref{sec:ren-vert} as 
\begin{align}
\Gamma_0(h\to {\cal VV}') &= \frac{|\hat{\Gamma}_{h{\cal VV}'}^1(m_{\cal V}^2,m_{\cal V'}^2,m_h^2)|^2}{8\pi m_h} \lambda^{1/2}\left(\frac{m_{\cal V}^2}{m_h^2},\frac{m_{{\cal V}'}^2}{m_h^2} \right),  \label{eq:loop-induced}
\end{align}
where Eq.~(\ref{ward}) is used. 
The analytic expressions for $\hat{\Gamma}_{h{\cal VV}'}^{1,2}$ are given in Appendix~\ref{sec:hgg}. 

Let us discuss QCD corrections to these loop induced decay rates at NLO. 
For $h\to gg$, there are two sources of {the QCD corrections:} virtual gluon exchanges in the quark loop diagrams and real gluon emissions. 
In the $\overline{\rm MS}$ scheme, the QCD corrected decay rate is given as~\cite{Djouadi:2005gi}
\begin{align}
\Gamma(h\to gg)=\Gamma_0(h\to gg)\left[1 + \frac{\alpha_s(\mu)}{\pi}\left(\frac{95}{4}-\frac{7}{6}N_f+\frac{33-2N_f}{6}\log\frac{\mu^2}{m_h^2}\right)\right], 
\end{align}
in the limit of $m_t\to \infty$ with $N_f$ being the number of light flavors. 
Numerically, the magnitude of NLO correction is about 70\% for $N_f = 5$ and $\mu = m_h$. 

For $h\to\gamma\gamma$ and $h\to Z\gamma$, quark loop diagrams are modified by QCD corrections. 
We only take into account the NLO QCD correction for the top loop contributions, because those to the other quark loops are negligible. 
In the limit of $m_t\to \infty$, the QCD correction is easily implemented in the $\overline{\rm MS}$ scheme as~\cite{Djouadi:2005gi} 
\begin{align}
\hat{\Gamma}_{h{\cal V}\gamma}^1(m_{\cal V}^2,0,m_h^2)_{t}  \to \hat{\Gamma}_{h{\cal V}\gamma}^1(m_{\cal V}^2,0,m_h^2)_t\left[1-\frac{\alpha_s(\mu)}{\pi}\right],~~~({\cal V}=\gamma,~Z), 
\end{align}
where $\hat{\Gamma}_{h{\cal V}\gamma}^1(m_{\cal V}^2,0,m_h^2)_t$ is the top quark loop contribution to the renormalized $h\gamma\gamma$ and $hZ\gamma$ vertices.
The typical magnitude of the QCD corrections is a few percent level with respect to the LO result. 

\subsection{New physics effects in loops\label{sec:newphys}}

\begin{table}[!t]
\begin{center}
{\renewcommand\arraystretch{1.2}
\begin{tabular}{cccccc}\hline\hline
$\Delta_{\text{EW}}^{b}$ & $\Delta_{\text{EW}}^{c}$ & $\Delta_{\text{EW}}^{\tau}$ & $\Delta_{\text{EW}}^{Z}$ & $\Delta_{\text{EW}}^{W}$  \\\hline
1.67\% & 1.78\% & 4.91\% & 6.87\% &  3.14\%  \\\hline\hline
\end{tabular}}
\caption{EW corrections $\Delta_{\text{EW}}^X$ ($X=b,c,\tau,Z,W$) in the SM. }
\label{tab:sm}
\end{center}
\end{table}

We show numerical values of the EW corrections $\Delta_{\text{EW}}^X$ including weak boson and scalar boson loop effects to the decay rates discussed in the previous subsections. 
The numerical values of our inputs shown in Eq.~(\ref{inputs}) are fixed to be the default values implemented in the {\tt H-COUP} code~\cite{Kanemura:2017gbi}.
In order to extract the new physics effects of the EW corrections to the partial decay rates, 
we introduce 
\begin{align}
\overline{\Delta}_{\text{EW}}^X = \Delta_{\text{EW}}^X\big|_{\rm NP} - \Delta_{\text{EW}}^X\big|_{\rm SM}, \label{eq:delbar}
\end{align}
where $\Delta_{\text{EW}}^X|_{\rm NP}$ $(\Delta_{\text{EW}}^X|_{\rm SM})$ denotes the prediction of $\Delta_{\text{EW}}^X$ in the models with the extended Higgs sectors (SM). 
Our results for $\Delta_{\text{EW}}^X|_{\rm SM}$ are summarized in Table~\ref{tab:sm}. 

It is important to mention here that the dominant contribution to $\overline{\Delta}_{\text{EW}}^X$ comes from the first term of Eqs.~(\ref{eq:delew}), (\ref{eq:del_ew_z}) and (\ref{eq:del_ew_w})
for $\overline{\Delta}_{\text{EW}}^f$, $\overline{\Delta}_{\text{EW}}^Z$ and $\overline{\Delta}_{\text{EW}}^W$, respectively. 
In the case with $\kappa_V^{} \simeq 1$ and $m_\varphi \gg  m_h$, additional Higgs boson loop effects of $\Delta_{\text{EW}}^X$ are approximately expressed as~\cite{Kanemura:2015mxa} 
\begin{align}
\overline{\Delta}_{\text{EW}}^X \simeq -\frac{1}{16\pi^2}\frac{1}{6}\sum_\varphi c_\varphi \frac{m_\varphi^2}{v^2}\left(1 - \frac{M^2}{m_\varphi^2} \right)^2 ,  \label{eq:nondec}
\end{align}
where $c_\varphi= 2(1)$ for additional charged (neutral) scalar loop contributions.\footnote{In the THDMs, the charged Higgs boson and top quark loop contribution can also be important for $\overline{\Delta}_{\rm EW}^b$. 
The analytic expression for the top quark mass dependence due to charged Higgs boson loops is found in Ref.~\cite{Kanemura:2015mxa}. }
The above equation indicates that scalar loop effects become non-decoupling when $m_\varphi$ is mostly determined by $v$, or equivalently the case with 
$M^2/v^2 \ll 1$. In such a non-decoupling case, 
the right-hand side of Eq.~(\ref{eq:nondec})
is nearly proportional to $m_\varphi^2$. Of course, there must be an upper limit on $m_\varphi$ from the unitarity bound, under which 
the magnitude of the non-decoupling effect can be typically a few percent level, {as we will see in the plots below.} 

\begin{figure}[!t]
 \begin{center}
 \includegraphics[scale=0.7]{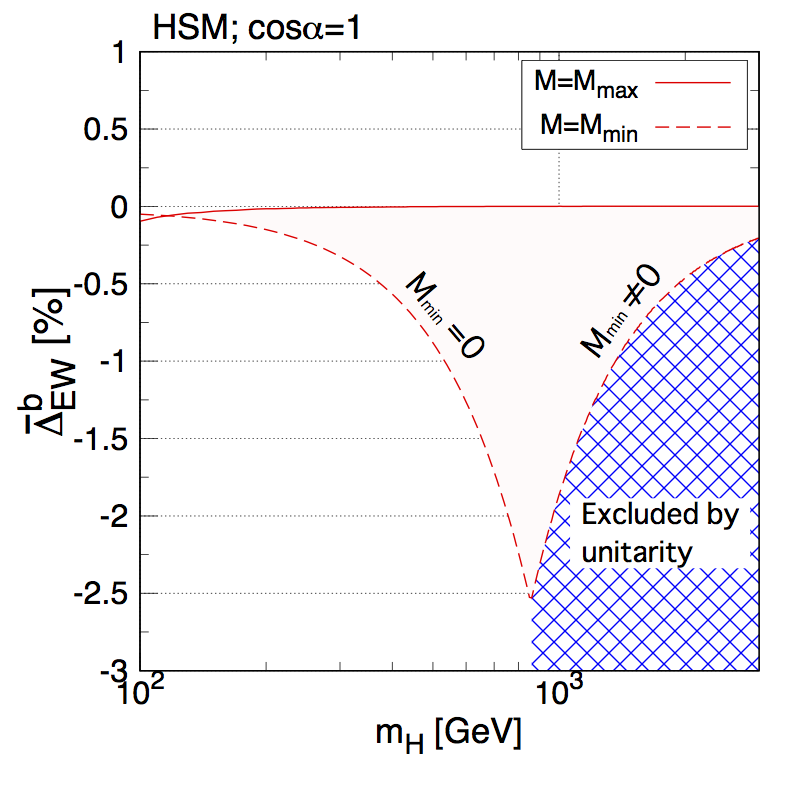} \hspace{3mm}
\includegraphics[scale=0.7]{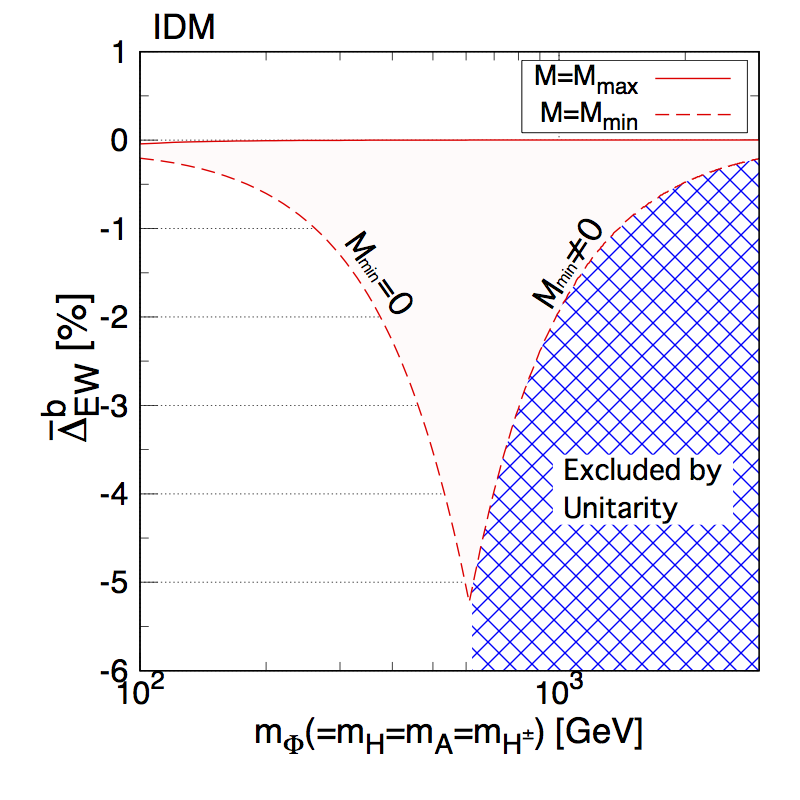} 
 \caption{New physics effects in the EW corrections $\overline{\Delta}_{\text{EW}}^{b}$ as a function of a mass of the extra scalar boson 
in the HSM (left) and the IDM (right). 
We take $c_\alpha = 1$, $\mu_S = 0$ and $\lambda_S = 0.1$ in the HSM, while set $m_H^{} = m_A^{} = m_{H^\pm}$ and $\lambda_2 = 0.1$ in the IDM. 
The solid (dashed) curves denote the case with the maximal (minimal) value of $M^2$ allowed by the perturbative unitarity, vacuum stability bounds and $S,~T$ parameters. }
 \label{fig:delb-hsm}
 \end{center}
 \end{figure}

Now, we show the plots of $\overline{\Delta}_{\text{EW}}^X$ in each model with the extended Higgs sectors discussed in Sec.~\ref{sec:model}. 
Although the quantity $\overline{\Delta}_{\text{EW}}^X$ cannot directly be measured at collider experiments, 
studying the prediction of $\overline{\Delta}_{\text{EW}}^X$ turns out to be important to understand the behavior of the deviation in branching ratios from the SM prediction, which will be discussed in the next section. 
In the following discussion, we impose the bounds from the perturbative unitarity, the vacuum stability, the conditions to avoid wrong vacua  (taking $M^2\geq 0$) and the $S,T$ parameters discussed in Sec.~\ref{sec:model}.
The flavor constraints discussed in Sec.~\ref{sec:2hdm} are also important to be taken into account particularly in the THDMs, but 
{we do not impose them here in order to study and} compare the behavior of $\overline{\Delta}_{\text{EW}}^X$ among the extended Higgs sectors.  
In the later discussion given in Sec.~\ref{sec:correlations}, we discuss the branching ratios imposing the flavor constraints as well. 

In Fig.~\ref{fig:delb-hsm}, predictions of $\overline{\Delta}_{\text{EW}}^b$ are shown in the HSM (left) and the IDM (right), 
{results for the other fermions $f$ look almost the same as what are shown in these figures.} 
In the HSM, we take $c_\alpha = 1$, $\mu_S = 0$ and $\lambda_S =0.1$, where $\overline{\Delta}_{\text{EW}}^b$ does not directly depend on $\lambda_S$, but it 
indirectly determines the allowed size of $\overline{\Delta}_{\text{EW}}$ via the unitarity and vacuum stability bounds. 
In the IDM, we take $m_H^{} = m_A^{} = m_{H^\pm}$ and $\lambda_2 = 0.1$, where $\overline{\Delta}_{\text{EW}}^b$ does not directly depend on $\lambda_2$. 
As we see from the plots, the magnitude of $\overline{\Delta}_{\text{EW}}^b$ becomes larger when the mass of the extra scalar boson is taken to be larger up to around 900 GeV and 600 GeV in the 
HSM and IDM, respectively. 
The maximal deviation is found to be $|\overline{\Delta}_{\text{EW}}^b| \simeq 2.5$ (5\%) in the HSM (IDM), which is given at $M^2 = 0$. 
The larger maximal amount of the deviation in the IDM as compared with the HSM is due to more than one additional scalar {boson} running in the loop in the IDM.
{In the case of larger values} of $m_H^{}$, the magnitude of $|\overline{\Delta}_{\text{EW}}^b|$ monotonically decreases, because $M^2 = 0$ cannot be taken due to the unitarity constraint. 
We then can see the decoupling behavior, $\overline{\Delta}_{\text{EW}}^b \to 0$, at the large mass limit {in both models}, where 
loop effects of additional Higgs bosons vanish. 
%
The behavior of $\overline{\Delta}_{\text{EW}}^Z$ and $\overline{\Delta}_{\text{EW}}^W$ in the HSM and the IDM is almost the same as that of $\overline{\Delta}_{\text{EW}}^b$ shown in Fig.~\ref{fig:delb-hsm}. 

\begin{figure}[!t]
 \begin{center}
 \includegraphics[scale=0.7]{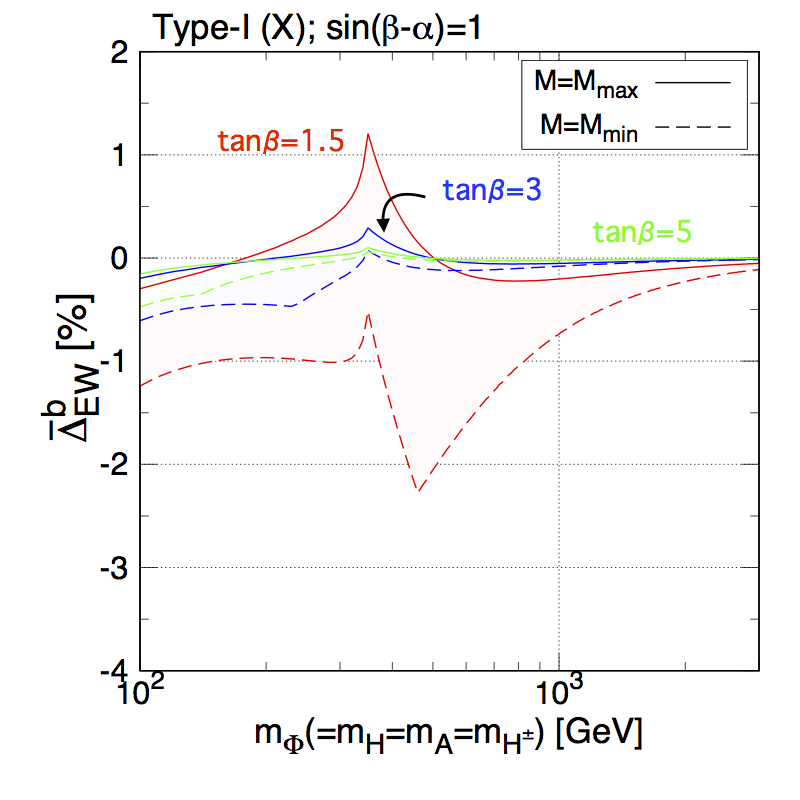} \\
 \includegraphics[scale=0.7]{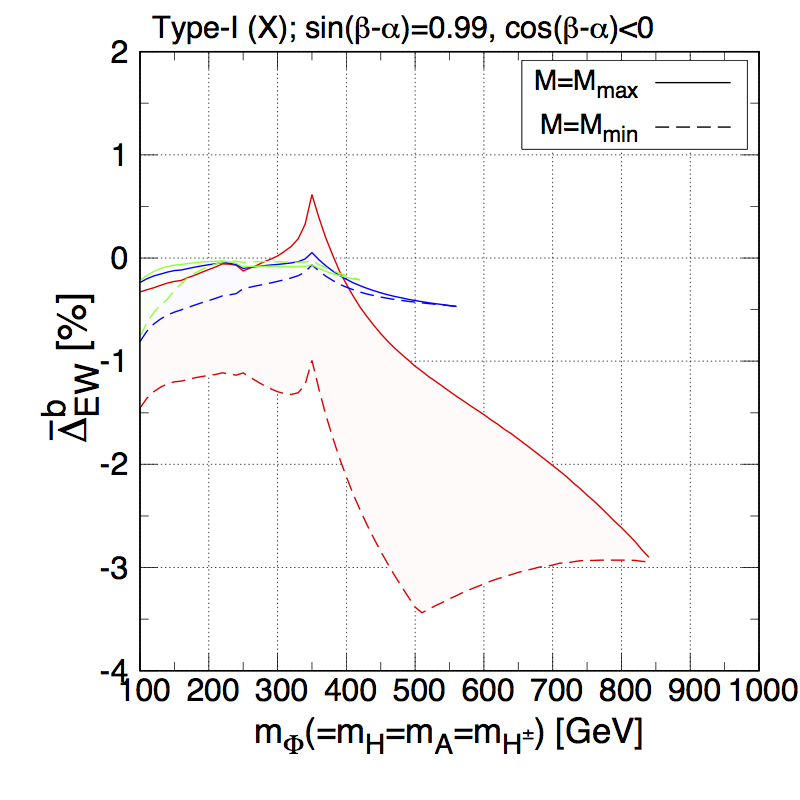}  \hspace{5mm}
 \includegraphics[scale=0.7]{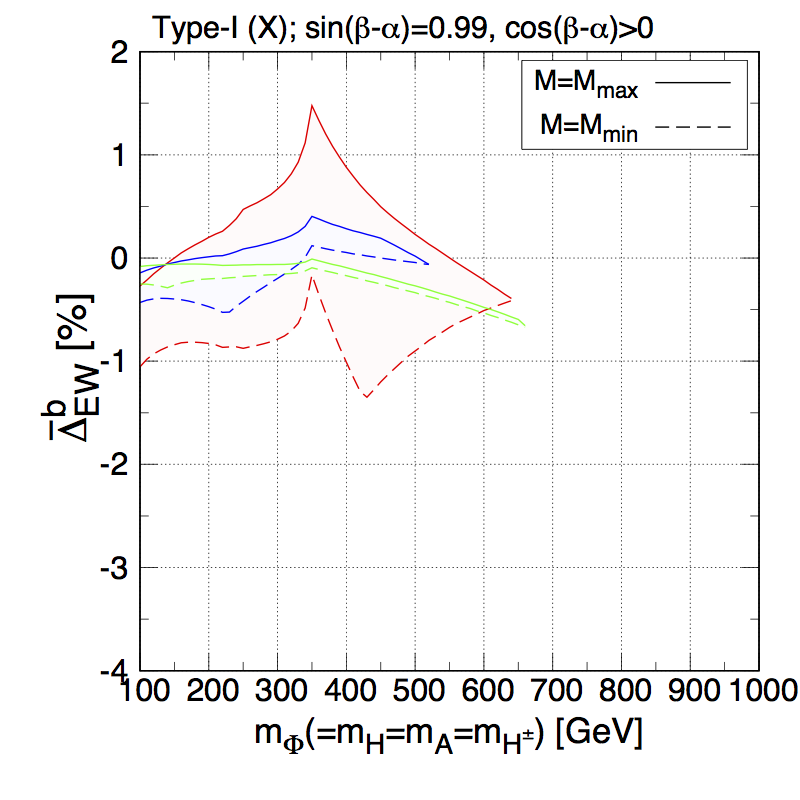} 
 \caption{New physics effects in the EW corrections $\overline{\Delta}_{\text{EW}}^{b}$
as a function of $m_\Phi (=m_H = m_A = m_{H^\pm})$ in the Type-I and Type-X THDMs with fixed values of $\tan\beta = 1.5$ (red), 3 (blue) and 5 (green). 
The upper panel shows the case with $s_{\beta-\alpha} = 1$ and the lower left (right) panel shows 
the case with $s_{\beta-\alpha} = 0.99$ and $c_{\beta-\alpha} <0$ ($c_{\beta-\alpha} > 0$). 
The solid (dashed) curves denote the case with the maximal (minimal) value of $M^2$ allowed by 
the perturbative unitarity, vacuum stability bounds and $S,~T$ parameters. }
 \label{fig:delb-1}
 \end{center}
 \end{figure}

\begin{figure}[!t]
 \begin{center}
 \includegraphics[scale=0.7]{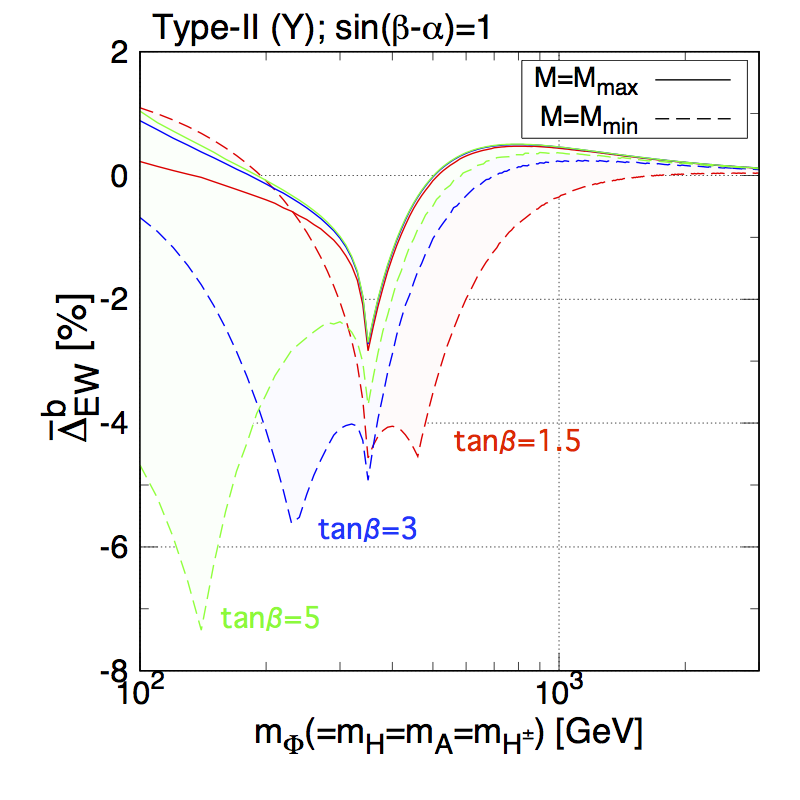} \\
 \includegraphics[scale=0.7]{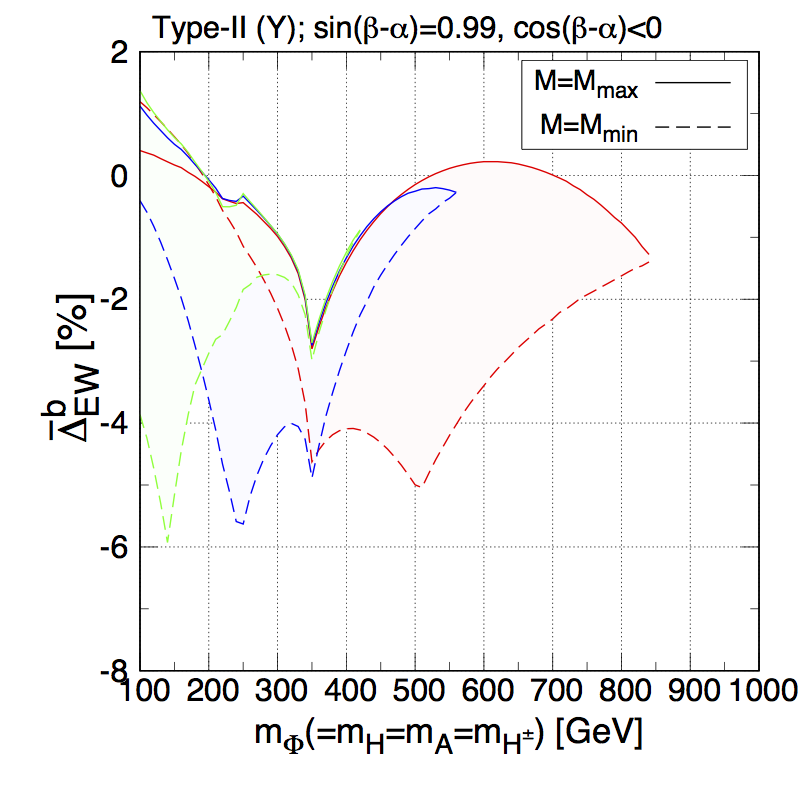}  \hspace{5mm}
 \includegraphics[scale=0.7]{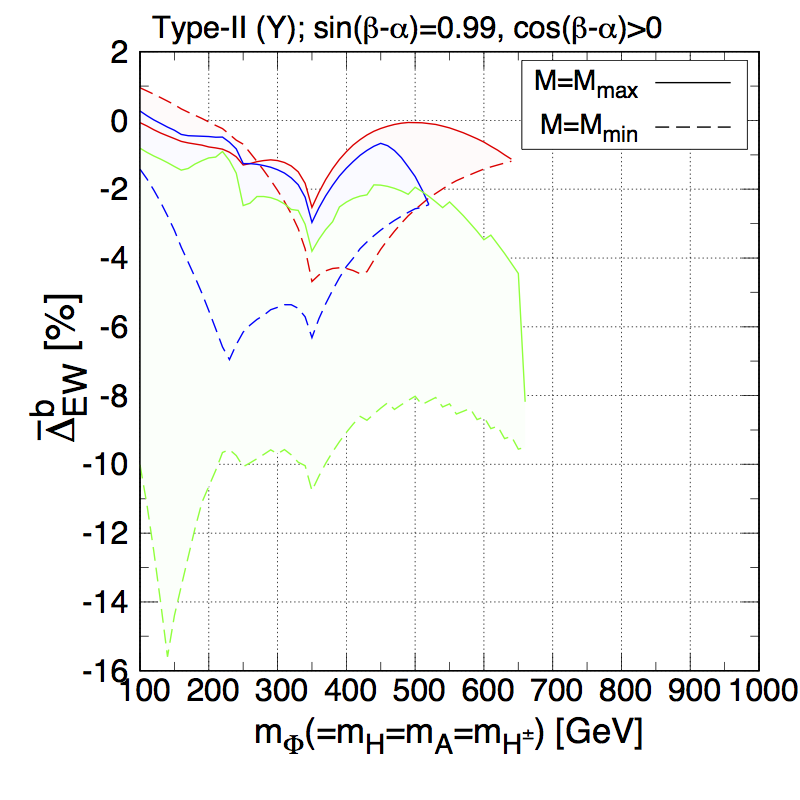} 
 \caption{Same as Fig.~\ref{fig:delb-1}, but for the Type-II and Type-Y THDMs. }
 \label{fig2}
 \end{center}
 \end{figure}

In Fig.~\ref{fig:delb-1}, the value of $\overline{\Delta}_{\text{EW}}^{b}$ is plotted as a function of the common additional Higgs boson mass $m_\Phi$ defined by 
$m_\Phi =m_H = m_A = m_{H^\pm}$ in the Type-I THDM with fixed values of $\tan\beta$. 
Here, we take $s_{\beta-\alpha} = 1$ (upper panel) and $s_{\beta-\alpha} = 0.99$ (lower panels), where the lower left and right panels show the cases of $c_{\beta-\alpha} < 0$ and $c_{\beta-\alpha} > 0$, respectively. 
We note that the results in the Type-X THDM are almost the same as those in the Type-I THDM. 
From the upper panel, the decoupling behavior can clearly be seen in the large mass region as in the HSM and the IDM. 
On the other hand, in the case with $s_{\beta-\alpha}=0.99$ shown in the lower panels, 
the decoupling limit cannot be taken, so that there appears the upper limit on the mass of the extra Higgs boson 
from the theoretical constraints depending on the value of $\tan\beta$ and the sign of $c_{\beta-\alpha}$. 
At $m_\Phi \sim 2m_t$, 
the threshold effects of $t\bar{t}$ appear, which push $\overline{\Delta}_{\text{EW}}^b$ into the positive direction. 
This peak comes from the top quark loop contribution to the $Z$--$A$ mixing diagram which appears in the counterterm of the $\beta$ parameter. 
More detailed discussions have been given in Ref.~\cite{Kanemura:2014dja}. 
We can also see the dip above the $t\bar{t}$ threshold for $\tan\beta = 1.5$. 
The origin of this dip can be explained {in} the same way as in Fig.~\ref{fig:delb-hsm}. 
Namely, {the point, where the dip appears,} corresponds to the maximal value of the mass of the extra Higgs boson with $M^2 = 0$ allowed by the unitarity bound. 
Similar to the results in the HSM and the IDM seen in Fig.~\ref{fig:delb-hsm}, the non-decoupling effect of the extra Higgs boson, which can be more significant 
for smaller $M^2$, pushes down the value of $\overline{\Delta}_{\text{EW}}^b$. 
For larger values of $\tan\beta$, allowed regions of $\overline{\Delta}_{\text{EW}}^{b}$ for a fixed value of $m_\Phi$ are significantly shrunk as compared to the case with $\tan\beta =1.5$, 
while {the behavior explained above} does not change so much. 

 \begin{figure}[!t]
 \begin{center}
 \includegraphics[scale=0.65]{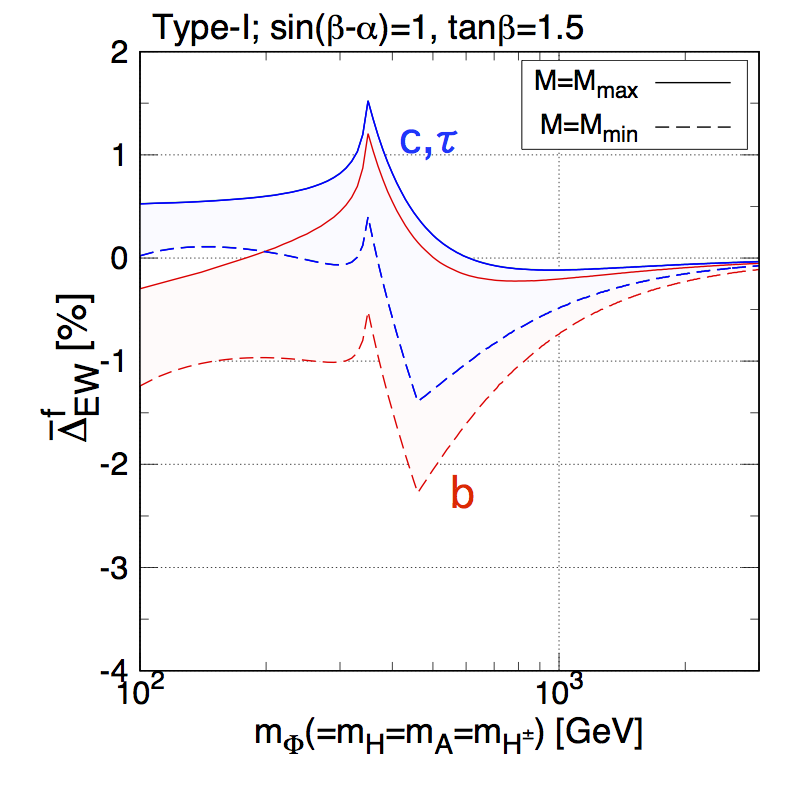} \hspace{5mm}
 \includegraphics[scale=0.65]{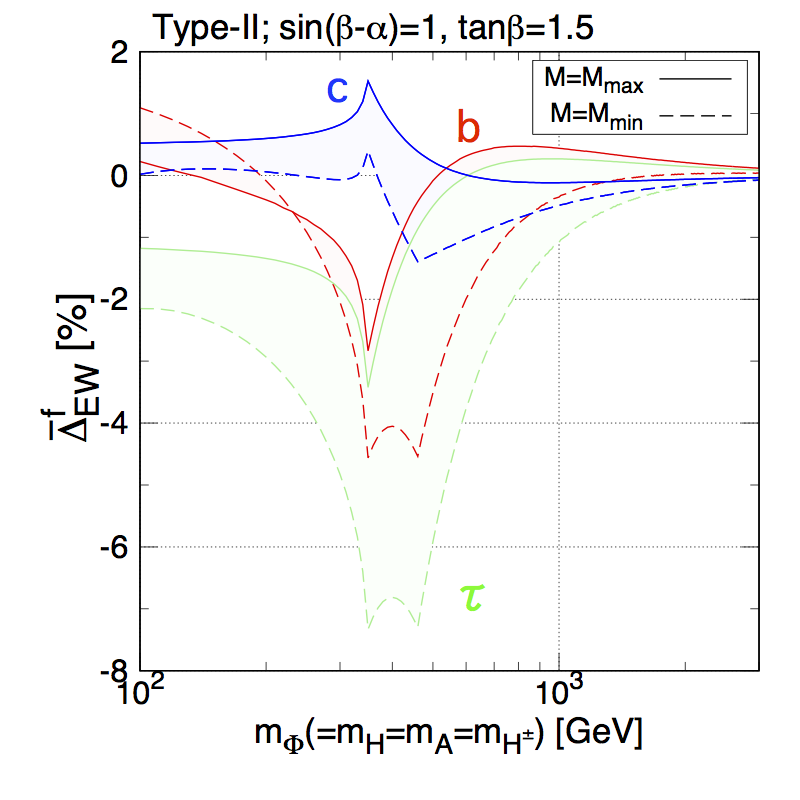}  \\
 \includegraphics[scale=0.65]{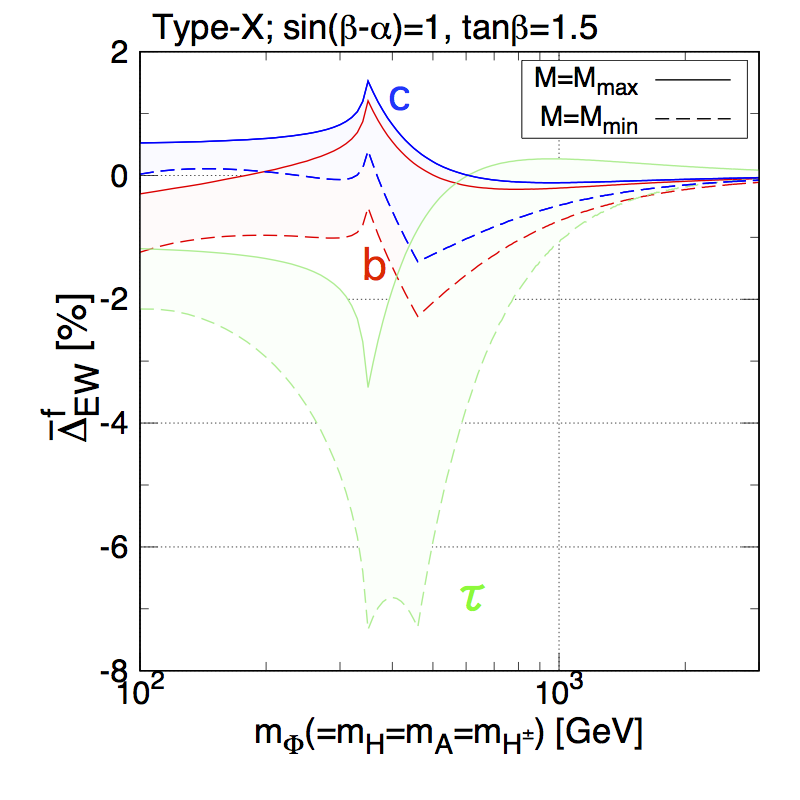} \hspace{5mm}
 \includegraphics[scale=0.65]{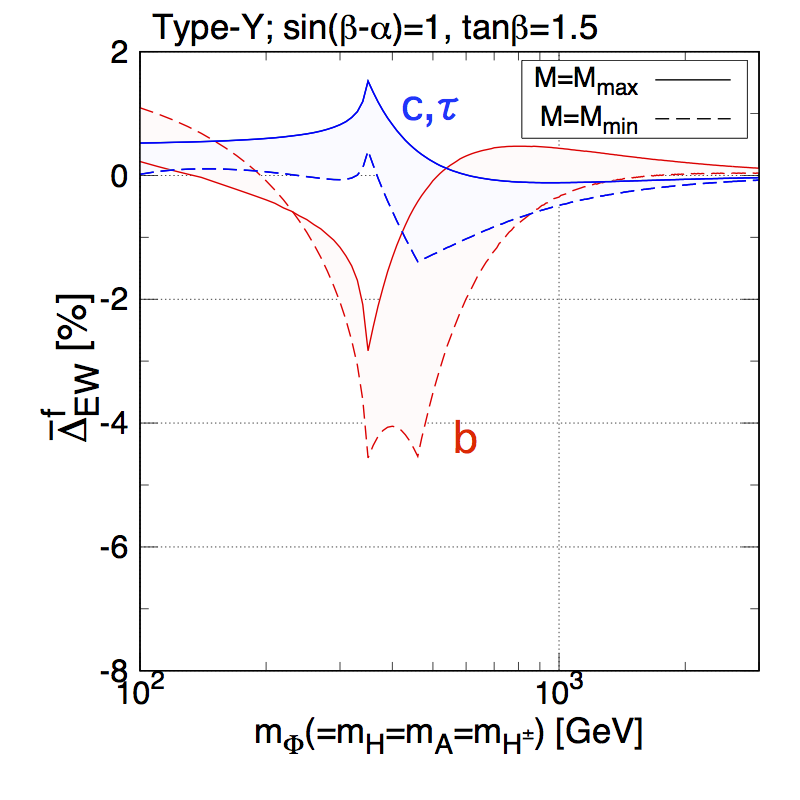}  
 \caption{New physics effects in the EW corrections $\overline{\Delta}_{\text{EW}}^{f}$ ($f=b,c,\tau$) 
as a function of $m_\Phi (=m_H = m_A = m_{H^\pm})$ in the Type-I (upper-left), Type-II (upper-right), Type-X (lower-left) 
and Type-Y (lower-right) THDM with $s_{\beta-\alpha} = 1$ and $\tan\beta = 1.5$. 
The solid (dashed) curves denote the case with the maximal (minimal) value of $M^2$ allowed by 
the perturbative unitarity, vacuum stability bounds and $S,~T$ parameters. }
 \label{fig:delbct}
 \end{center}
 \end{figure}

\begin{figure}[!t]
\begin{center}
\includegraphics[scale=0.65]{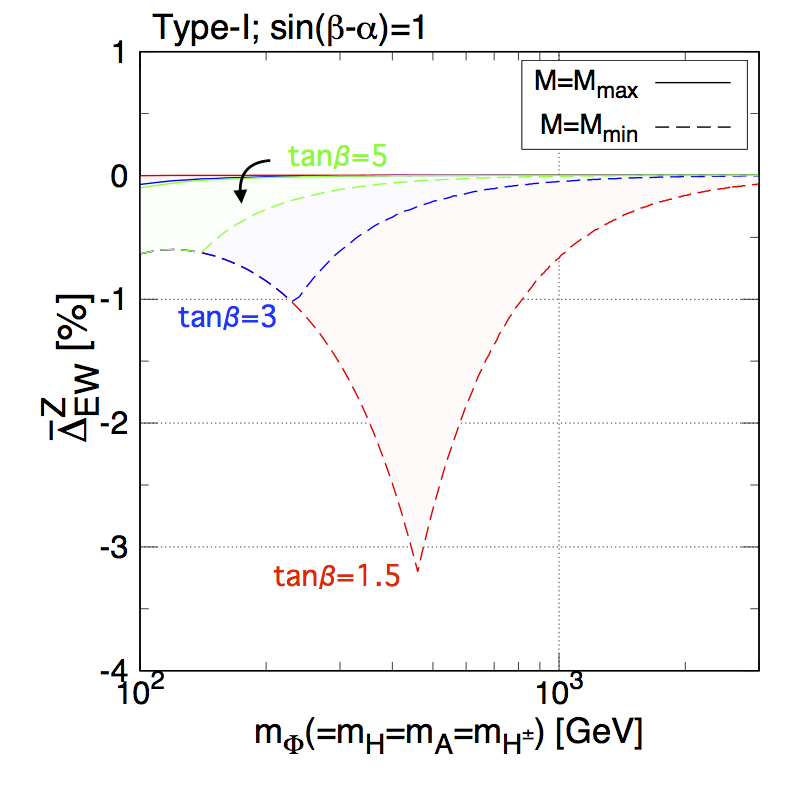} \\
\includegraphics[scale=0.65]{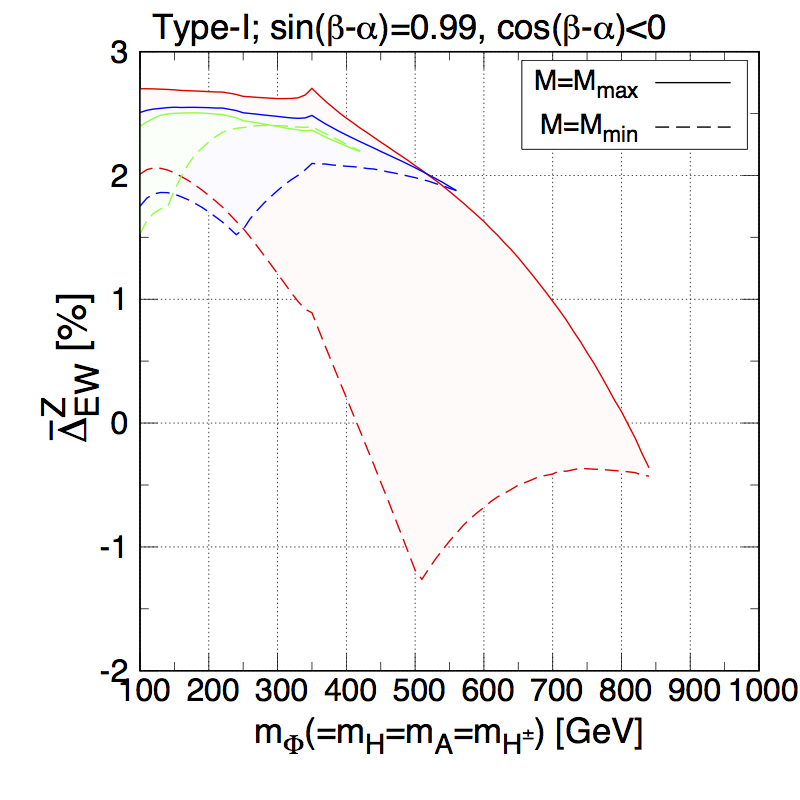} \hspace{5mm}
\includegraphics[scale=0.65]{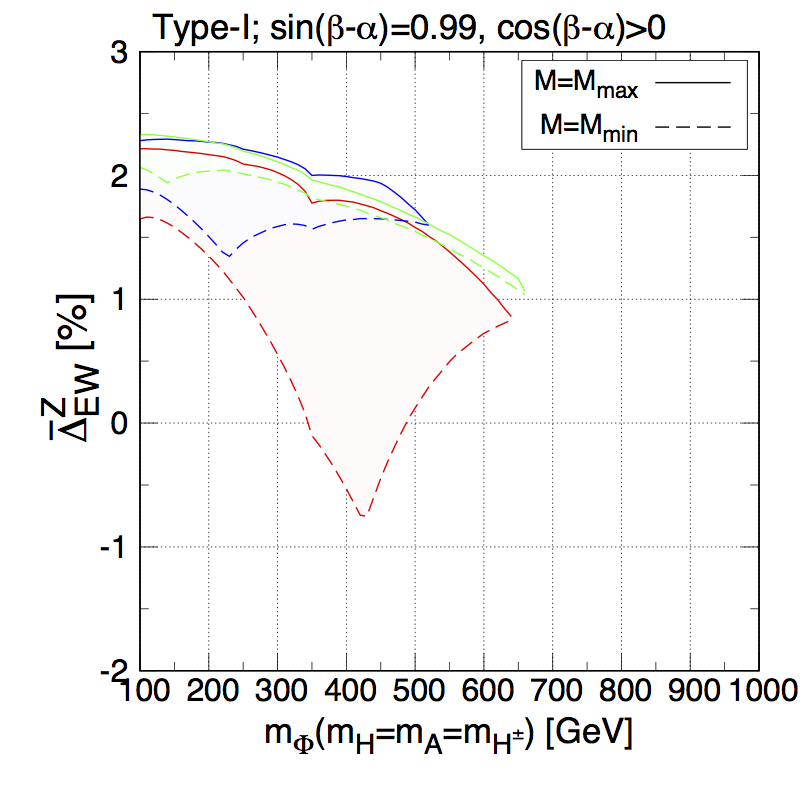}
\caption{New physics effects in the EW corrections $\overline{\Delta}_{\text{EW}}^{Z}$ as a function of $m_\Phi (=m_H = m_A = m_{H^\pm})$ in the Type-I THDM with fixed values of $\tan\beta = 1.5$ (red), 3 (blue) and 5 (green). 
The upper panel shows the case with $s_{\beta-\alpha} = 1$ and the lower left (right) panel shows 
the case with $s_{\beta-\alpha} = 0.99$ and $c_{\beta-\alpha} <0$ ($c_{\beta-\alpha} > 0$). 
The solid (dashed) curves denote the case with the maximal (minimal) value of $M^2$ allowed by 
the perturbative unitarity, vacuum stability bounds and $S,~T$ parameters.  }
\label{fig:delz}
\end{center}
\end{figure}

In Fig.~\ref{fig2}, {we show similar plots as in Fig.~\ref{fig:delb-1}, however, for the case {of} the Type-II THDM.}
The results in the Type-Y THDM are almost the same as those in the Type-II THDM. 
Again, we can see the decoupling behavior for $s_{\beta-\alpha} = 1$, and observe the upper limit on $m_\Phi$ for $s_{\beta-\alpha} = 0.99$, where 
the value of the upper limit does not depend on the types of Yukawa interaction. 
Although the behavior of the additional Higgs boson loop contribution{,} i.e., pushing down the value of $\overline{\Delta}_{\text{EW}}^b$, can also be seen as in the Type-I case, 
the effect of the $t\bar{t}$ threshold appears in the opposite direction as compared to the case {of} the Type-I THDM. 
This can be understood by the difference of the $\tan\beta$ dependence {on} the $\zeta_f$ factor, see Table~\ref{tab:z2}. 
In addition, for larger values of $\tan\beta$, the magnitude of $\overline{\Delta}_{\text{EW}}^b$ tends to get larger. 
For example, for $s_{\beta-\alpha} = 1$,  the maximally allowed value of $|\overline{\Delta}_{\text{EW}}^b|$ 
is about 4.5, 5.5 and 7.5\% for $\tan\beta = 1.5$, 3 and 5, respectively.

Differently from the HSM and IDM, the value of $\overline{\Delta}_{\text{EW}}^{f}$ can be drastically changed depending  not only on the type of Yukawa interaction but also the type of fermion. 
Thus, in Fig.~\ref{fig:delbct} we show the results for $\overline{\Delta}_{\text{EW}}^{b}$, $\overline{\Delta}_{\text{EW}}^{c}$ and $\overline{\Delta}_{\text{EW}}^{\tau}$ in four types of the THDMs. 
Here, we show the case of $s_{\beta-\alpha} = 1$ and $\tan\beta = 1.5$ for all the types of the THDMs. 
It is seen that the direction of the peak at around $m_\Phi^{} = 2m_t$ is determined to be positive (negative) if $\zeta_f = \cot\beta$ ($-\tan\beta$), see Table~\ref{tab:z2}. 
The behavior of $\overline{\Delta}_{\text{EW}}^{c}$ and $\overline{\Delta}_{\text{EW}}^{\tau}$ is also classified by the factor of $\zeta_f${,} e.g. that of $\overline{\Delta}_{\text{EW}}^{\tau}$ in the Type-II THDM
is common to the Type-X THDM.
{Concerning $\overline{\Delta}_{\text{EW}}^{b}$, the} behavior is different from e.g. $\overline{\Delta}_{\text{EW}}^{\tau}$ in the Type-II THDM even though these two depend on the same factor of $\zeta_f$. 
This can be understood by the fact that the top mass dependence enters in $\overline{\Delta}_{\text{EW}}^{b}$ when charged Higgs bosons run in the loop, while for the other $\overline{\Delta}_{\text{EW}}^{f}$
a dependence {on} fermion masses in loops is negligibly small. 
Detailed discussions have been given in Ref.~\cite{Kanemura:2014dja} for the behavior of the radiative correction to the Yukawa couplings in the THDMs. 

\begin{figure}[!t]
\begin{center}
\includegraphics[scale=0.65]{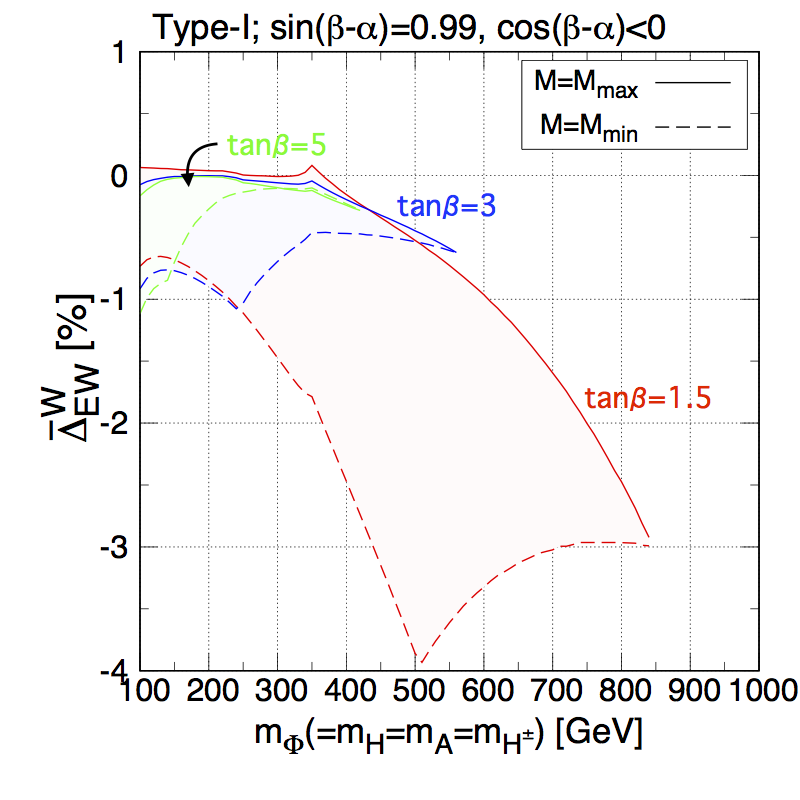} \hspace{5mm}
\includegraphics[scale=0.65]{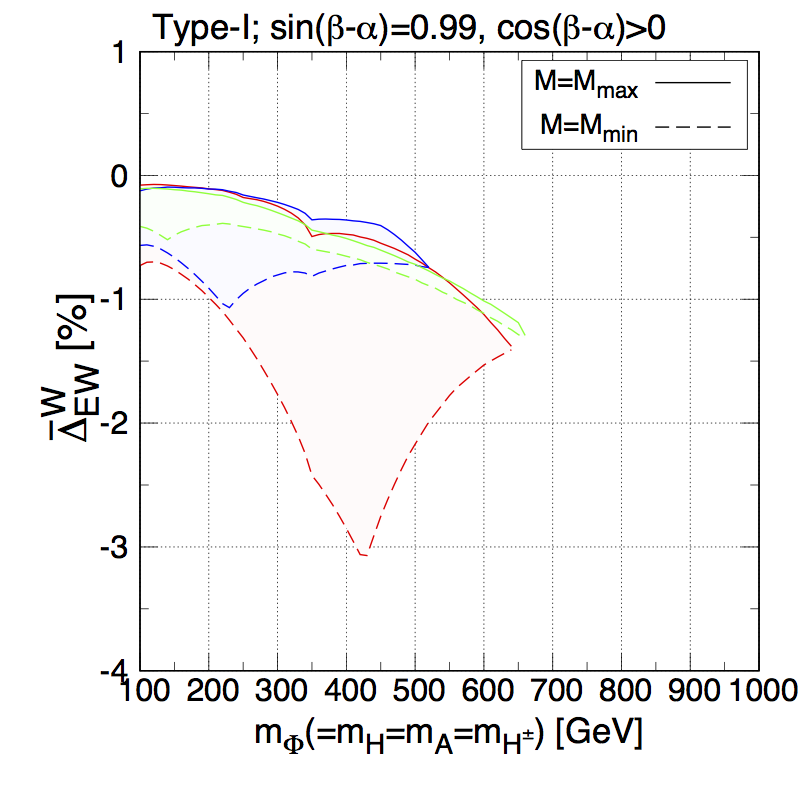}
\caption{New physics effects in the EW corrections $\overline{\Delta}_{\text{EW}}^{W}$ as a function of $m_\Phi (=m_H = m_A = m_{H^\pm})$ in the Type-I THDM with fixed values of $\tan\beta = 1.5$ (red), 3 (blue) and 5 (green). 
The left (right) panel shows
the case with $s_{\beta-\alpha} = 0.99$ and $c_{\beta-\alpha} <0$ $(c_{\beta-\alpha} > 0)$. 
The solid (dashed) curves denote the case with the maximal (minimal) value of $M^2$ allowed by 
the perturbative unitarity, vacuum stability bounds and $S,~T$ parameters. }
\label{fig:delw}
\end{center}
\end{figure}

In Fig.~\ref{fig:delz}, we show the value of $\overline{\Delta}_{\text{EW}}^{Z}$ in the Type-I THDM with a fixed value of $\tan\beta$. 
The results in all the other types of the THDMs are almost the same as those in the Type-I THDM. 
Similar to Fig.~\ref{fig:delb-1}, we can see the decoupling behavior for $s_{\beta-\alpha} = 1$ (upper panel) at the large mass region, while 
for $s_{\beta-\alpha} = 0.99$ (lower panels), there appears the upper limit on the additional Higgs boson mass $m_\Phi$ depending on $\tan\beta$. 
In addition, the position of {the dip at $m_\Phi \simeq 500$, 250 and 150 GeV} for $s_{\beta-\alpha} = 1$ with $\tan\beta = 1.5$, 3 and 5, respectively, is the same 
as that shown in the upper panel of Fig.~\ref{fig:delb-1}, because it is determined by the unitarity bound. 
It is seen that for $s_{\beta-\alpha} = 1$, the possible values of $\overline{\Delta}_{\text{EW}}^{Z}$ with larger $\tan\beta$  are included in those with smaller $\tan\beta$. 
This is simply because the unitarity bound more strongly constrains the possible non-decoupling effect for a larger value of $\tan\beta$. 
The lowest value of $\overline{\Delta}_{\text{EW}}^{Z}$ is found to be around $-3\%$, $-1\%$ and $-0.5\%$ for $\tan\beta = 1.5$, 3 and 5, respectively, 
and the largest value corresponds to the SM prediction{, i.e.} $\overline{\Delta}_{\text{EW}}^{Z}\simeq 0$. 
For $s_{\beta-\alpha} = 0.99$, we find that $\overline{\Delta}_{\text{EW}}^{Z}$ can be positive. 
This is because of the contribution from the virtual photon propagation shown as the diagram (a) in Fig.~\ref{FIG:hzff}, which
is proportional to $\hat{\Gamma}_{hZ\gamma}^{1,\text{loop}}/\hat{\Gamma}_{hZZ}^{1,\text{tree}}$, see Eq.~(\ref{eq:del_ew_z}). 
Because the tree level $hZZ$ vertex  $\hat{\Gamma}_{hZZ}^{1,\text{tree}}$ is now suppressed by the $\kappa_V^{}(=s_{\beta-\alpha})$ factor, 
this contribution can be larger than the case with $s_{\beta-\alpha} = 1$.  

This behavior should be compared with the results for $\overline{\Delta}_{\text{EW}}^{W}$ shown in Fig.~\ref{fig:delw}. 
Because there is no {virtual-photon-propagation diagram} in the $h \to WW^*$ process{,} as seen in Fig.~\ref{FIG:hzff}, 
the value of $\overline{\Delta}_{\text{EW}}^{W}$ is negative.  
For $s_{\beta-\alpha} = 1$, the value of $\overline{\Delta}_{\text{EW}}^{W}$ is almost the same as that of $\overline{\Delta}_{\text{EW}}^{Z}$, 
so that we do not show the result of $\overline{\Delta}_{\text{EW}}^{W}$ with $s_{\beta-\alpha} = 1$. 

\section{Numerical results for the Higgs boson decay rates\label{sec:numerical}}

In this section, we numerically show predictions of the total width and the branching ratios of the Higgs boson at NLO in the HSM, the THDMs and the IDM. 
After we show these quantities, we demonstrate if these extended Higgs models can be distinguished by the difference of the pattern of 
deviations in the branching ratios from those of the SM predictions. 
Similar to the analysis in Sec.~\ref{sec:newphys}, we take into account the constraints from the unitarity, the vacuum stability, the conditions to avoid wrong vacua and the $S$, $T$ parameters. 
Except for Sec.~\ref{sec:correlations}, we dare not to impose the flavor constraints in order to just see the predictions of deviations in total width and branching ratios. 
For the THDMs, we introduce the common mass of the additional Higgs bosons $m_\Phi^{}$; i.e.,  $m_\Phi^{} = m_H^{} = m_A^{} = m_{H^\pm}^{}$. 

\subsection{Total widths}

 \begin{figure}[!t]
 \begin{center}
 \includegraphics[scale=0.8]{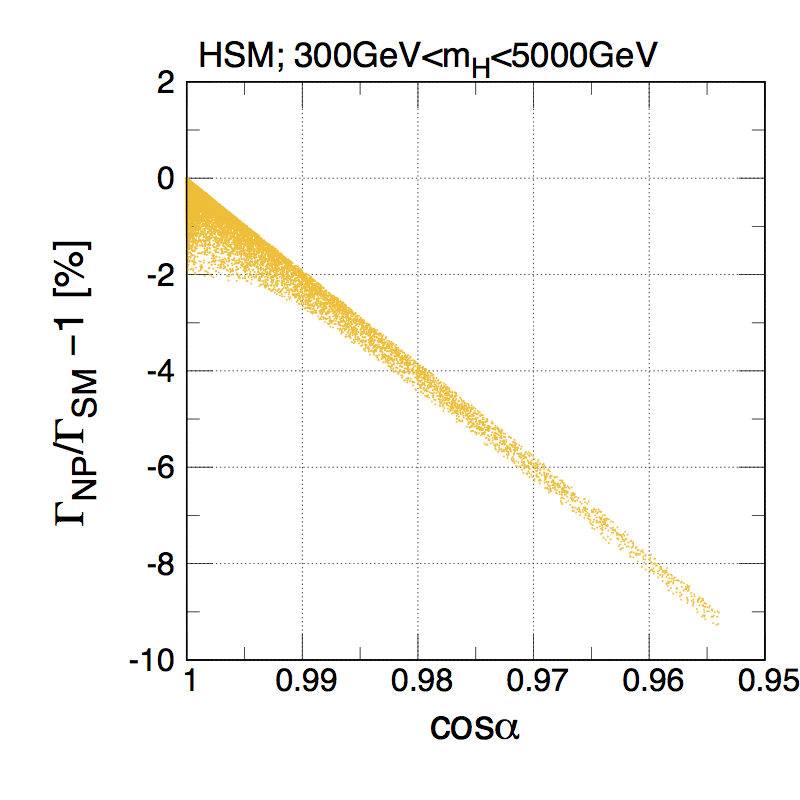} 
 \includegraphics[scale=0.8]{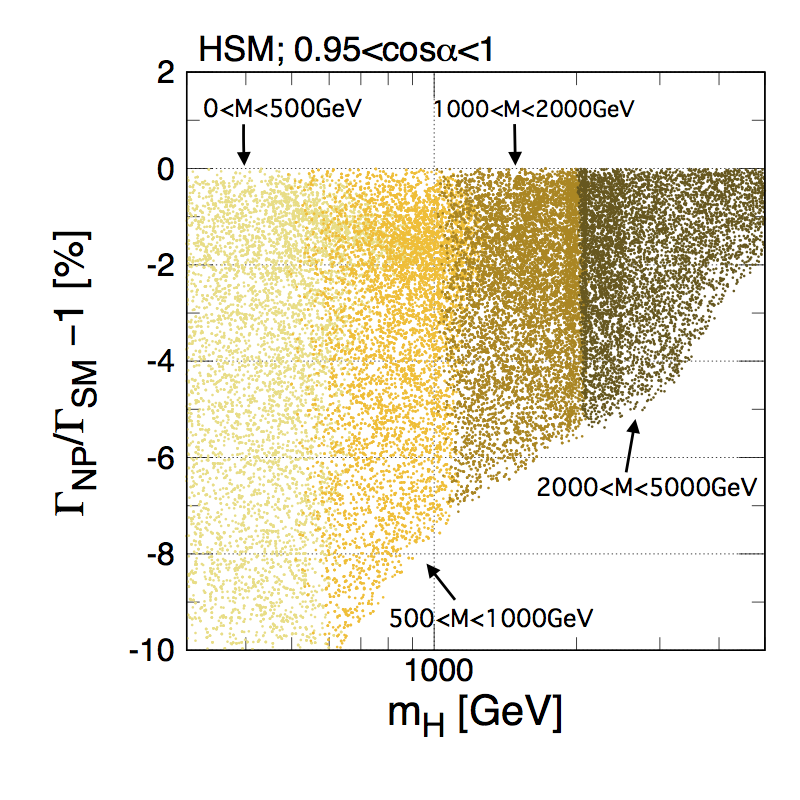} 
 \caption{Deviation in the total width from the SM prediction as a function of $c_\alpha$ (left) and $m_H^{}$ (right) in the HSM with $\mu_S = 0$ and $\lambda_S = 0.1$. 
The values of $c_\alpha$, $m_H$ and $M^2$ are scanned within $0.95 < c_\alpha < 1$, $300\leq m_H \leq 5000$ GeV and $0 \leq M^2 \leq m_H^2$, respectively. }
 \label{fig:width_hsm}
 \end{center}
 \end{figure}

We first {discuss} the total width of the Higgs boson $h$. 
In Fig.~\ref{fig:width_hsm}, we show the deviation in the total width from the SM prediction in the HSM. 
We scan the parameters $c_\alpha$, $m_H$ and $M^2$ within $0.95 < c_\alpha < 1$, $300\leq m_H \leq 5000$ GeV and $0 \leq M^2 \leq m_H^2$, respectively.
The dependences on $c_\alpha$ and $m_H^{}$ are then displayed in the left and right panels, respectively. 
At tree level, the deviation in the width is determined by $s_\alpha^2$, and it almost corresponds to the upper edge of 
the distribution in the left panel. 
The loop effects typically reduce the width by at most about 2\% level. 
In the right panel, it is seen that the magnitude of allowed deviations becomes smaller for larger mass regions, because the large mixing is excluded by the theoretical bounds. 
We note that the information of the width is important to identify the HSM, because 
the branching ratios of the Higgs boson are almost the same as those of the SM due to nearly universal suppression of the partial decay rates in the HSM. 

 \begin{figure}[t]
 \begin{center}
 \includegraphics[scale=0.8]{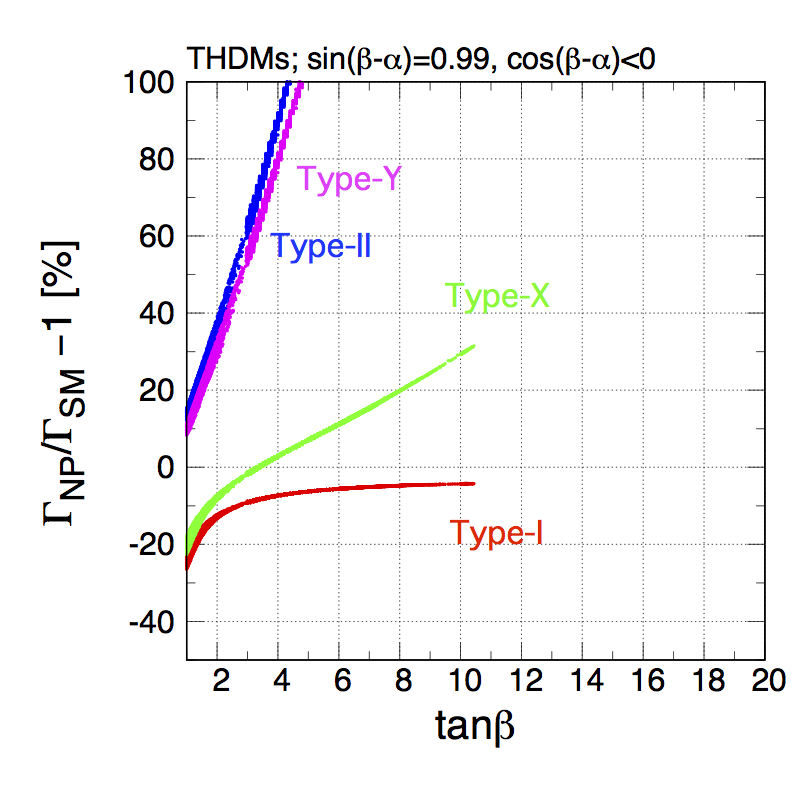} 
 \includegraphics[scale=0.8]{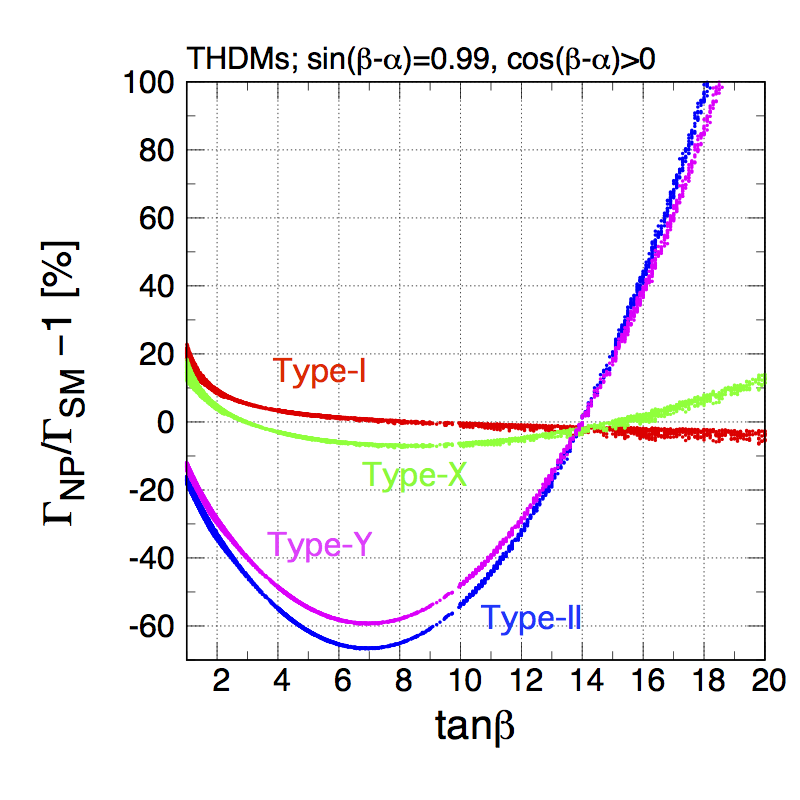} 
 \caption{Deviation in the total width from the SM prediction in four types of the THDMs with 
$s_{\beta-\alpha} = 0.99$ as a function of $\tan\beta$.
The left and right panels show the case of $c_{\beta-\alpha} < 0$ and $c_{\beta-\alpha} > 0$, respectively. 
The values of $m_\Phi^{}$ and $M^2$ are scanned within  $300 \leq m_\Phi  \leq  1000$ GeV and $0 \leq M^2  \leq m_\Phi^2$, respectively.  
}
 \label{fig:width_thdm}
 \end{center}
 \end{figure}

 \begin{figure}[t]
 \begin{center}
 \includegraphics[scale=0.7]{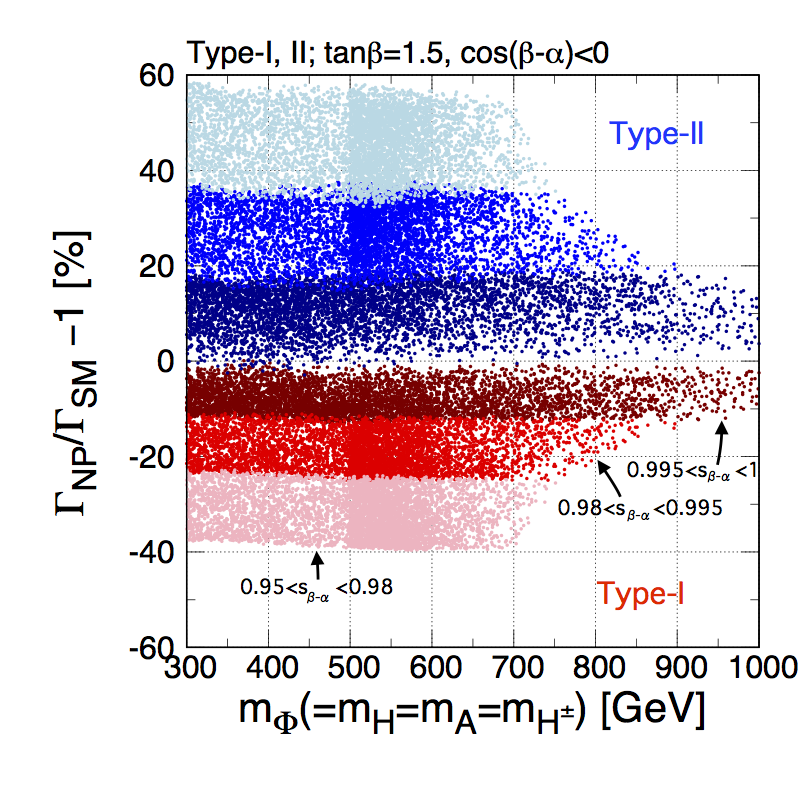} \hspace{3mm}
 \includegraphics[scale=0.7]{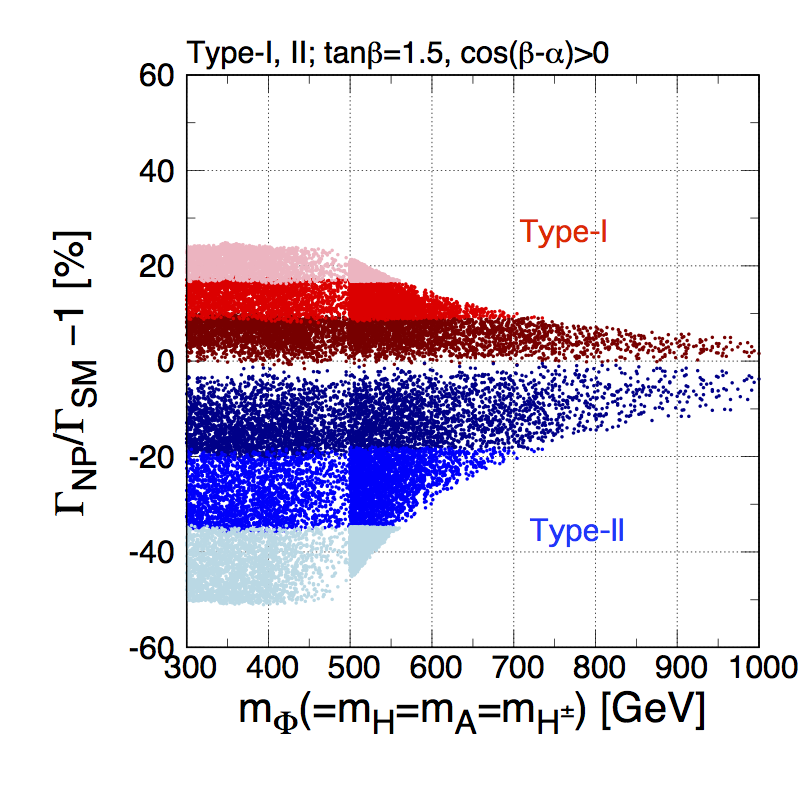} \\
 \includegraphics[scale=0.7]{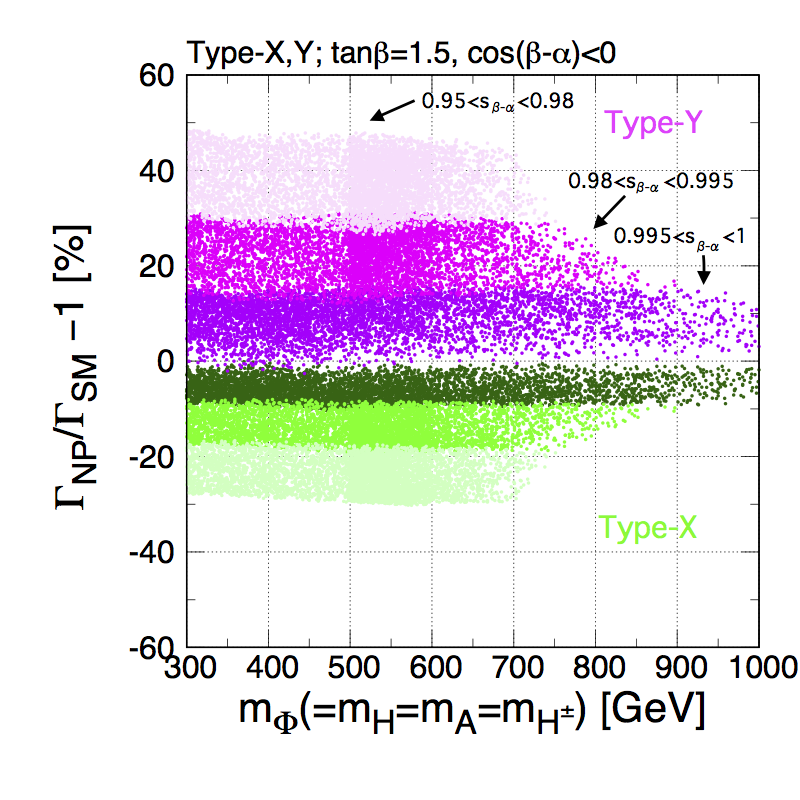} \hspace{3mm}
 \includegraphics[scale=0.7]{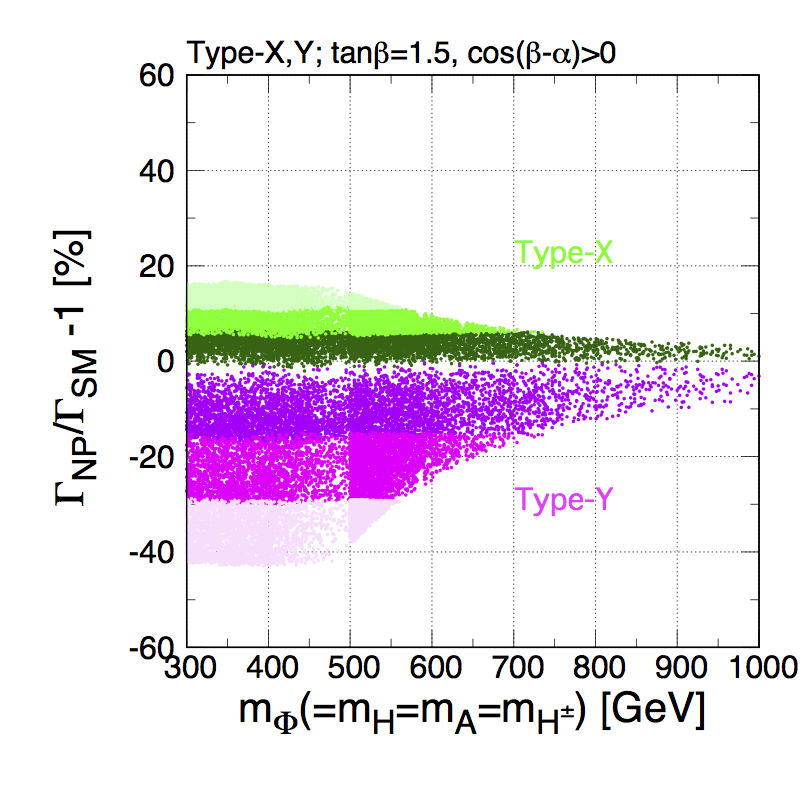} 
 \caption{Deviation in the total width from the SM prediction in the Type-I and Type-II THDMs (upper panels) and in the Type-X and Type-Y THDMs (lower panels) with 
$\tan\beta = 1.5$  as a function of $m_\Phi (=m_H = m_A = m_{H^\pm})$.
The left and right panels show the case of $c_{\beta-\alpha} < 0$ and $c_{\beta-\alpha} > 0$, respectively. 
The values of $M^2$ and $s_{\beta-\alpha}$ are scanned with the ranges of $0 \leq M^2  \leq m_\Phi^{2}$ and $0.95 \leq s_{\beta-\alpha} \leq 1$. 
}
 \label{fig:width_thdm2}
 \end{center}
 \end{figure}

In Fig.~\ref{fig:width_thdm}, the deviation in the total width is shown as a function of $\tan\beta$ in four types of the THDMs with 
$s_{\beta-\alpha} = 0.99$ and $c_{\beta-\alpha} < 0~(>0)$ in the left (right) panel. 
We scan the values of $M^2$ and $m_\Phi^{}$. 
In the left plot, it is seen that except for the Type-I THDM, the width becomes larger as $\tan\beta$ increases, because 
some of the partial widths have a $\tan\beta$ enhancement{,} e.g., the $h \to b\bar{b}$ ($h \to \tau \bar{\tau}$) mode in the Type-II and Type-Y (Type-II and Type-X) THDMs{. 
In} the Type-I THDM on the contrary, the total width approaches to the SM prediction, more precisely $s_{\beta-\alpha}^2\Gamma_{\rm SM}$, at the large $\tan\beta$ region. 
We note that the curves are truncated at around $\tan\beta = 11$ (the same thing also happens in the Type-II and Type-Y THDMs), because of the theoretical constraints. 
In the case with $c_{\beta-\alpha}>0$ (the right panel), 
the situation is drastically different from the case with $c_{\beta-\alpha}<0$. 
The total width has the minimal value at $\tan\beta \sim 7$ in the Type-II, Type-X and Type-Y THDMs, due to 
the cancellation between the $s_{\beta-\alpha}$ term and the $c_{\beta-\alpha}$ term in $\kappa_f$, see Table~\ref{tab:kappa}.  
This behavior is remarkably observed in the Type-II and Type-Y THDMs, because the $h \to b\bar{b}$ mode, which is the biggest partial width of $h$ in the SM, 
follows the behavior explained above.  
We can also see that at $\tan\beta \simeq 14$, the deviation in the total width becomes zero, as we have $\kappa_f^2 \simeq 1$ for all the types of Yukawa interaction. 
In the Type-I THDM, the width approaches to the SM value at a large value of $\tan\beta$ as also seen in the case with $c_{\beta-\alpha} < 0$. 
The typical amount of the loop corrections to the total width is a few percent level, which is shown by a width of each curve. 

In Fig.~\ref{fig:width_thdm2}, we show the $m_\Phi$ dependence on the deviation in the total width in four types of the THDMs. 
{Here, we scan} the values of $M^2$ and $s_{\beta-\alpha}$, {while we fix} $\tan\beta$ to be 1.5. 
For $c_{\beta-\alpha} < 0$ (left panels), 
the deviation is distributed in the positive (negative) direction in the Type-II and Type-Y (Type-I and Type-X) THDMs, while 
for $c_{\beta-\alpha} > 0$ (right panels), the situation is opposite. 
This can be understood by focusing on the deviation in the decay rate of $h \to b\bar{b}$ which is expressed 
by $\kappa_b^2 -1 = 2\zeta_b s_{\beta-\alpha}c_{\beta-\alpha} + c_{\beta-\alpha}^2(\zeta_b^2 -1)$ at tree level. 
As we are considering $s_{\beta-\alpha} \sim 1$, the $2\zeta_b s_{\beta-\alpha}c_{\beta-\alpha}$ term dominantly determines the behavior. 
We see that the allowed magnitude of the deviation is shrunk at around $m_\Phi = 700 ~(450)$ GeV for $c_{\beta-\alpha}< 0$ ($c_{\beta-\alpha}> 0$), {because 
in the region above $m_\Phi = 700 ~(450)$ GeV the unitarity and/or vacuum stability bounds constrain $s_{\beta-\alpha}$ to be closer to 1.}
As it is expected from the decoupling theorem, for larger $m_\Phi$ the magnitude of the deviation is getting smaller, but it can still be ${\cal O}(10)\%$ level at around $m_\Phi =1$ TeV.  

 \begin {figure}[!t]
 \begin{center}
 \includegraphics[scale=0.9]{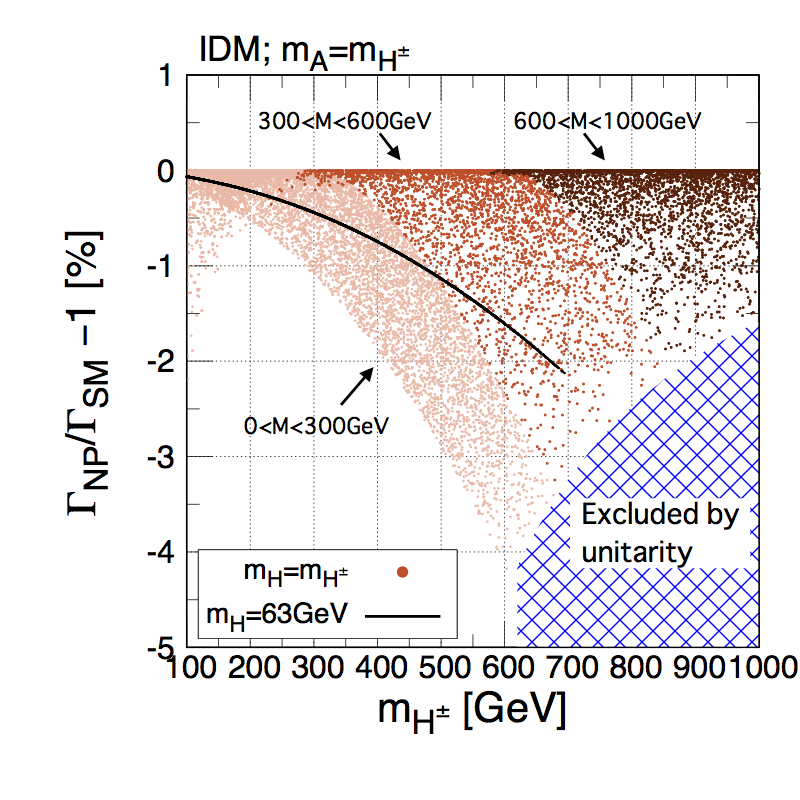} 
 \caption{Deviation in the total width from the SM prediction in the IDM with $\lambda_2 = 0.1$. 
The value $M^2$ is scanned within $0\leq M^2 \leq m_{H^\pm}^2$ under the constraints from the perturbative unitarity and the vacuum stability for the case of $m_H^{} =m_{H^\pm}^{}$ shown by dots. 
The black curve shows the case for $m_H^{} = 63$ GeV. 
 }
 \label{fig:idm}
 \end{center}
 \end{figure}

Finally in Fig.~\ref{fig:idm}, we show the total width in the IDM as a function of the charged scalar boson mass $m_{H^\pm}^{}$ with $m_A^{} = m_{H^\pm}^{}$. 
We here take two cases; i.e., (i) $m_H^{}$ is fixed to 63 GeV and (ii) $m_H = m_{H^\pm}$. 
The case (i) is motivated by the dark matter physics~\cite{Ilnicka:2015jba,Belyaev:2016lok,Ilnicka:2018def,Belyaev:2018ext}, where $H$ can be a dark matter candidate. 
In this case, the value of $M^2$ is taken such that the $HHh$ coupling normalized to $v$ becomes around $10^{-3}$ to avoid constraints from dark matter direct detection experiments.  
In the IDM, the total width does not change from the SM value at tree level, so that any deviation is purely due to loop effects. 
We can see that in the case (i), the total width monotonically decreases and the deviation is larger as $m_{H^\pm}$ is getting larger. 
The black curve is truncated at around $m_{H^\pm} = 700 $ GeV, because of the unitarity constraint. 
In the case (ii), the magnitude of the deviation becomes larger up to $m_{H^\pm} \simeq  600$ GeV, while 
it becomes smaller above $m_{H^\pm} \simeq  600$ GeV. 
The maximal deviation is given at $M^2 = 0$ for $m_{H^\pm} <  600$ GeV, while the unitarity constrains the minimal value of $M^2$ above $m_{H^\pm}\simeq$ 600 GeV
and possible deviations become smaller.

 \begin{figure}[!t]
 \begin{center}
 \includegraphics[scale=0.4]{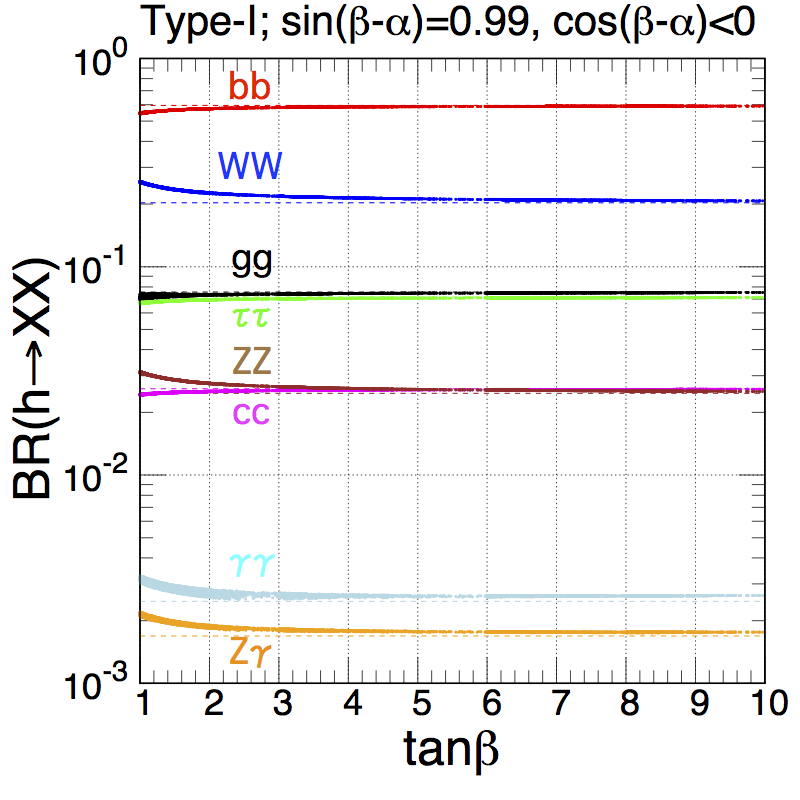} \hspace{-3mm}
 \includegraphics[scale=0.4]{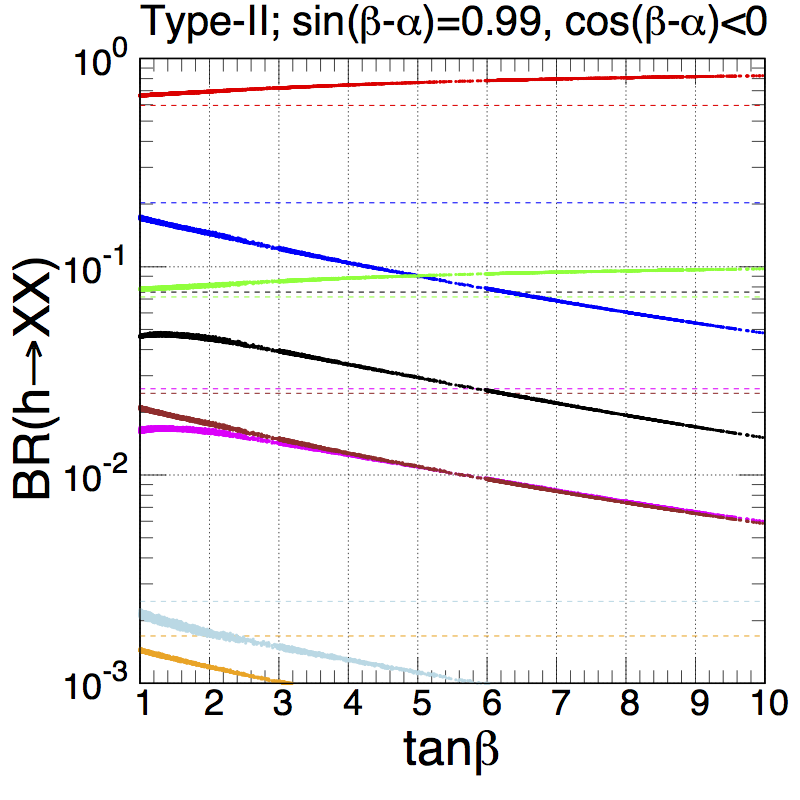} \hspace{-3mm}
 \includegraphics[scale=0.4]{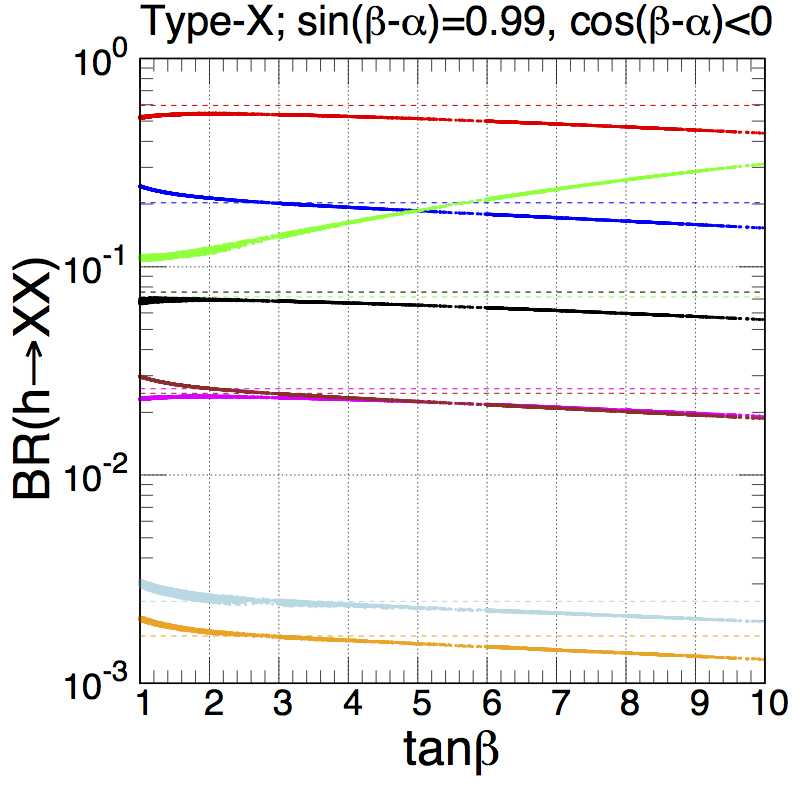} \hspace{-3mm}
 \includegraphics[scale=0.4]{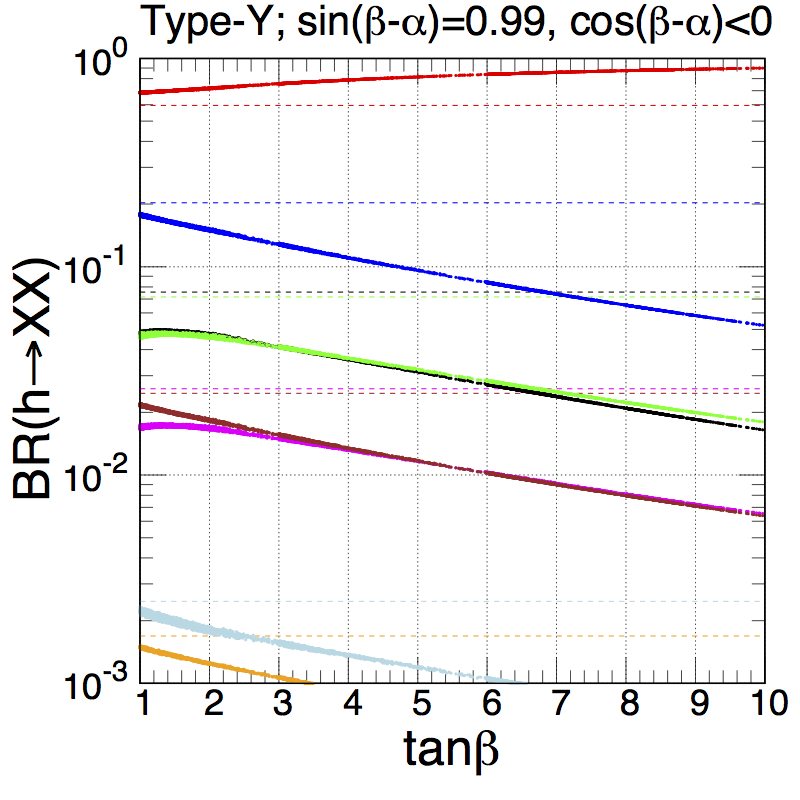} \\
 \includegraphics[scale=0.4]{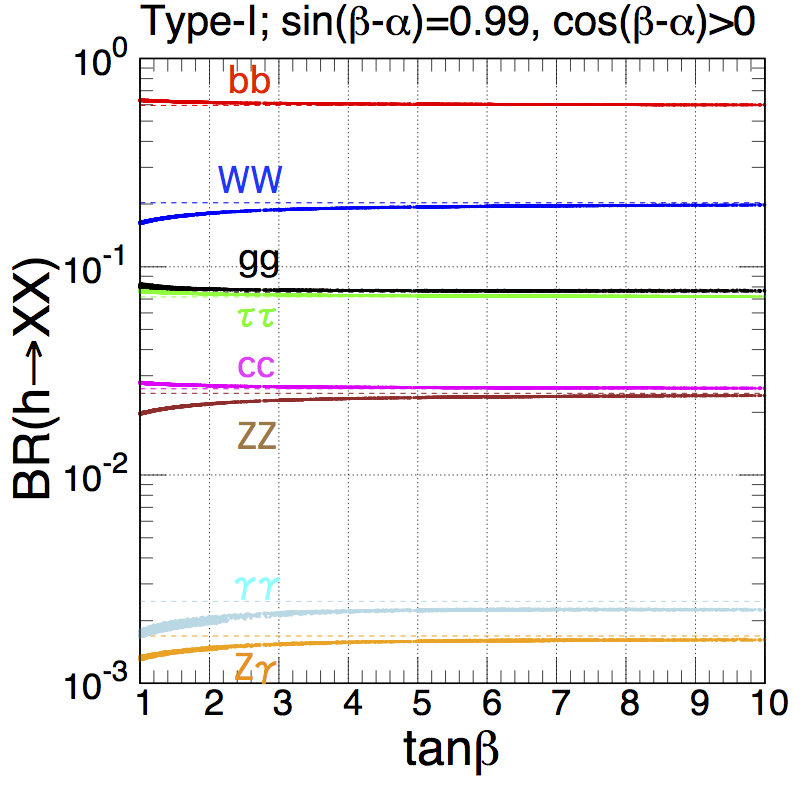} \hspace{-3mm}
 \includegraphics[scale=0.4]{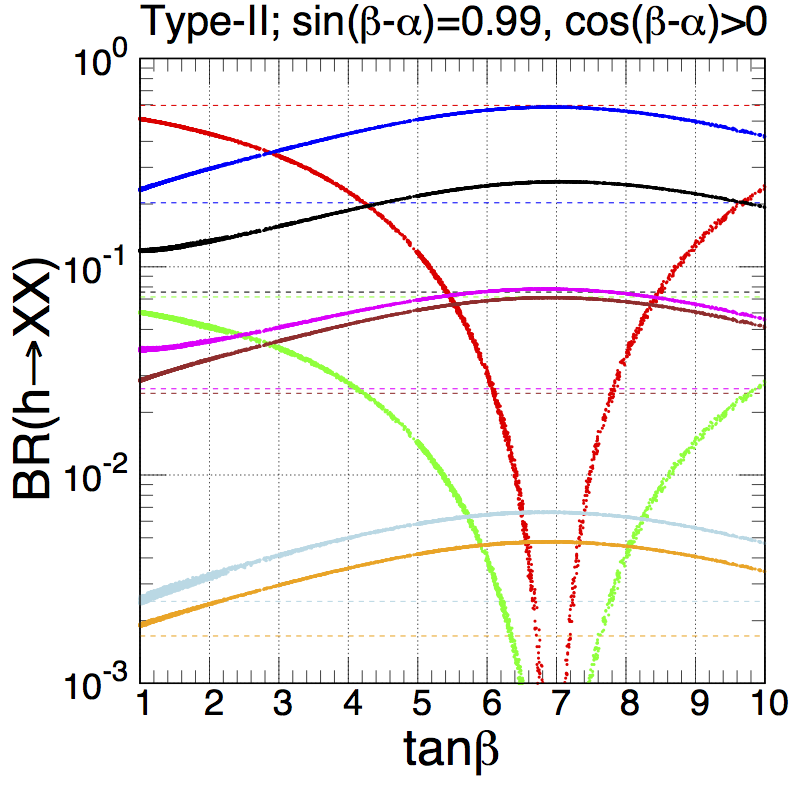}\hspace{-3mm}
 \includegraphics[scale=0.4]{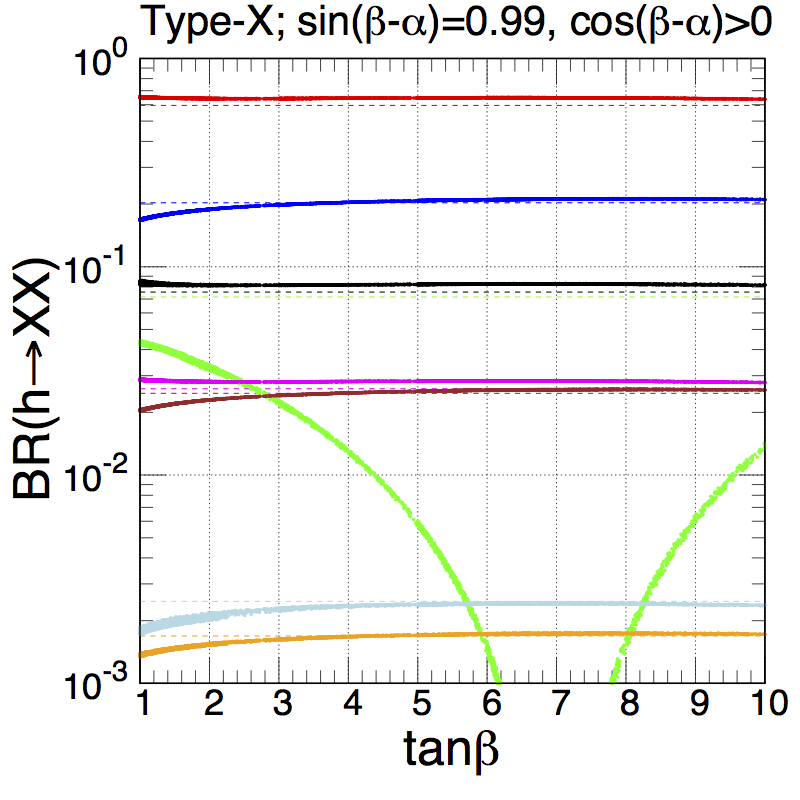} \hspace{-3mm}
 \includegraphics[scale=0.4]{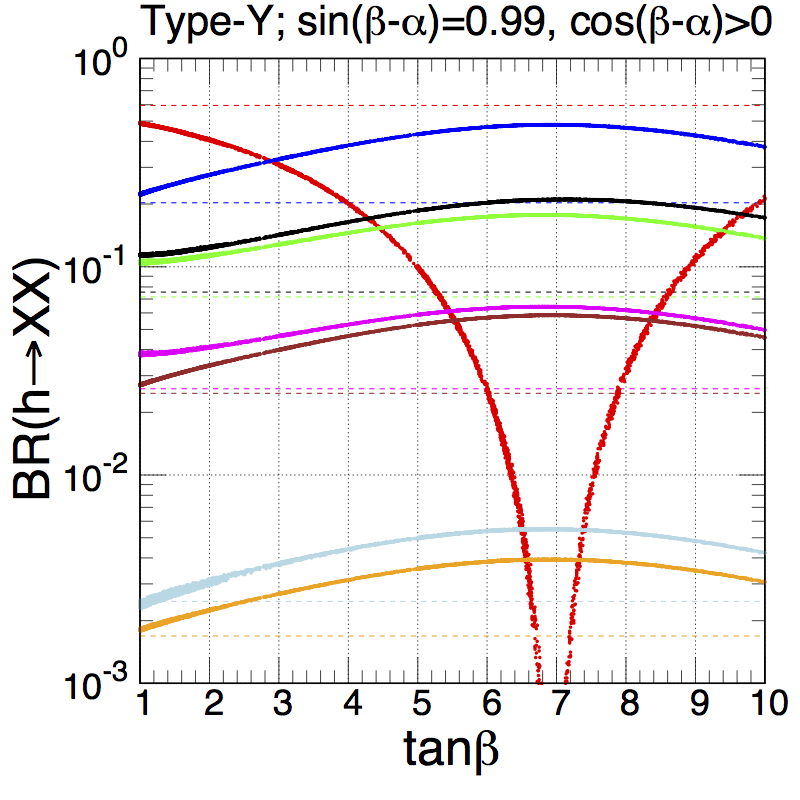} 
 \caption{Branching ratios as a function of $\tan\beta$ 
in the Type-I, Type-II, Type-X and Type-Y THDMs (from left to right) with $s_{\beta-\alpha} = 0.99$. 
We take $c_{\beta-\alpha} < 0$ $(c_{\beta-\alpha} > 0)$ for the upper (lower) panels. 
The values of $M^2$ and $m_\Phi (=m_H = m_A = m_{H^\pm})$ are scanned with the ranges of $0 \leq M^2  \leq m_\Phi^{2}$ and $300 \leq m_\Phi \leq 1000$ GeV. 
The dashed lines show the predictions in the SM.}
 \label{fig:br1}
 \end{center}
 \end{figure}

\subsection{Branching ratios \label{sec:br}}

\begin{table}[!t]
\begin{center}
{\renewcommand\arraystretch{1.2}
\begin{tabular}{cccccccc}\hline\hline
BR($h \to b\bar{b}$) & BR($h \to c\bar{c}$) & BR($h \to \tau\bar{\tau}$) & BR($h \to WW^*$) & BR($h \to ZZ^*$)  \\\hline
59.5\% & 2.60\% & 7.16\% &20.3\% &2.47\%  \\\hline\hline
\end{tabular}}
\caption{The SM predictions of the branching ratios of the Higgs boson $h$ at NLO. }
\label{tab:smbr}
\end{center}
\end{table}

We move on to the discussion of the branching ratios of the Higgs boson $h$ at NLO. 
For reference, in Table~\ref{tab:smbr} we give our results for the branching ratios in the SM. 

In the HSM and the IDM, the branching ratios for $h$ are almost the same as those in the SM predictions, because the partial decay rates are 
universally suppressed by the radiative corrections and the mixing, where the latter does not happen in the IDM. 
Thus, in the following discussion, we concentrate on the THDMs. 
In the THDMs, branching ratios can be modified from those in the SM {by both tree-level mixing effects} parameterized by the scaling factor $\kappa_X^{}$ and loop effects. 
When $s_{\beta-\alpha} \neq 1$ is taken, branching ratios can significantly be modified from the SM predictions due to the tree level mixing effects, and 
the pattern of deviations strongly depends on the type of Yukawa interactions. 
In this case, we may be able to determine the type from the pattern of deviations. 
On the other hand, loop contributions to deviations in the branching ratios are relatively smaller than the tree level mixing effects, so that 
it would be relatively difficult to extract the loop effects.
When $s_{\beta-\alpha} = 1$, however, the pure loop effect can be extracted, because the tree level mixing vanishes. 
Therefore, in the following discussion, we first show the predictions of branching ratios in the THDMs with $s_{\beta-\alpha} \neq 1$ to see 
how the mixing effects modify them. 
We then show those with $s_{\beta-\alpha} = 1$ in order to extract the size of loop effects. 

In Fig.~\ref{fig:br1}, we show the branching ratios as a function of $\tan\beta$ in four types of the THDMs
with $s_{\beta-\alpha} = 0.99$ and $c_{\beta-\alpha} < 0$ ($c_{\beta-\alpha} > 0$) 
in the upper (lower) panels. The values of $M^2$ and $m_\Phi$ are scanned. 
The typical behavior can be explained by the tree level results, see {e.g.} Ref.~\cite{Kanemura:2014bqa}. 
For example, except for the Type-I THDM, some of the branching ratios fall down at $\tan\beta\simeq 7$ with $c_{\beta-\alpha} > 0$,  
because of the fact that some $\kappa_f$ factors {become zero, e.g. $\kappa_b$} in the Type-II and Type-Y THDMs, and 
it makes the value of the total width to be minimal as we saw it in the right panel of Fig.~\ref{fig:width_thdm}. 
{Loop effects due to additional Higgs bosons} appear as a width of each curve. 
%

In order to see the deviation in the ratio of the branching ratio from the SM prediction, 
we introduce the following quantity for the $h \to XX$ mode
\begin{align}
\Delta\mu_{XX}^{} \equiv \frac{\text{BR}(h \to XX)_{\text{NP}}}{\text{BR}(h \to XX)_{\text{SM}}} -1.   \label{delta_nlo}
\end{align}
Using the formulae of the partial decay rates at NLO discussed in Sec.~\ref{sec:decay}, 
$\Delta \mu_{XX}$ can be written in terms of the EW corrections $\overline{\Delta}_{\text{EW}}^X$ defined in Eq.~(\ref{eq:delbar}) in the alignment limit as 
\begin{align}
\Delta \mu_{XX}^{} &\simeq  \overline{\Delta}_{\text{EW}}^X- \sum_f\text{BR}^0(h \to f \bar{f})\overline{\Delta}_{\text{EW}}^f   - \sum_V\text{BR}^0(h \to VV^*)\overline{\Delta}_{\text{EW}}^V , \label{muxx_app}
\end{align}
where $\text{BR}^0$ denotes the tree level branching ratio in the SM.
We note that the second term of the right hand side is dominantly determined by $\overline{\Delta}_{\text{EW}}^b$ because the branching ratio of $h \to b\bar{b}$ typically has the 
largest value among all the decay modes. 
This expression is helpful to understand the behavior of some plots which will be shown below. 

\begin{figure}[!t]
\begin{center}
\includegraphics[scale=0.65]{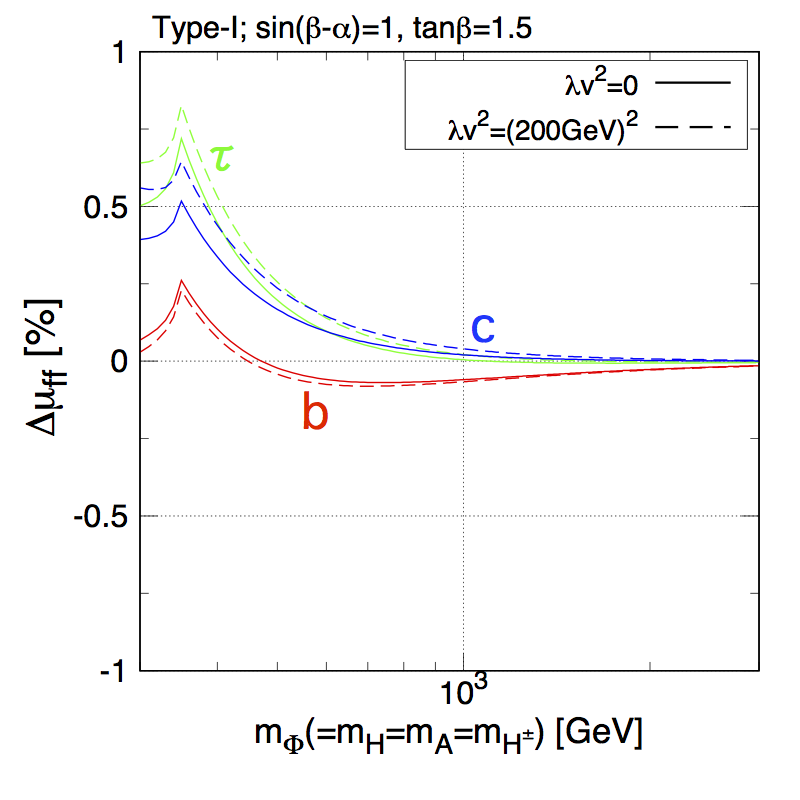} \hspace{3mm}
\includegraphics[scale=0.65]{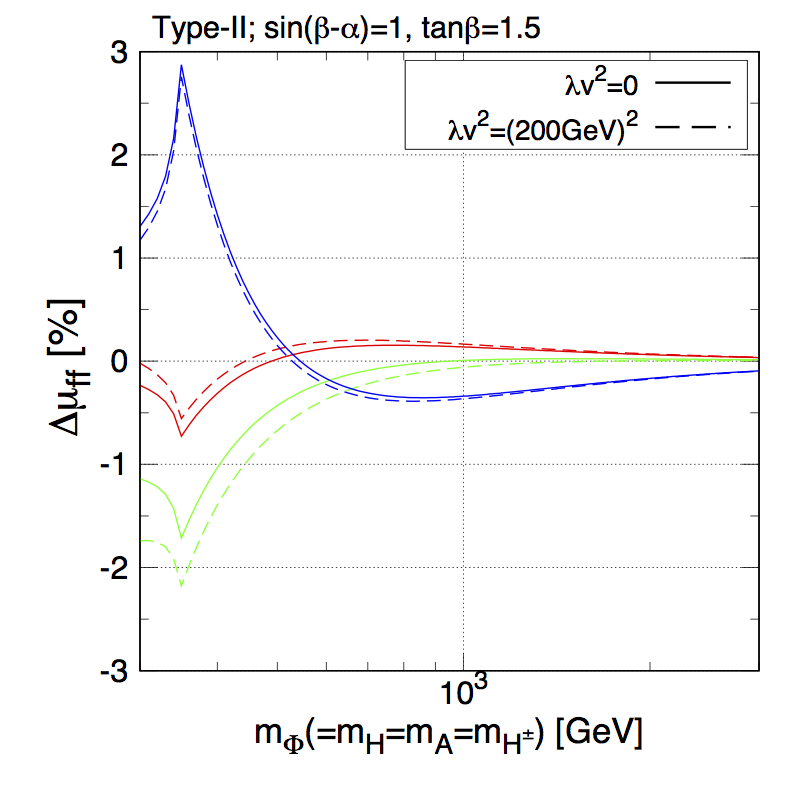} \\
\includegraphics[scale=0.65]{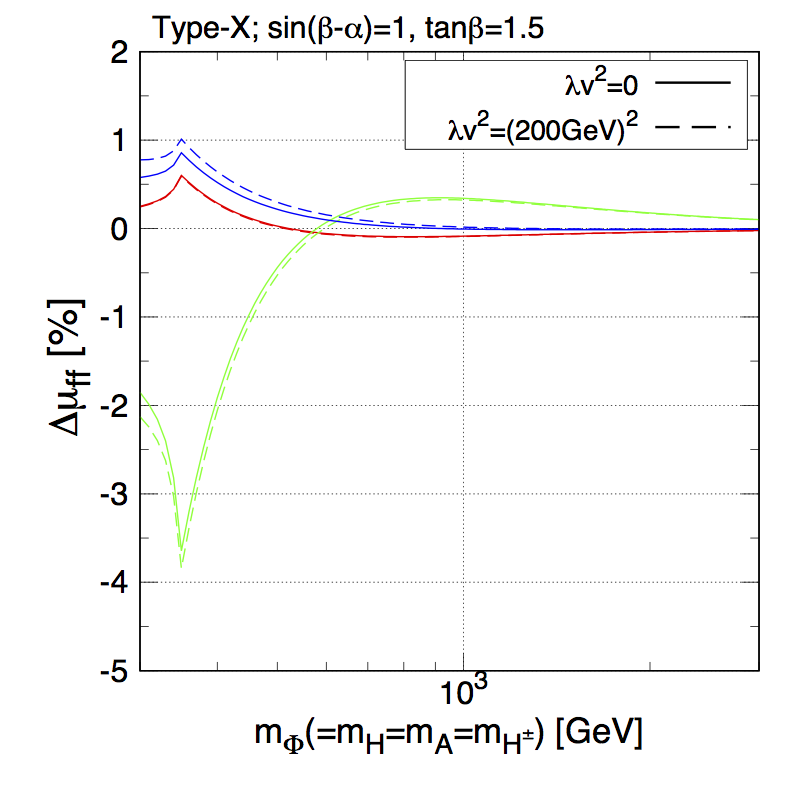} \hspace{3mm}
\includegraphics[scale=0.65]{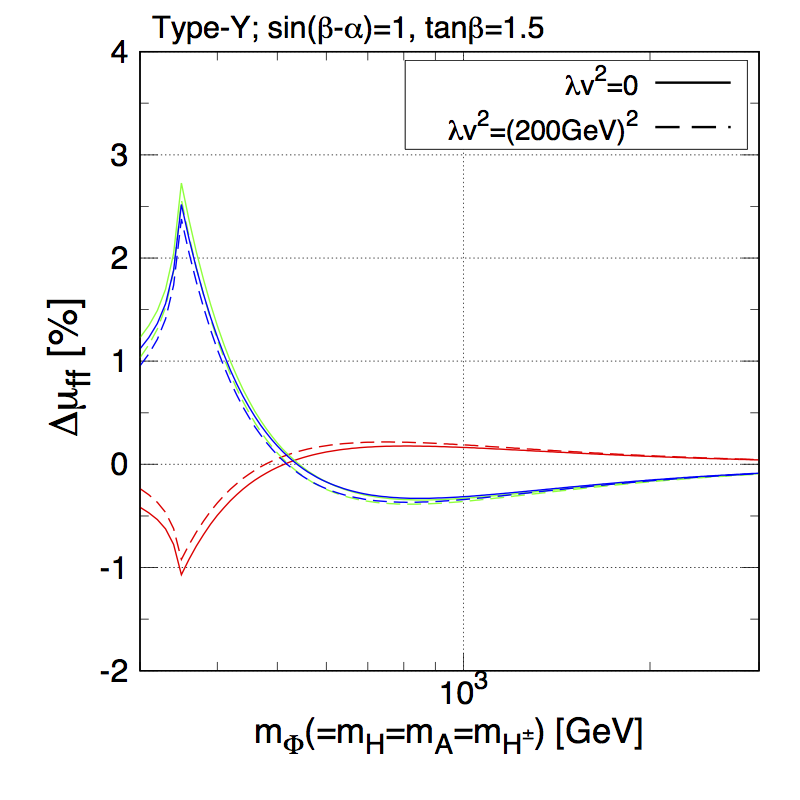} 
\caption{Predictions of $\Delta\mu_{ff}$ ($f=b,c,\tau$) defined in Eq.~(\ref{delta_nlo})
in the Type-I (upper-left), Type-II (upper-right), Type-X (lower-left) and Type-Y (lower-right) THDMs with $s_{\beta-\alpha} = 1$ and 
$\tan\beta = 1.5$ as a function of $m_\Phi (=m_H = m_A = m_{H^\pm})$. 
The solid and dashed curves show the case with $\lambda v^2 = 0$ and $(200~\text{GeV})^2$, respectively, where $\lambda v^2 = m_\Phi^2 - M^2$. }
\label{fig:dmu3}
\end{center}
\end{figure}

In Fig.~\ref{fig:dmu3}, we show $\Delta\mu_{ff}$ ($f = b,c,\tau$) as a function of $m_\Phi$ in four types of the THDMs with $s_{\beta-\alpha} = 1$ and $\tan\beta = 1.5$. 
{Here, we fix the value of $\lambda v^2 = m_\Phi^2 - M^2$ to be 0} (solid curves) and (200 GeV)$^2$ (dashed curves). 
We see the decoupling behavior in the large mass region, and observe the peak at around $m_\Phi = 2m_t$ depending on the type of Yukawa interaction and the type of fermions, where 
the direction of the peak is the same as that for the plots of $\overline{\Delta}_{\rm EW}^f$ shown in Fig.~\ref{fig:delbct}. 
Notice that the peak appearing at $m_\Phi > 2m_t$ in Fig.~\ref{fig:delbct} does not appear in this plot, as we here fix the value of $\lambda v^2$. 

\begin{figure}[!t]
\begin{center}
\includegraphics[scale=0.65]{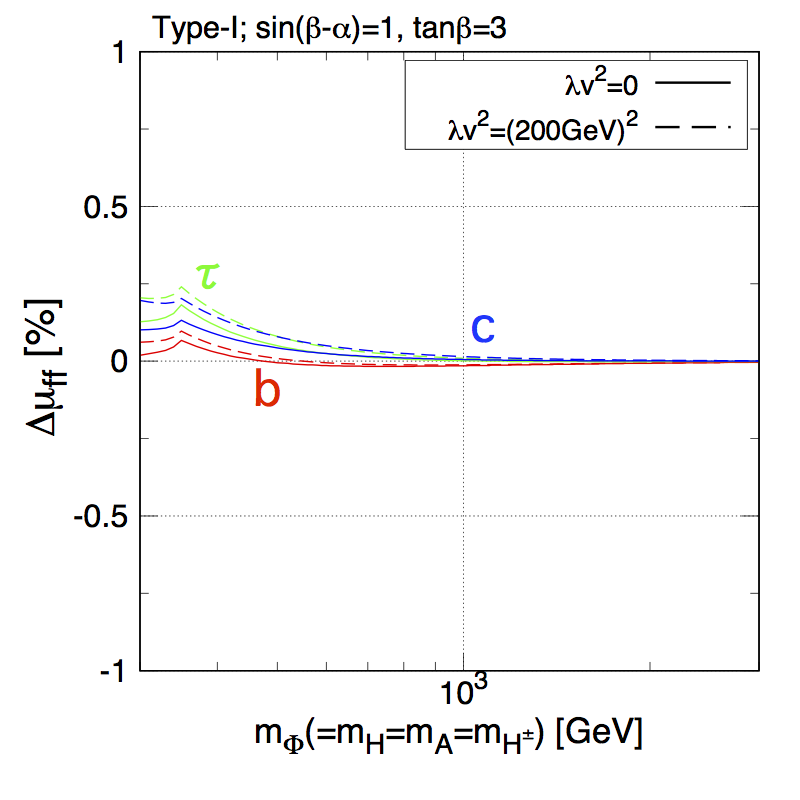} \hspace{3mm}
\includegraphics[scale=0.65]{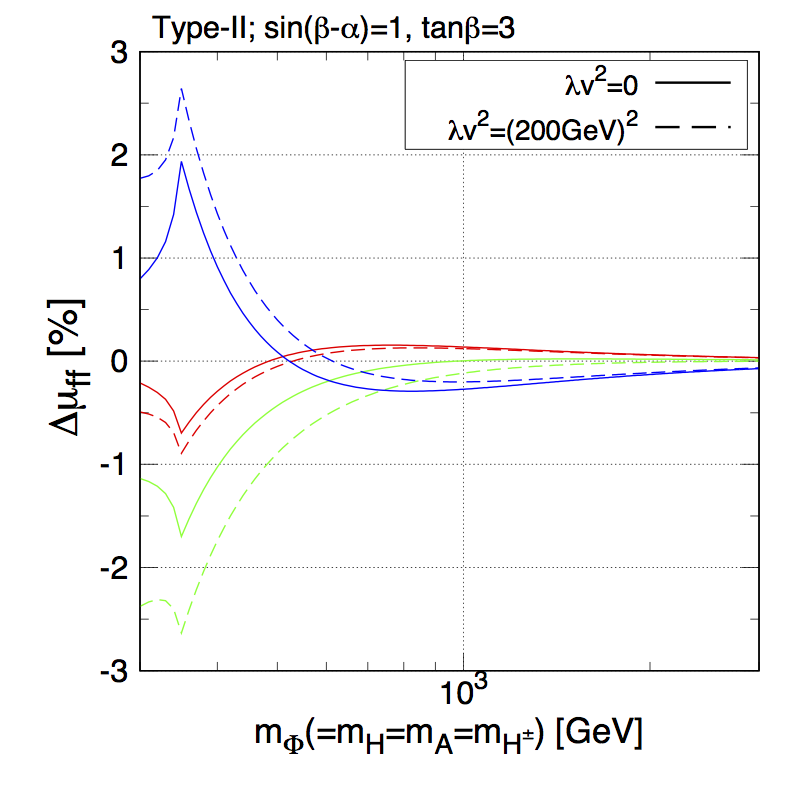} \\
\includegraphics[scale=0.65]{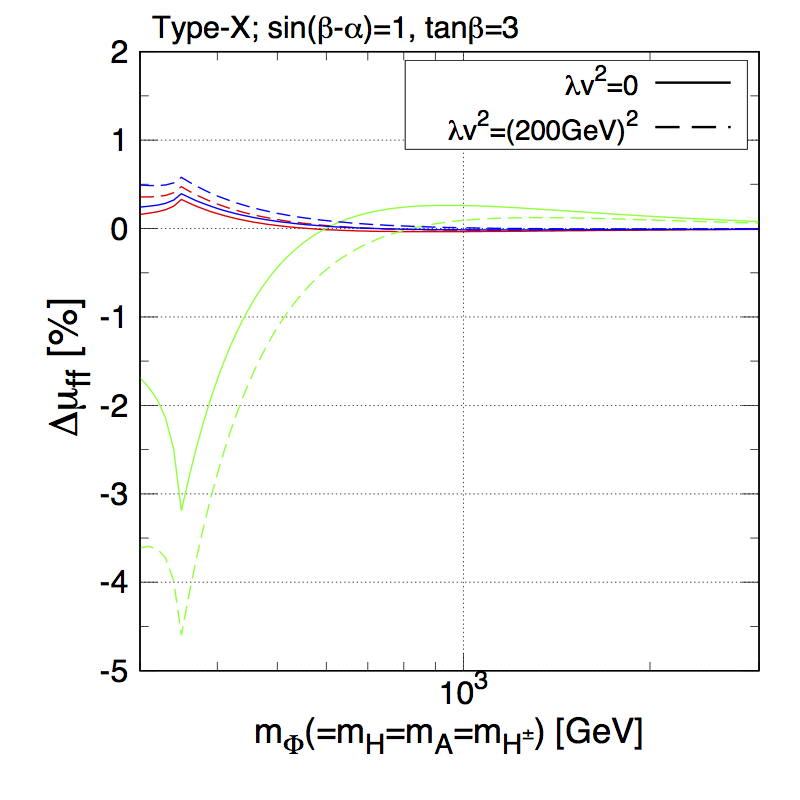} \hspace{3mm}
\includegraphics[scale=0.65]{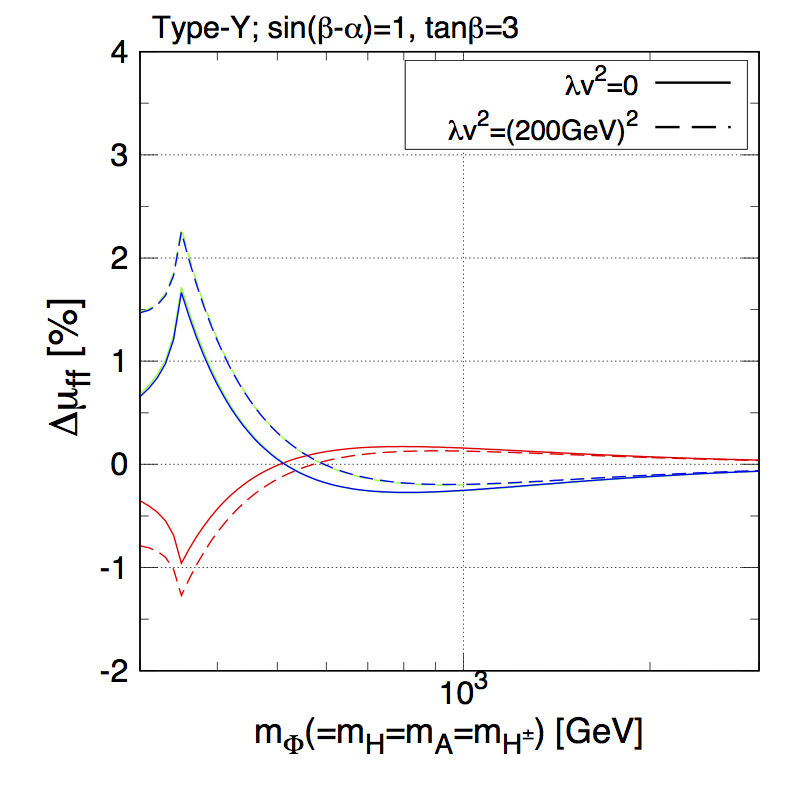} 
\caption{Same as Fig.~\ref{fig:dmu3}, but for $\tan\beta = 3$. }
\label{fig:dmu4}
\end{center}
\end{figure}

Fig.~\ref{fig:dmu4} shows the case for $\tan\beta = 3$, and all the other choices are the same as in Fig.~\ref{fig:dmu3}. 
We see that some of $\Delta\mu_{ff}$ values with $\lambda v^2 = (200~\text{GeV})^2$  are largely different from {those} with $\lambda v^2 = 0$. 
For example, the dashed curve for $\Delta\mu_{\tau\tau}$ (green) in the Type-X THDM is located lower than the corresponding solid curve 
because $\overline{\Delta}_{\rm EW}^\tau$ appearing in the first term of Eq.~(\ref{muxx_app}) has a smaller value in the case with $\lambda v^2 = (200~\text{GeV})^2$. 

\begin{figure}[!t]
\begin{center}
\includegraphics[scale=0.65]{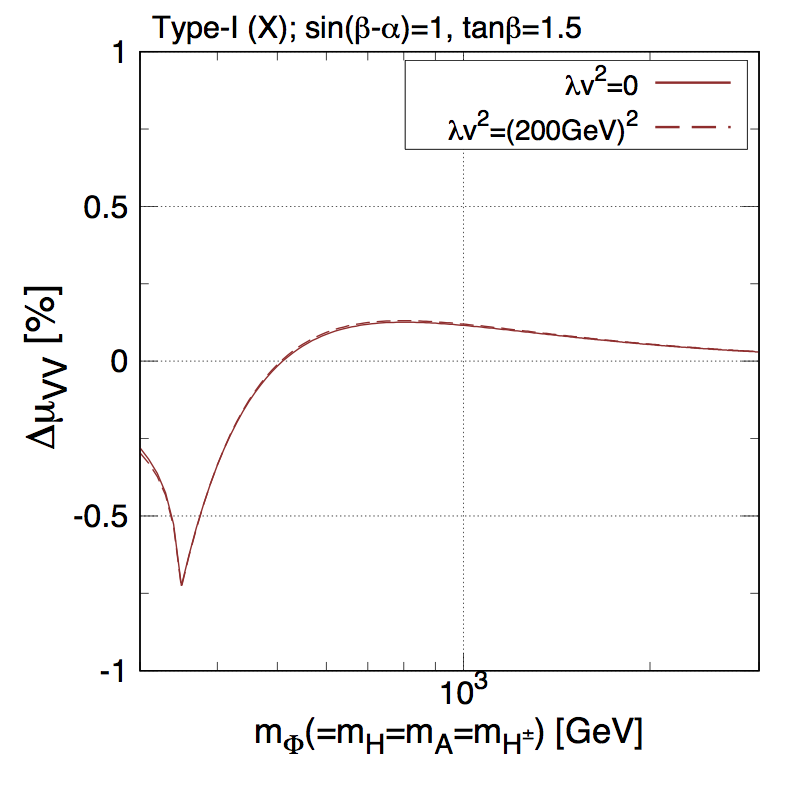} \hspace{3mm}
\includegraphics[scale=0.65]{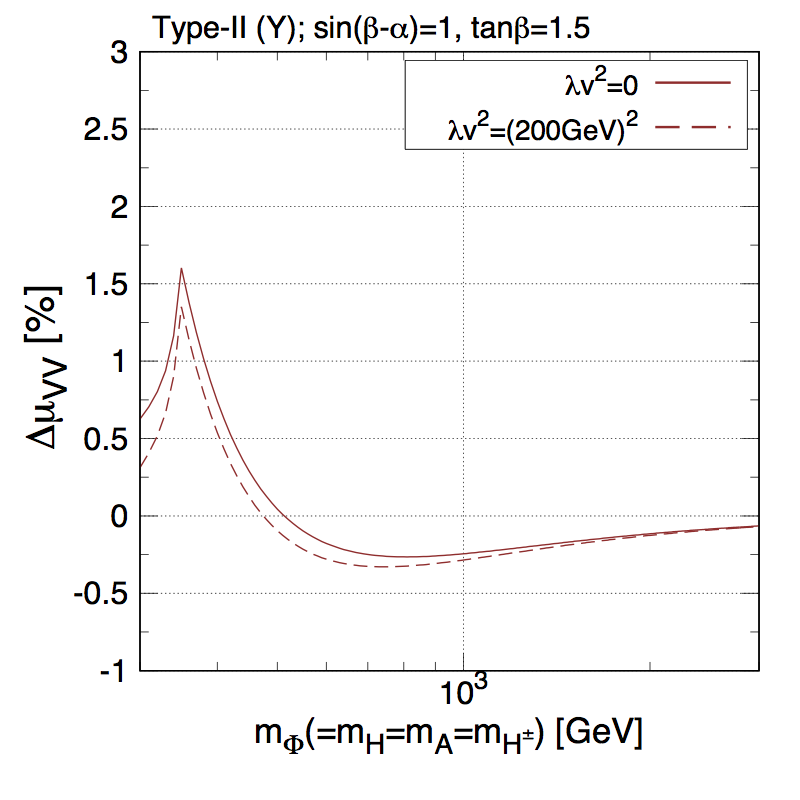} 
\caption{Predictions of $\Delta\mu_{VV}$ ($V=W,Z$) defined in Eq.~(\ref{delta_nlo})
in the Type-I (left) and Type-II (right) THDMs with $s_{\beta-\alpha} = 1$ and $\tan\beta = 1.5$ as a function of $m_\Phi (=m_H = m_A = m_{H^\pm})$. 
The solid and dashed curves show the case with $\lambda v^2 = 0$ and $(200~\text{GeV})^2$, where $\lambda v^2 = m_\Phi^2 - M^2$. }
 \label{fig:dmu1}
\vspace{3mm}
\includegraphics[scale=0.65]{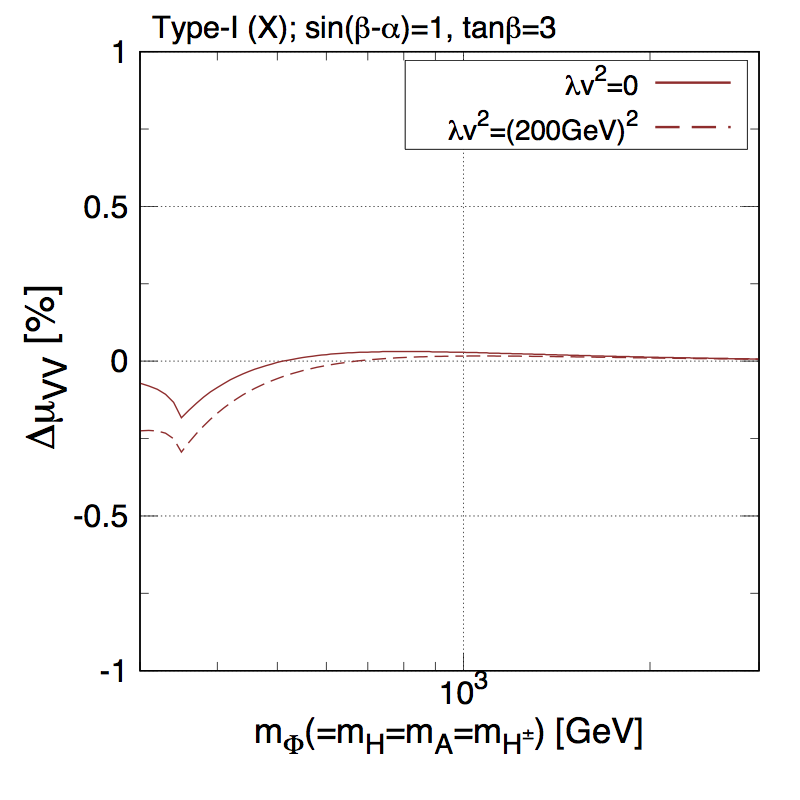} \hspace{3mm}
\includegraphics[scale=0.65]{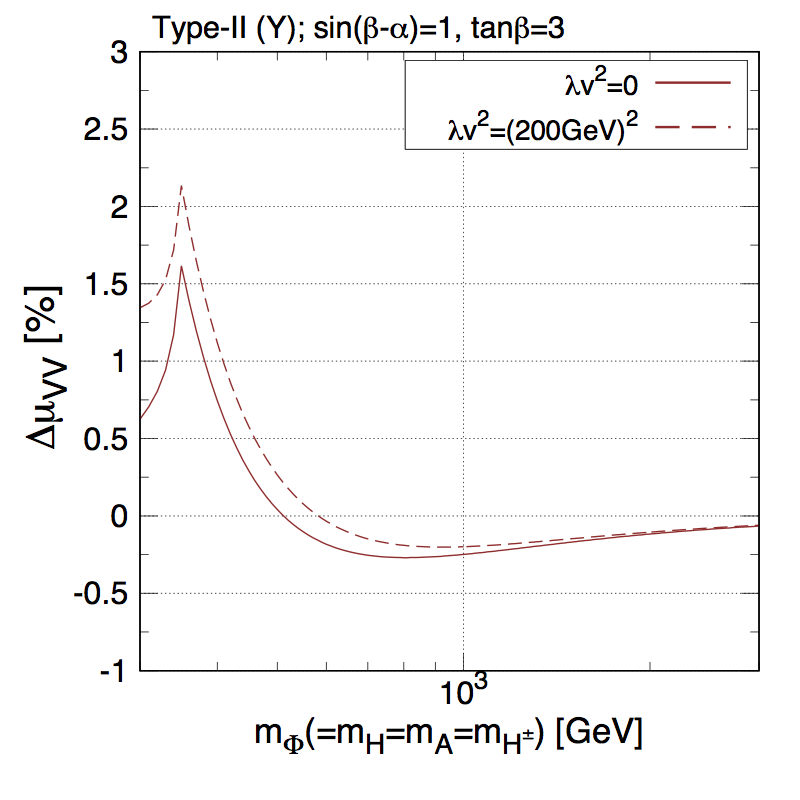} 
\caption{Same as Fig.~\ref{fig:dmu1}, but for $\tan\beta=3$. }
 \label{fig:dmu11}
 \end{center}
 \end{figure}

The $m_\Phi$ dependence of $\Delta\mu_{VV}$ $(V=W,Z)$ is shown in the THDMs with $s_{\beta-\alpha} = 1$ and $\tan\beta = 1.5$ (Fig.~\ref{fig:dmu1}) and 3 (Fig.~\ref{fig:dmu11}), 
where the values of $\Delta\mu_{WW}$ and $\Delta\mu_{ZZ}$ are almost the same {of} each other in this case. 
As in Fig.~\ref{fig:dmu3}, the value of $\lambda v^2$ is fixed to 0 (solid curves) and (200 GeV)$^2$ (dashed curves). 
The left and right panels show the results in the Type-I and Type-II THDMs, respectively, while the results in the Type-X (Type-Y) THDM
are almost the same as those in the Type-I (Type-II) THDM. 
In all the panels, the value of $\Delta \mu_{VV}$ approaches to zero in the large $m_\Phi$ region, because of the decoupling property of the additional Higgs bosons.  
In the left (right) panel, a peak appears at around $m_\Phi = 2m_t$, because the EW correction to the partial width of $h \to b\bar{b}$ mode has a peak 
in the Type-I and Type-X (Type-II and Type-Y) THDMs, see Figs.~\ref{fig:delb-1} and \ref{fig2}.  
We can also see that in the Type-I THDM with $\tan\beta = 1.5$
the value of $\Delta\mu_{VV}$ with $\lambda v^2=(200~\text{GeV})^2$ 
is almost the same as that with $\lambda v^2 = 0$, because the change of $\overline{\Delta}_{\text{EW}}^b$ due to taking different values of $\lambda v^2$ is 
accidentally cancelled by that of $\overline{\Delta}_{\text{EW}}^V$. 
For  $\tan\beta = 3$, 
the value of $\Delta\mu_{VV}$ with $\lambda v^2=(200~\text{GeV})^2$ is slightly smaller than that with $\lambda v^2=0$, because the change
of $\overline{\Delta}_{\text{EW}}^b$ becomes smaller than that for $\tan\beta = 1.5$, while $\overline{\Delta}_{\text{EW}}^V$ does not depend on $\tan\beta$ so much. 
On the other hand, in the Type-II THDM with $\tan\beta = 3$
the value of $\Delta\mu_{VV}$ with $\lambda v^2=(200~\text{GeV})^2$ is larger than that for $\lambda v^2=0$, 
because of the larger negative shift of $\overline{\Delta}_{\text{EW}}^b$.

\subsection{Correlations \label{sec:correlations}}

\begin{table}[!t]
\begin{center}
{\renewcommand\arraystretch{1.2}
\begin{tabular}{cccccccc}\hline\hline
$h \to b\bar{b}$ & $h \to c\bar{c}$ & $h \to \tau\bar{\tau}$ & $h \to WW^*$ & $h \to ZZ^*$ & $h \to gg$ & $h \to \gamma \gamma$& $h \to \mu\mu$\\\hline
0.89\% & 3.2\% & 1.4\% & 1.9\% & 6.7\% & 2.7\% & 13\% & 27\%  \\\hline\hline
\end{tabular}}
\caption{Expected 1$\sigma$ uncertainty for the measurements of the branching ratios of the Higgs boson $h$ at ILC250~~\cite{Fujii:2017vwa}. }
\label{tab:br-ilc}
\end{center}
\end{table}

 \begin{figure}[!t]
  \begin{center}
  \includegraphics[scale=0.75]{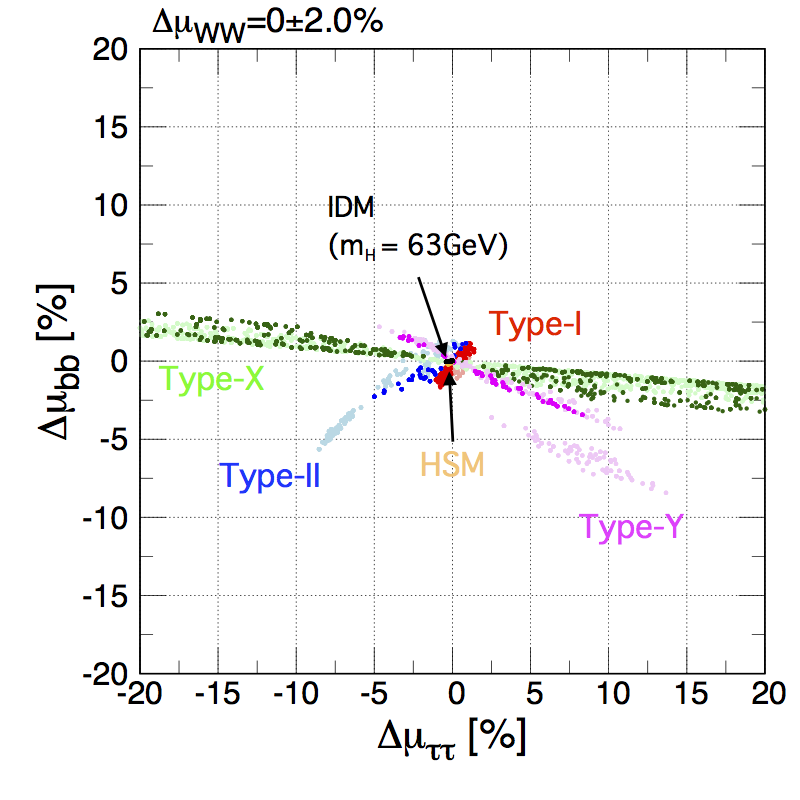} \hspace{3mm}
  \includegraphics[scale=0.75]{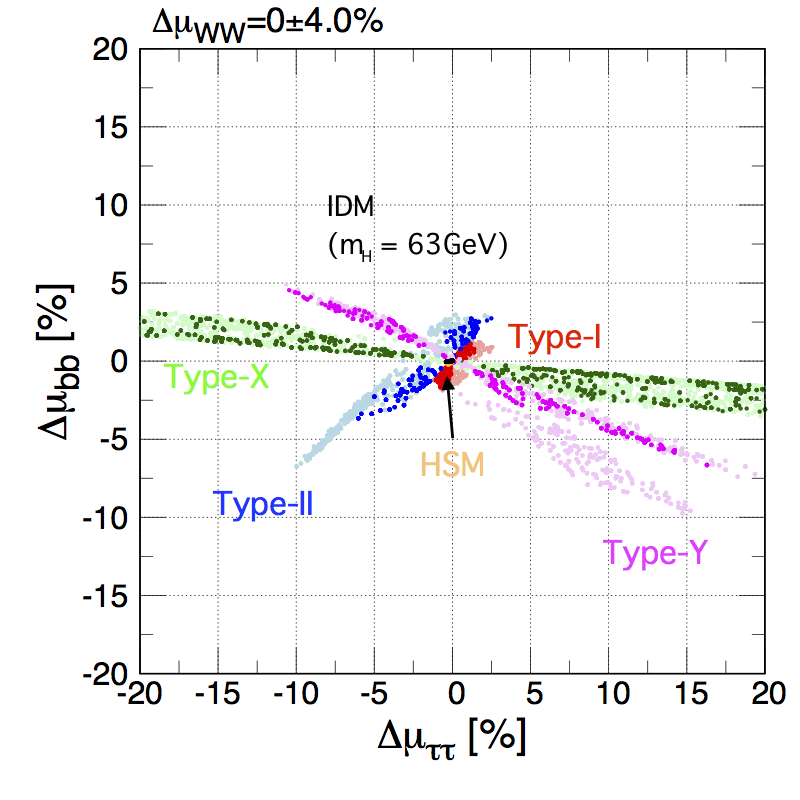} 
  \caption{Correlation between $\Delta\mu_{\tau\tau}^{}$ and $\Delta\mu_{bb}^{}$ in the Type-I (red), Type-II (blue), 
 Type-X (green), Type-Y (magenta) THDMs, the HSM (orange) and the IDM (black). 
 The left (right) panel shows the case with $\Delta\mu_{WW}=0 \pm 2\%$ $(0 \pm 4\%)$. 
 In the THDMs, we scan $1.5 \leq \tan\beta \leq 10$, $0\leq M^2 \leq m_\Phi^2$ and $300~(600)\leq m_\Phi \leq 1000$ GeV for (darker) colored points.
 In the HSM, we scan $300 \leq m_H \leq 5000$ GeV and $0\leq M^2 \leq m_H^2$, 
while in the IDM we fix $m_H$ to be 63 GeV and scan $100 \leq m_A^{}(=m_{H^\pm}^{}) \leq 1000$ GeV. }
  \label{fig:corr1} 
  \end{center}
  \end{figure}

\begin{figure}[!t]
 \begin{center}
 \includegraphics[scale=0.75]{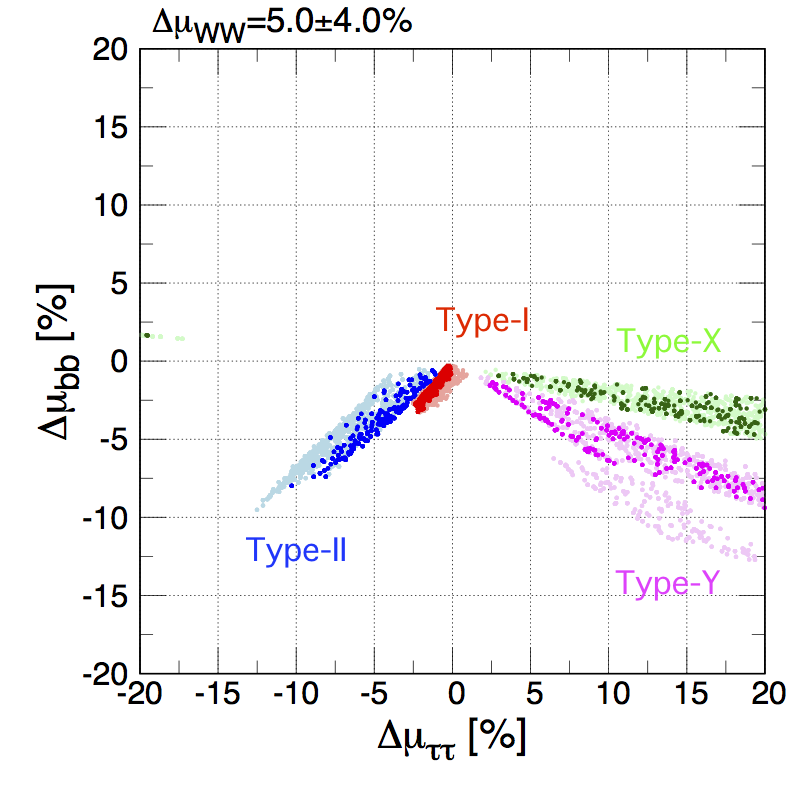}  \hspace{3mm}
 \includegraphics[scale=0.75]{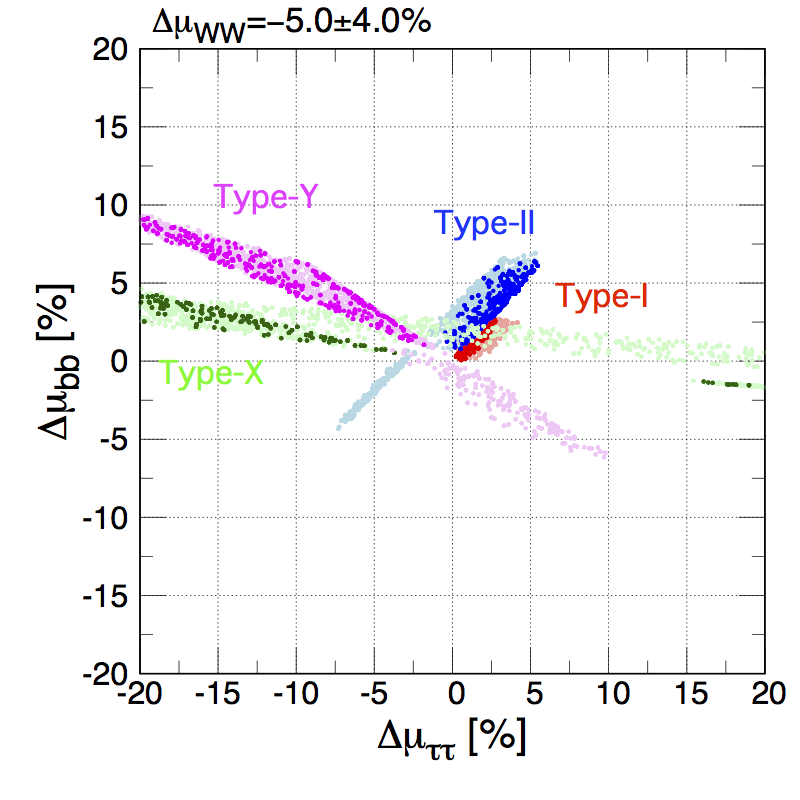} 
 \caption{Same as Fig.~\ref{fig:corr1}, but for the case with 
$\Delta\mu_{WW}$ to be $5 \pm 4\%$ (left) and $-5 \pm 4\%$ (right).  }
 \label{fig:corr2}
 \end{center}
 \end{figure}

In the discussions so far, we have seen the deviation in the total width and those in branching ratios in each extended Higgs sector. 
We now see correlations among the deviations in branching ratios in all the extended Higgs sectors discussed in this paper in order to clarify how we can distinguish 
extended Higgs sectors from the precise measurements of the branching ratios. 

The branching ratios of the Higgs boson will be measured as accurately as possible at future collider experiments. 
In particular at the ILC, we can measure the cross section of $e^+e^- \to Zh$ without depending on the decay of $h$ by using the recoil method~\cite{Yan:2016xyx,Asner:2013psa}. 
This makes the measurements of the branching ratios possible without depending on the cross section. 
At the ILC with the collision energy of 250 GeV 
and the integrated luminosity of 2 ab$^{-1}$ (ILC250), the branching ratios are expected to be measured as shown in Table~\ref{tab:br-ilc}. 
We thus consider the situation where the branching ratios are measured to some extent at ILC250, 
namely we impose the further constraint on the value of $\Delta \mu_{XX}$ with a given central value and an error in addition to the theoretical constraints which are imposed in the discussion above. 

In order to take into account the constraints from $B_s \to X_s \gamma$~\cite{Misiak:2017bgg,Haller:2018nnx}, 
we consider the case with larger masses of extra Higgs bosons, i.e, $m_\Phi \geq 600$ GeV 
in the THDMs as well as that for $m_\Phi \geq 300$ GeV. 
As discussed in Sec.~\ref{sec:model}, the lower bound on $m_{H^\pm}^{}$ is about 600 GeV in the Type-II and Type-Y THDMs. 

\begin{figure}[!t]
 \begin{center}
 \includegraphics[scale=0.75]{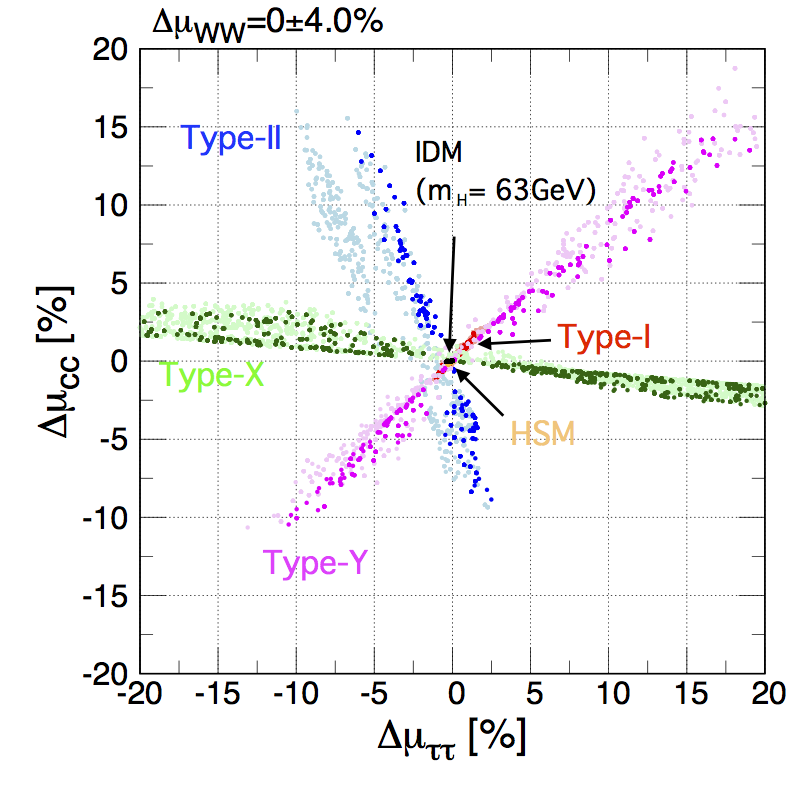} \\
 \includegraphics[scale=0.75]{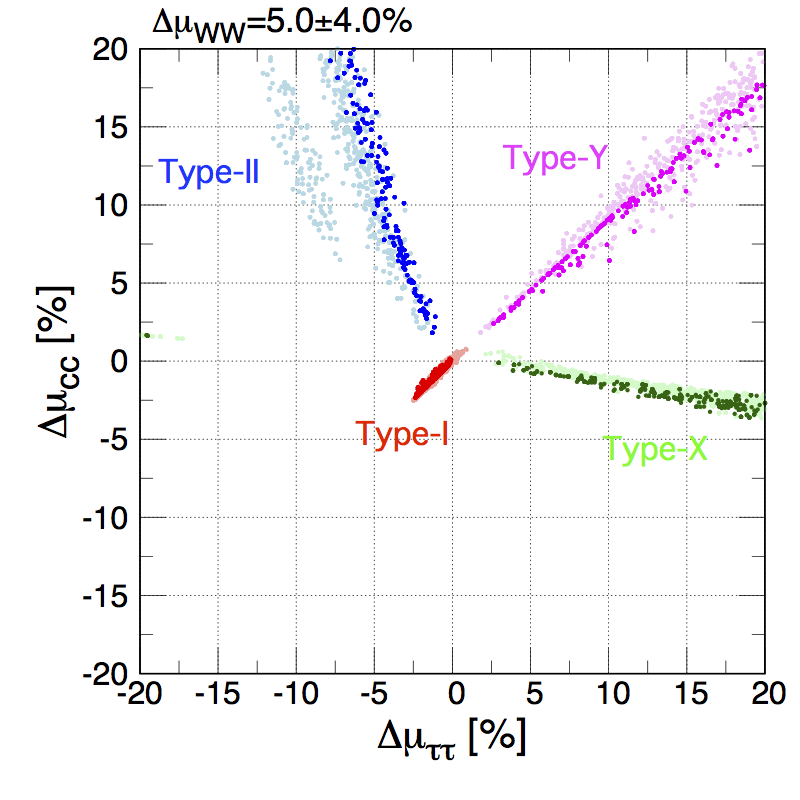} \hspace{3mm} 
 \includegraphics[scale=0.75]{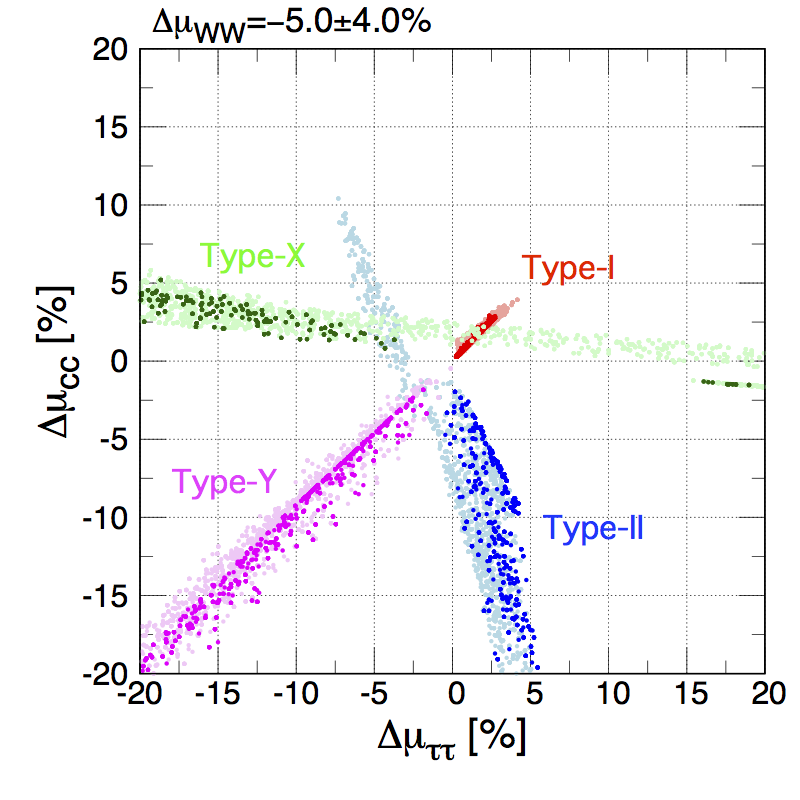}
 \caption{Correlation between $\Delta\mu_{\tau\tau}^{}$ and {$\Delta\mu_{cc}^{}$} in the Type-I (red), 
Type-II (blue), Type-X (green) and Type-Y (magenta) THDMs. The upper panel shows the case with 
$\Delta\mu_{WW}^{}=0\pm 4\%$, while the lower left (right) panel shows the case with 
$\Delta\mu_{WW}^{} = 5 \pm 4\%$ ( $\Delta\mu_{WW}^{} = -5 \pm 4\%$). 
The ranges of scanning parameters are the same as those of Fig.~\ref{fig:corr1}.}
 \label{fig:corr3}
 \end{center}
 \end{figure}

In Fig.~\ref{fig:corr1}, we show the correlation between $\Delta\mu_{\tau\tau}^{}$ and $\Delta\mu_{bb}^{}$ in the HSM, four types of the THDMs and the IDM under the additional constraint from $\Delta \mu_{WW} = 0\pm 2\%$ (left) and $\Delta \mu_{WW} = 0\pm 4\%$ (right). 
The errors of 2\% and 4\% are taken to consider about 1$\sigma$ and 2$\sigma$ level region at ILC250, respectively~\cite{Fujii:2017vwa}. 
Parameters of each model are scanned as it is described in the caption. 
We see that the predictions in the HSM and the IDM are given almost at the origin of this plane, because in these models the partial decay rates are almost universally suppressed as we already mentioned in Sec.~\ref{sec:br}. 
On the other hand, the predictions of the THDMs are spread out into the different directions depending on the type of Yukawa interactions. 
In the Type-Y THDM, two allowed regions appear in the fourth quadrant {if we choose values for $m_\Phi$ starting from $300$ GeV} (lighter points). 
This can be explained from Fig.~\ref{fig:br1} (lower-rightmost panel), where we can find the two values of $\tan\beta$ providing the same value of BR$(h \to \tau\bar{\tau})$ while different values of BR$(h \to b\bar{b})$. 
This, however, vanishes when we take $m_\Phi \geq 600$ GeV (darker points), as such configuration is favored by the flavor experiments, particularly in the Type-II and Type-Y THDMs. 
Remarkably, only in the Type-X THDM, the allowed points are distributed in the wide (small) range of $\Delta\mu_{\tau\tau}$ $(\Delta\mu_{bb})$.
This can also be understood from the third panels of Fig.~\ref{fig:br1}, where only the BR($h \to \tau\bar{\tau}$) mode can significantly be changed depending on $\tan\beta$, while all the other branching ratios 
do not change so much. 
{In contrast in the Type-II THDM large variations of  BR$(h\to\tau\tau)$ appear together with large variations of BR$(h\to bb)$.}
Thus, the other decay branching ratios, particularly the $h \to WW^*$, also strongly vary at the same time, {and then 
such configurations are constrained by the bound on $\Delta \mu_{WW}$. }
From this figure, we find that if $\Delta \mu_{\tau\tau}$ is found to be a several percent, we can distinguish the models considered in this paper. 
In the following discussion, we focus on the case with $\Delta \mu_{WW}$ to be constrained at the 2$\sigma$ level{,} i.e., allowing 4\% uncertainty. 

Let us also consider the case where the central value of $\Delta\mu_{WW}$ is found to be nonzero, and $\Delta\mu_{WW} = 0$ is excluded {at} the 2$\sigma$ level. 
In Fig.~\ref{fig:corr2}, we show {such situations with $\Delta\mu_{WW} = 5.0\pm 4.0\%$ (left) and $\Delta\mu_{WW} = -5.0\pm 4.0\%$ (right).} 
In this setup, predictions of $\Delta\mu_{WW}$ in the HSM and the IDM are almost zero, so that these models are 
excluded, while four types of THDMs can explain such a deviation. 
If the value of $\Delta\mu_{WW}$ is given to be $5.0 \pm 4.0\%$ (left), then four types of Yukawa interactions are {well separated of one another}, so 
that we can determine the type by measuring $\Delta\mu_{\tau\tau}$ and $\Delta\mu_{bb}$ in addition to $\Delta\mu_{WW}$. 
We note that the positive value of $\Delta\mu_{WW}$ essentially comes from the reduction of the other decay rates, especially the $h \to b\bar{b}$ mode, 
because the partial width of $h \to WW^*$ reduces by the tree level scaling factor $\kappa_V^{} \leq 1$ and the one-loop effect as seen from Fig.~\ref{fig:delw}. 
In fact, in the left panel of Fig.~\ref{fig:corr2}, the allowed points mainly appear in the regions with 
$\Delta\mu_{bb} < 0$.

For the case with  $\Delta\mu_{WW} = -5.0\pm 4.0\%$, some of the THDMs are overlapped in this plane. 
This is because the negative value of $\Delta\mu_{WW}$ can be explained by either decreasing the partial width of the $h \to WW^*$ mode
or increasing the other partial widths. 
Therefore, the branching ratio of $h \to b\bar{b}$ can be either larger or smaller than the SM prediction, as seen in the right panel. 
This makes discrimination among the four types of Yukawa interactions difficult as compared to the case with positive $\Delta \mu_{WW}$. 

In Fig.~\ref{fig:corr3}, we show the correlation between $\Delta \mu_{\tau\tau}$ and $\Delta \mu_{cc}$ under the constraint on $\Delta\mu_{WW}$ with the 4\% uncertainty. 
The central value of $\Delta\mu_{WW}$ is supposed to be 0 in the upper panel, and to be $+5\%$ and $-5\%$ in the lower left and right panels, respectively. 
As compared to Fig.~\ref{fig:corr1} (right), the allowed points on the upper panel are widely distributed in the $\Delta \mu_{\tau\tau}$ and $\Delta \mu_{cc}$ plane, because 
the branching ratio of $h \to c\bar{c}$ typically has a smaller portion of the total width than that of $h \to b\bar{b}$. 
If we only look at the correlation shown in the upper panel, it seems difficult to distinguish the models unless $\Delta\mu_{\tau\tau}$ is given to several percent level. 
However, we would like to emphasize that by looking also at the corresponding plot shown in Fig.~\ref{fig:corr1} (right), 
we can distinguish the models. 
For example, if $\Delta \mu_{\tau\tau}$ and $\Delta \mu_{cc}$ are measured {to be small negative and positive respectively, i.e., the second quadrant in this plane, 
both the Type-II and Type-X THDMs} can explain such situation, but these two models may not be distinguished {from} each other. 
However, by looking at the correlation between $\Delta \mu_{\tau\tau}$ and $\Delta \mu_{bb}$ with a negative value of $\Delta \mu_{\tau\tau}$, the Type-II and Type-X THDMs have allowed points {in the different directions compared to each other}. 
Therefore, we can distinguish all the four THDMs by using the combination of these correlations even if the central value of $\Delta \mu_{WW}$ is measured to be close to zero. 

Similar to Fig.~\ref{fig:corr2}, we show the case for nonzero $\Delta \mu_{WW}$ at the 2$\sigma$ level in the lower two plots in Fig.~\ref{fig:corr3}. 
We see that for the case with the central value of $\Delta \mu_{WW}$ to be $+5\%$, 
four types of THDMs are clearly {separated from one another}, while the case with the central value of $\Delta \mu_{WW}$ to be $-5\%$ 
two models are overlapping at some regions of this plane. 
However, again  the correlation between $\Delta \mu_{\tau\tau}$ and $\Delta \mu_{bb}$ helps for further discrimination of the models.  

\begin{figure}[!t]
 \begin{center}
 \includegraphics[scale=0.75]{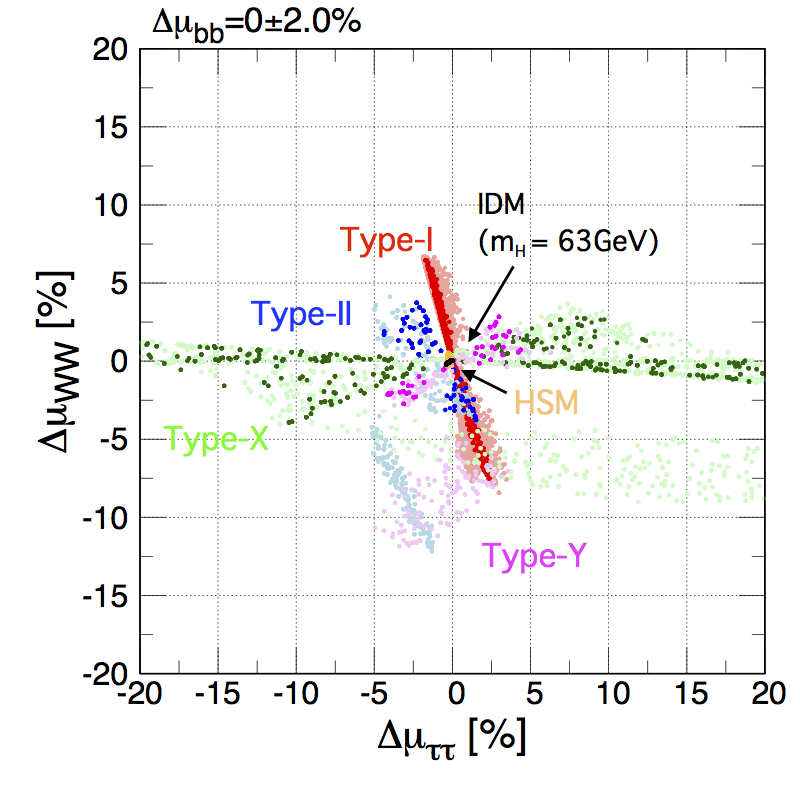} \\
 \includegraphics[scale=0.75]{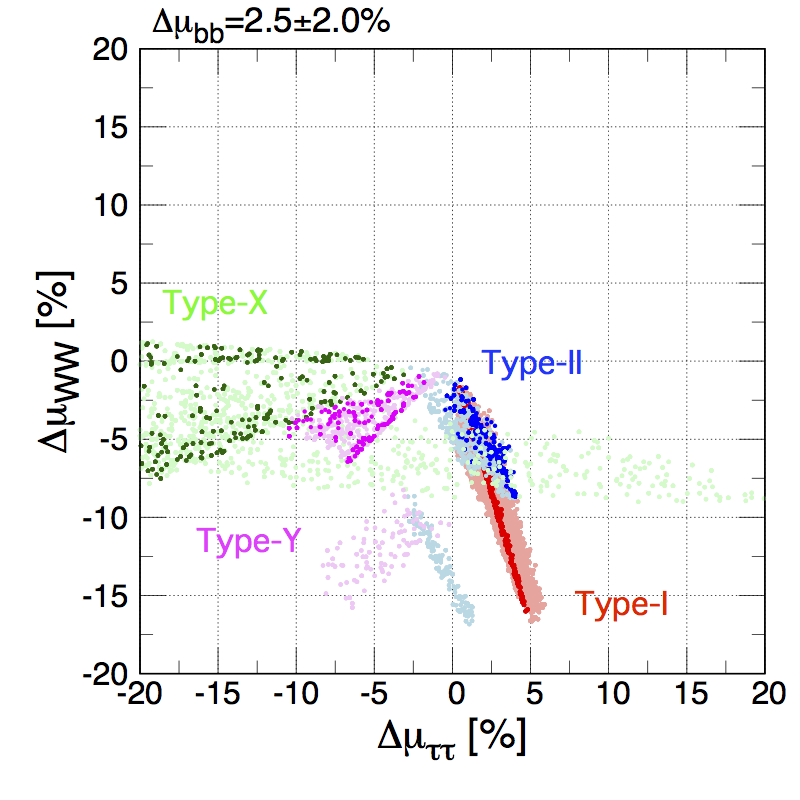} \hspace{3mm}
 \includegraphics[scale=0.75]{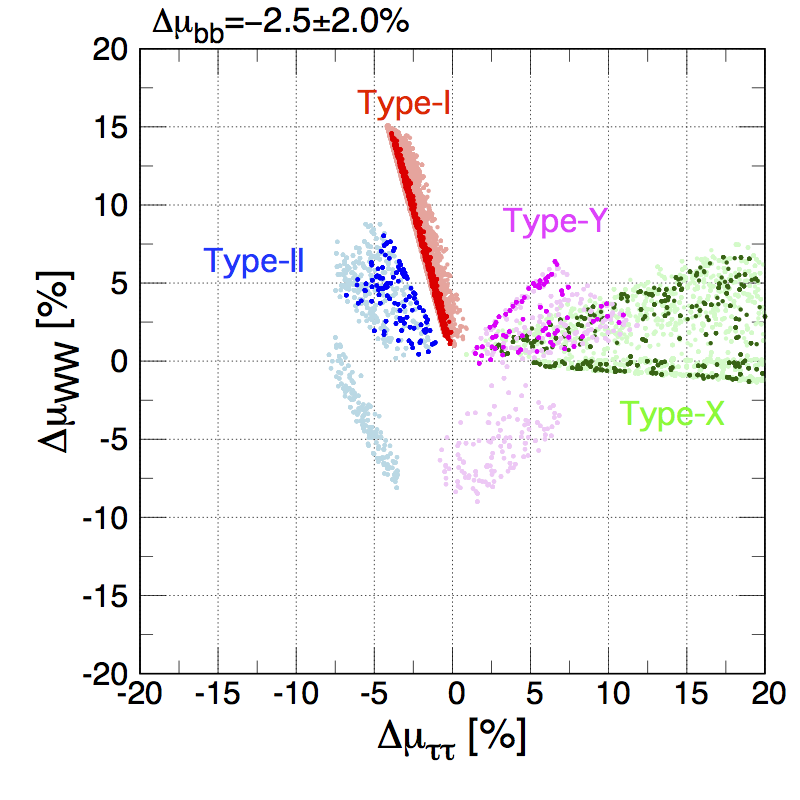}
 \caption{Correlations between $\Delta\mu_{\tau\tau}^{}$ and $\Delta\mu_{WW}^{}$ in the Type-I (red), Type-II (blue), 
Type-X (green) and Type-Y (magenta) THDMs. 
The upper panel shows the case with $\Delta\mu_{bb}=0 \pm 2\%$, while the lower-left (right) panel shows the 
case with  $\Delta\mu_{bb}=2.5 \pm 2\%$ ($\Delta\mu_{bb}=-2.5 \pm 2\%$). 
The values of $\tan\beta$, $M^2$ and $m_\Phi (=m_H = m_A = m_{H^\pm})$ are scanned with the ranges of 
$1.5 \leq \tan\beta \leq 10$, $0\leq M^2 \leq m_\Phi^2$  and $300 \leq m_\Phi \leq 1000$ GeV, respectively, 
under the constraints from the perturbative unitarity, the vacuum stability and the $S$, $T$ parameters. }
 \label{fig:corr4}
 \end{center}
 \end{figure}

In the above analyses, we constrained the value of $\Delta \mu_{WW}$. 
We now constrain $\Delta \mu_{bb}$ instead of $\Delta \mu_{WW}$. 
In Fig.~\ref{fig:corr4}, we show the correlation between $\Delta \mu_{\tau\tau}$ and $\Delta \mu_{WW}$ under the constraint on 
$\Delta \mu_{bb} = 0\pm 2\%$ (upper panel), 
$\Delta \mu_{bb} = 2.5\pm 2\%$ (lower left panel) and 
$\Delta \mu_{bb} = -2.5\pm 2\%$ (lower right panel). 
The error of 2\% is taken to consider about the 2$\sigma$ level region at ILC250. 
For the case with $\Delta \mu_{bb} = 0\pm 2\%$, if $\Delta \mu_{\tau\tau}$ is measured to be a few percent level, 
it might be difficult to distinguish the models shown in this figure, particularly for the case with $m_\Phi > 600$ GeV shown as darker points. 
If we consider the case with $300 < m_\Phi < 600$ GeV, it is seen that  the points with $-10\% \lesssim \Delta \mu_{WW} \lesssim -5\%$ are also allowed in 
all the four types of THDMs.  
When we consider the case with $\Delta \mu_{bb} = 2.5\pm 2\%$, the situation is drastically changed, where most of the allowed points 
are distributed in the region with $\Delta \mu_{WW} <0$, because of the compensation of the positive deviation in the branching ratio of the $h \to b\bar{b}$ mode. 
The opposite situation can be seen in the lower right panel showing the case with $\Delta \mu_{bb} = -2.5\pm 2\%$. 

\begin{figure}[!t]
 \begin{center}
 \includegraphics[scale=0.75]{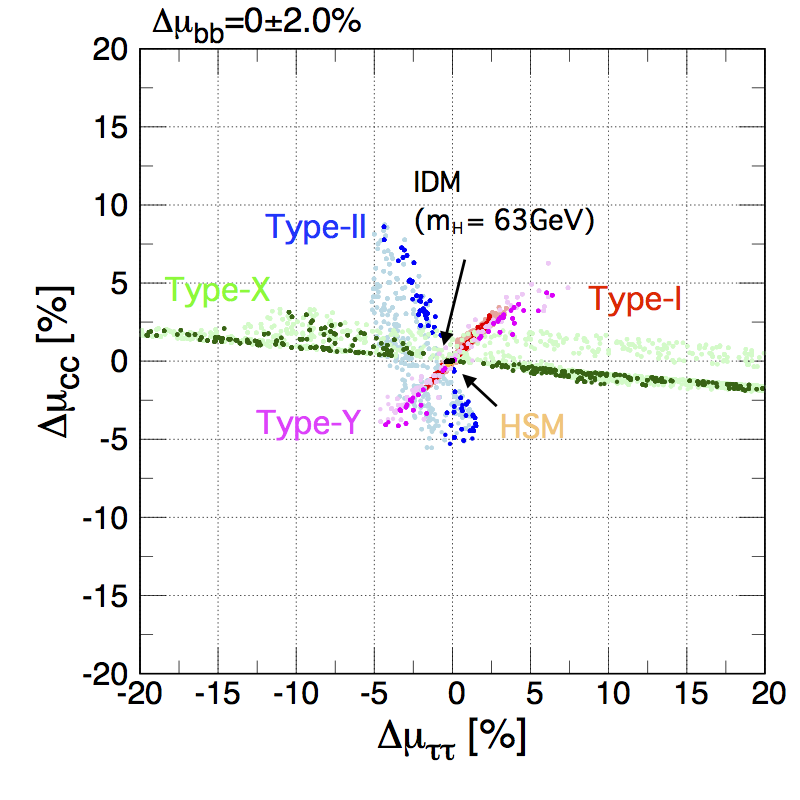} \\
 \includegraphics[scale=0.75]{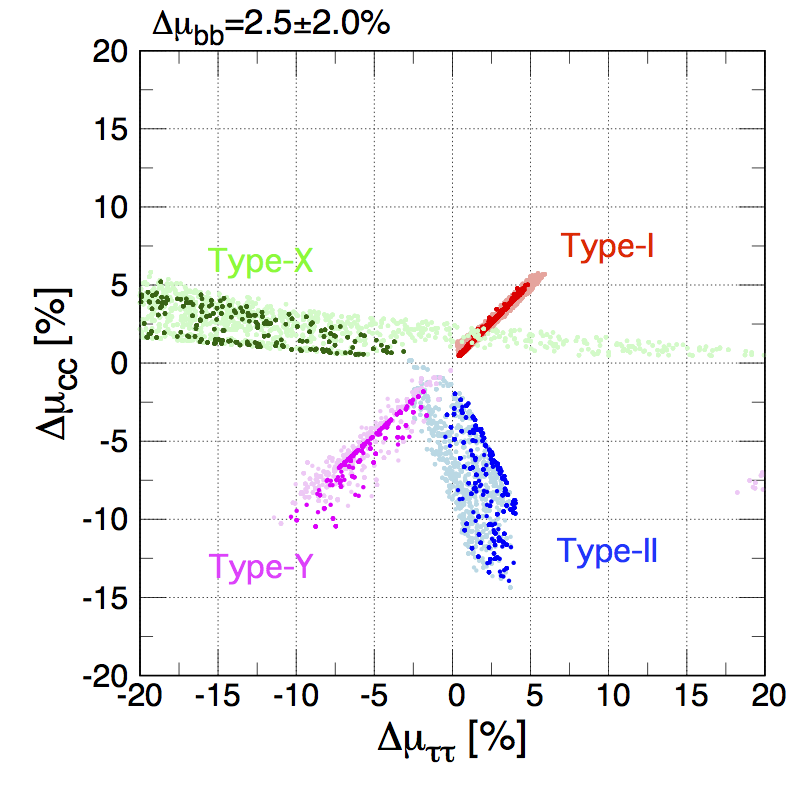} \hspace{3mm}
 \includegraphics[scale=0.75]{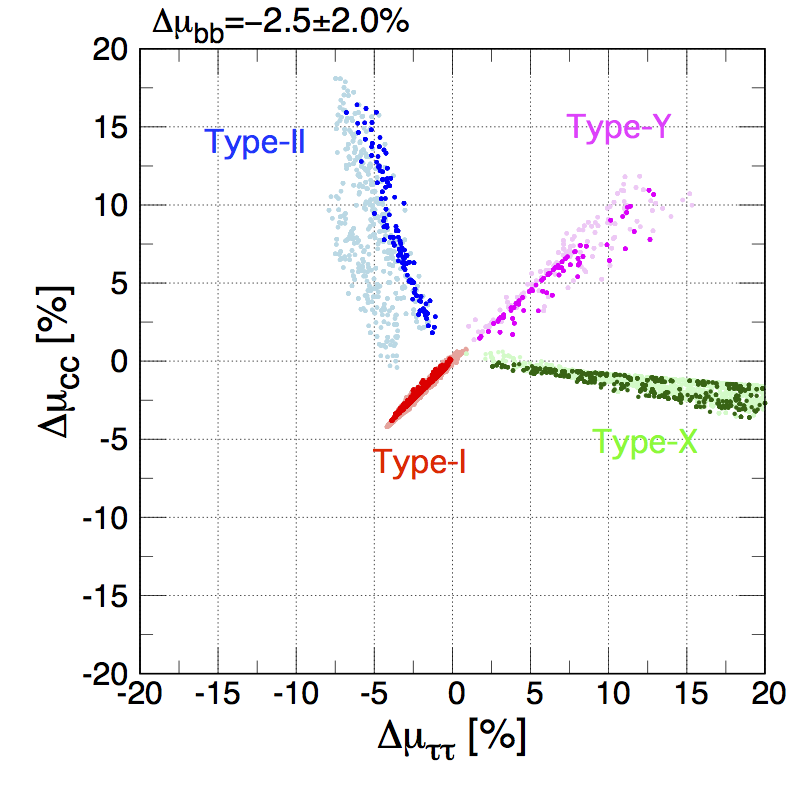}
 \caption{Same as Fig.~\ref{fig:corr4}, but for the correlation between $\Delta\mu_{\tau\tau}$ and $\Delta\mu_{cc}$. 
}
 \label{fig:corr5}
 \end{center}
 \end{figure}

Finally, we show the correlation between $\Delta \mu_{\tau\tau}$ and $\Delta \mu_{cc}$ in Fig.~\ref{fig:corr5} using {the same setup as in} Fig.~\ref{fig:corr4}. 
The shape of the upper panel looks similar to that seen in the upper panel of Fig.~\ref{fig:corr3}, while the allowed points in Fig.~\ref{fig:corr5} are distributed in 
smaller regions than those shown in Fig.~\ref{fig:corr3}. 
This is simply because the foreseen uncertainty of the measurements of $\Delta \mu_{bb}$ is smaller as compared to that of  $\Delta \mu_{WW}$. 
Interestingly, for both the cases of $\Delta\mu_{bb}= 2.5 \pm 2\%$ (left) and $\Delta\mu_{bb}= -2.5 \pm 2\%$ (right)
four types of the THDMs are well separated, and it becomes clearer when $m_\Phi$ is taken to be greater than 600 GeV. 

In this subsection, we have discussed various correlations between the deviations in branching ratios from the SM predictions at NLO. 
First, if we observe a percent level deviation in one of the decay modes of $h$, 
then the HSM and IDM could be ruled out as the branching ratios are almost the same as the SM predictions in these models.
Second, the discrimination of four types of the THDMs strongly depends on the situation. 
If we observe a positive deviation in the branching ratio for the $h \to WW^*$ mode,  
the discrimination is possible by looking at the correlation between $\Delta \mu_{\tau\tau}$ and $\Delta \mu_{bb}$ or 
$\Delta \mu_{\tau\tau}$ and $\Delta \mu_{cc}$. 
On the contrary, if we measure a negative deviation in the branching ratio of the $h \to WW^*$ mode, then 
the discrimination of four types becomes more complicated as two of four models can overlap with each other 
in the correlation of $\Delta \mu_{\tau\tau}$ and $\Delta \mu_{bb}$ or 
$\Delta \mu_{\tau\tau}$ and $\Delta \mu_{cc}$. 
However, using three observables $\Delta \mu_{\tau\tau}$, $\Delta \mu_{bb}$ and $\Delta \mu_{cc}$ with the results from the direct searches of 
additional scalar bosons and from flavor experiments, 
we may be able to separate four types of the THDMs. 

\section{Conclusions\label{sec:conc}}

We have discussed the total width and the branching ratios of the 125 GeV Higgs boson $h$ at NLO in EW and QCD
in the HSM, four types of the THDMs and the IDM. 
These quantities can be measured at collider experiments as precisely as possible under a given machine performance.
Thus, accurate calculations for the total width and branching ratios are quite important to compare them with future precision data{,} e.g. at the HL-LHC and the ILC. 
For the one-loop computation, we systematically applied the on-shell renormalization scheme for each model, in which 
we adopted the {\tt H-COUP} program to evaluate numerical values of the renormalized Higgs boson vertices. 
The analytic expressions for the decay rates of $h \to f\bar{f}$, $h \to ZZ^* \to Zf\bar{f}$ and $h \to WW^* \to Wf\bar{f}'$ are 
presented at NLO, among which the $h \to WW^* \to Wf\bar{f}'$ mode is newly computed in this paper. 
We also provided the decay rates of the loop induced processes; i.e., $h \to \gamma\gamma$, $h \to Z\gamma$ and $h \to gg$ with QCD corrections at NLO. 

We have shown that in the HSM and the IDM, all the partial decay rates are almost universally suppressed by both the tree level mixing (for the HSM) and the one-loop effect, 
so that the branching ratios remain almost the same values as those in the SM. 
Thus, if deviations in the branching ratio from the {SM prediction (denoted as $\Delta \mu_{XX}^{}$ for the decay $h\to XX$)} are found, then we may be able to exclude the HSM and IDM. 
On the contrary, when we observe the deviation in the total width but not in the branching ratios, then it could be a smoking gun signature to identify these two models. 
We also have found that if we observe a positive deviation in the branching ratio of the $h \to WW^*$ mode, four types of the THDMs 
can be well {separated from one another} from the correlation between $\Delta \mu_{\tau\tau}$ and $\Delta \mu_{bb}$ or
$\Delta \mu_{\tau\tau}$ and $\Delta \mu_{cc}$. 
If we observe a negative deviation in the branching ratio of the $h \to WW^*$ mode, 
some of the THDMs can overlap in the $\Delta \mu_{\tau\tau}$ and $\Delta \mu_{bb}$ plane, but this can be disentangled by further looking at another correlation, such as $\Delta \mu_{\tau\tau}$ and $\Delta \mu_{cc}$. 

While the branching ratios are measured with a percent level at future precision experiments,
direct searches for additional Higgs bosons are expected to make progress at the LHC Run-III and the HL-LHC.
If additional Higgs bosons are discovered, we can give stronger predictions of the correlation among the branching ratios
by using their masses as inputs.
Even if additional Higgs bosons are not directly discovered, stronger mass bounds obtained from the direct searches provide narrower allowed regions
in the correlations.
On the other hand, if deviations in the Higgs boson couplings, widths and/or branching ratios from the SM predictions are found at future precision experiments, 
we can obtain upper limits on masses of additional Higgs bosons; see e.g., Figs.~\ref{fig:width_hsm}, \ref{fig:width_thdm2} and \ref{fig:idm}.
Therefore, indirect searches for extended Higgs models using deviations in the Higgs boson properties play the complementary role to the direct searches as well as flavor constraints.
{Using the synergy between the direct and indirect searches, the parameter space of {extended Higgs models} can be effectively narrow down. }

Finally, we would like to mention 
that the {\tt H-COUP} version 2.0, where all the NLO computations for the decay rates presented
in this paper are implemented, will be publicly available in near future~\cite{future}.

\vspace*{-4mm}

\begin{acknowledgments}
This work is supported in part by the Grant-in-Aid on Innovative Areas, the Ministry of Education, Culture, 
Sports, Science and Technology, No.~16H06492 and No.~18H04587, and Grant H2020-MSCA-RISE-2014 No.~645722 (Non-Minimal Higgs) [S.K.],  
JSPS KAKENHI Grant No.~18K03648 [K.M.], 
JSPS KAKENHI Grant No.~18J12866 [K.S.] and Early-Career Scientists, No.~19K14714 [K.Y.]. 

\end{acknowledgments}

\begin{appendix}

\section{Loop induced $h\gamma\gamma$, $hZ\gamma$, and $hgg$ vertices \label{sec:hgg}}

In this section, we give analytic expressions of loop induced vertices{,} i.e., $h\gamma\gamma$, $hZ\gamma $ and $hgg$ at one-loop level, 
which are required to calculate not only the decay rates of the loop induced processes given in Eq.~(\ref{eq:loop-induced}) but also those of $h\to ZZ^* \to Vf\bar{f}$ at NLO given in Eq.~(\ref{eq:del_ew_z}). 
{Here, we} present the formulae for the THDMs and the IDM. 
Those for the HSM are simply obtained by removing the charged scalar loop contribution denoted as 
$\hat{\Gamma}_{h\gamma\gamma/hZ\gamma}^i(p_1^2,p_2^2,q^2)_S$.

The loop induced $h\gamma\gamma$ and $hZ\gamma$ vertices can be decomposed into the contribution from charged scalar loops, fermion loops and weak boson loops as follows: 
\begin{align}
\hat{\Gamma}_{h\gamma\gamma}^i(p_1^2,p_2^2,q^2) &= 
\hat{\Gamma}_{h\gamma\gamma}^i(p_1^2,p_2^2,q^2)_S
+ \sum_F\kappa_F \hat{\Gamma}_{h\gamma\gamma}^i(p_1^2,p_2^2,q^2)_F
+ \kappa_V^{} \hat{\Gamma}_{h\gamma\gamma}^i(p_1^2,p_2^2,q^2)_V, \\
\hat{\Gamma}_{hZ\gamma}^i(p_1^2,p_2^2,q^2) &= 
\hat{\Gamma}_{hZ\gamma}^i(p_1^2,p_2^2,q^2)_S
+ \sum_F\kappa_F\hat{\Gamma}_{hZ\gamma}^i(p_1^2,p_2^2,q^2)_F 
+ \kappa_V^{}\hat{\Gamma}_{hZ\gamma}^i(p_1^2,p_2^2,q^2)_V. 
\end{align}
The analytic expressions for each contribution to the $h  Z \gamma$ vertex are given by 
\begin{align}
\hat{\Gamma}_{hZ\gamma}^1(p_1^2,p_2^2,q^2)_S &= -\frac{eg_Z^{}}{16\pi^2}(c_W^2-s_W^2)\lambda_{hH^+H^-}[4C_{24}(m_{H^\pm}^{}) - B_0(q^2; m_{H^\pm},m_{H^\pm})],  \\
\hat{\Gamma}_{hZ\gamma}^2(p_1^2,p_2^2,q^2)_S &= -\frac{4eg_Z^{}}{16\pi^2}(c_W^2-s_W^2)\lambda_{hH^+H^-}q^2C_{1223}(m_{H^\pm}^{}), \\
\hat{\Gamma}_{hZ\gamma}^1(p_1^2,p_2^2,q^2)_F &= -\frac{4eg_Z^{}}{16\pi^2}\frac{m_F^2}{v}N_c^Fv_FQ_F  [8C_{24}(m_F^{})-2B_0(q^2;m_F^{},m_F^{}) \notag\\
&\hspace{3cm} + (p_1^2+p_2^2 -q^2)C_0(m_F^{}) ], \\
\hat{\Gamma}_{hZ\gamma}^2(p_1^2,p_2^2,q^2)_F &= -\frac{8eg_Z^{}}{16\pi^2}\frac{m_F^{2}}{v}N_c^Fv_FQ_Fq^2\left[C_0(m_F^{}) + 4C_{1223}(m_F^{})  \right], \\
\hat{\Gamma}_{hZ\gamma}^1(p_1^2,p_2^2,q^2)_V&=\frac{2eg_Z^{}}{16\pi^2}\frac{m_W^2}{v}\Big\{
 c_W^2\Big[3(m_W^2 +p_1^2 + 2p_2^2-2q^2)C_0(m_W) \notag\\
&+5[4C_{24}(m_W) - B_0(q^2;m_W,m_W)] +2B_0(p_2^2;m_W,m_W)-2B_0(0;m_W,m_W)\Big]\notag\\
& + s_W^2[(3m_W^2 -2p_1^2 + p_2^2 +q^2)C_0(m_W^{}) +2B_0(p_2^2;m_W^{},m_W^{}) \notag\\
& -2B_0(0;m_W^{},m_W^{})+ B_0(q^2;m_W^{},m_W^{}) - 4C_{24}(m_W^{})]\notag\\
&-\frac{m_h^2(c_W^2-s_W^2)}{2m_W^2}[B_0(q^2,m_W^{},m_W^{})-4C_{24}(m_W^{})] +m_h^2s_W^2C_0(m_W^{}) \Big\}, \label{jjj} \\
\hat{\Gamma}_{hZ\gamma}^2(p_1^2,p_2^2,q^2)_V&=\frac{8eg_Z^{}}{16\pi^2}\frac{m_W^2}{v} q^2\Big\{
c_W^2[3C_0(m_W^{}) + 5C_{1223}(m_W^{})] - s_W^2[C_0(m_W^{}) + C_{1223}(m_W^{})] \notag\\
& + \frac{m_h^2(c_W^2-s_W^2)}{2m_W^2} C_{1223}(m_W^{}) \Big\}, \label{kkk}
\end{align}
where $B_i$, $C_{i}$ and $C_{ij}$ are the Passarino-Veltman's functions~\cite{Passarino:1978jh}. In this paper, 
we follow the convention of these functions given in Ref.~\cite{Kanemura:2015mxa}. 
{Here, we} use the shorthand notation for the $C$ functions defined by $C_{i,ij}(m) \equiv C_{i,ij}(p_1^2,p_2^2,q^2;m,m,m)$ and 
$C_{1223} \equiv C_{12} + C_{23}$. 
The form factors for the renormalized $h\gamma\gamma$ vertex is obtained from the expressions of the $h Z\gamma$ vertex by the replacement of $(g_Z^{},c_W^2,s_W^2,v_F) \to (e,1,-1,Q_F)$. 
Finally, the $hgg$ vertex is induced only from the quark loop. Thus, it is expressed  by 
\begin{align}
\hat{\Gamma}_{hgg}^1(p_1^2,p_2^2,q^2) &= -\frac{\alpha_s}{\pi}\frac{m_q^2}{v}  \left[8C_{24}(m_q)-2B_0(q^2;m_q,m_q) + (p_1^2+p_2^2 -q^2)C_0(m_q) \right]\delta^{ab}, \\
\hat{\Gamma}_{hgg}^2(p_1^2,p_2^2,q^2) &= -\frac{\alpha_s}{\pi}\frac{m_q^2}{v}   \left[C_0(m_q) + 4C_{1223}(m_q)  \right]\delta^{ab},  
\end{align}
where $a$ and $b$ represent the color index.

\section{Renormalized $Vf\bar{f}$ vertices \label{sec:vff}}

We give analytic formulae of the renormalized $Vf\bar{f}$ ($V=Z,W$) vertices, which appear in the decay rates of $h\to VV^* \to Vf\bar{f}$ at NLO 
given in Eq.~(\ref{eq:del_ew_z}) and (\ref{eq:del_ew_w}).   
{In the limit of massless external fermions,} expressions of these vertices are common to those in the SM.

The renormalized $Vf\bar{f}$ ($V=Z$, $W$) vertices can be decomposed in the massless limit for external fermions as 
\begin{align}
\hat{\Gamma}_{Vff}^\mu(p_1^2,p_2^2,q^2)&= g_V^{}\gamma^\mu (\hat{\Gamma}_{Vff}^V - \gamma_5 \hat{\Gamma}_{Vff}^A), 
\end{align}
where $p_1^\mu$ $(p_2^\mu)$ is the incoming four-momentum of the fermion (anti-fermion), 
and $q^\mu $ is the outgoing four-momentum of the gauge boson. 
The gauge coupling $g_V^{}$ is $g_Z$ and $g/\sqrt{2}$ for $Z$ and $W$, respectively.
Similar to Eqs.~(\ref{eq:form-loop}) and (\ref{gam_hff_loop}), we can further decompose these vertices into the tree level part and 1-loop part: 
\begin{align}
\hat{\Gamma}_{Vff}^i = \Gamma_{Vff}^{i,\text{tree}} + \Gamma_{Vff}^{i,\text{loop}},~~\text{with}~~
\Gamma_{Vff}^{i,\text{loop}} = \Gamma_{Vff}^{i,\text{1PI}} + \delta \Gamma_{Vff}^i,~~(i=V,A). 
\end{align}
The tree level contribution is given by 
\begin{align}
&\Gamma^{V,{\rm tree}}_{Zff} = v_f = \frac{I_f}{2} -s_W^2 Q_f,\quad 
\Gamma^{A,{\rm tree}}_{Zff} = a_f = \frac{I_f}{2}, \quad
\Gamma^{V,{\rm tree}}_{Wff} = \Gamma^{A,{\rm tree}}_{Wff} = \frac{1}{2}. \label{eq:vff_tree}
\end{align}
The counterterm contribution is determined by imposing the on-shell renormalization condition as   
 \begin{align}
\delta \Gamma_{Zff}^V &= \frac{1}{16\pi^2}
\Big[ e^2Q_f^2v_f(2B_1 + 1)(m_f^2;m_f,m_\gamma)
+g_Z^2v_f(v_f^2 + 3a_f^2)(2B_1 + 1)(m_f^2;m_f,m_Z) \notag\\
& +\frac{g^2}{4}(v_f+a_f)(2B_1 + 1)(m_f^2;m_{f'},m_W)
-g^2I_fc_W^2B_0(0;m_W,m_W)\Big], \label{delgamv} \\
\delta \Gamma_{Zff}^A &= \frac{1}{16\pi^2}
\Big[ e^2Q_f^2a_f(2B_1 + 1)(m_f^2;m_f,m_\gamma)
+g_Z^2v_f(3v_f^2 + a_f^2)(2B_1 + 1)(m_f^2;m_f,m_Z) \notag\\
& +\frac{g^2}{4}(v_f+a_f)(2B_1 + 1)(m_f^2;m_{f'},m_W)
-g^2I_fc_W^2B_0(0;m_W,m_W)\Big], \label{delgama} \\
\delta \Gamma_{Wff}^V &= \delta \Gamma_{Wff}^A 
= \frac{1}{16\pi^2}\Big[\frac{e^2}{4}Q_f^2(2B_1 + 1)(m_f^2;m_f,m_\gamma) + \frac{g_Z^2}{4}(v_f + a_f)^2(2B_1 + 1)(m_f^2;m_f,m_\gamma)\notag\\
  &+ \frac{g^2}{8}(2B_1 + 1)(m_f^2;m_{f'},m_W) \Big] + (f \leftrightarrow f') -\frac{g^2}{16\pi^2}B_0(0;m_W,m_W), 
 \end{align} 
where $m_f = m_{f'} = 0$. 
The 1PI diagram contributions to these vertices are calculated as 
\begin{align}
\Gamma_{Zff}^{V,\text{1PI}} = \frac{1}{16\pi^2} &\Big[v_fe^2 Q_f^2 F_{FVF}(f,\gamma,f) + g_Z^2v_f(v_f^2 + 3a_f^2)F_{FVF}(f,Z,f) \notag\\
&                         + \frac{g^2}{4}(v_{f'} + a_{f'})F_{FVF}(f',W,f') 
                         + I_f g^2c_W^2F_{VFV}(W,f',W) \Big], \\
\Gamma_{Zff}^{A,\text{1PI}} = \frac{1}{16\pi^2} & \Big[a_fe^2 Q_f^2 F_{FVF}(f,\gamma,f) + g_Z^2a_f(3v_f^2 + a_f^2)F_{FVF}(f,Z,f) \notag\\
&+ \frac{g^2}{4}(v_{f'} + a_{f'})F_{FVF}(f',W,f') 
 + I_f g^2c_W^2F_{VFV}(W,f',W) \Big], \\
\Gamma_{Wff}^{V,\text{1PI}} = \frac{1}{16\pi^2} &\Big\{\frac{e^2}{2}Q_fQ_{f'}F_{FVF}(f,\gamma,f') + \frac{g_Z^2}{2}(v_f + a_f)(v_{f'} + a_{f'})F_{FVF}(f,Z,f') \notag\\
& + \frac{g^2}{2}[F_{VFV}(W,f',Z) + F_{VFV}(Z,f,W)]\notag\\
& + 2e^2 Q_fI_f[F_{VFV}(\gamma,f,W) - F_{VFV}(Z,f,W)] \notag\\
& + 2e^2 Q_{f'}I_{f'}[F_{VFV}(W,f',\gamma) - F_{VFV}(W,f',Z)] \Big\}, \\
\Gamma_{Wff}^{A,\text{1PI}} = \Gamma_{Wff}^{V,\text{1PI}}, 
\end{align}
where
\begin{align}
F_{FVF}(X,Y,Z) &= 2q^2[C_{11} + C_{23}](p_1^2,p_2^2,q^2;m_X,m_Y,m_Z) \notag\\
&+ 4C_{24}(p_1^2,p_2^2,q^2;m_X,m_Y,m_Z) -2 , \\
F_{VFV}(X,Y,Z) &= q^2[C_0 + C_{11} + C_{23}](p_1^2,p_2^2,q^2;m_X,m_Y,m_Z) \notag\\
&+ 6C_{24}(p_1^2,p_2^2,q^2;m_X,m_Y,m_Z) -1.  
\end{align}

\section{Contributions from $hf\bar{f}$ vertex corrections and box diagrams \label{sec:box}}

In the calculation of the decay rate of $h \to VV^* \to Vf\bar{f}$, 
there are contributions from $hf\bar{f}$ vertex corrections denoted as $T_{hff}^V$ and box diagrams denoted as $B_V$ in Eqs.~(\ref{eq:del_ew_z}) and (\ref{eq:del_ew_w}). 
The analytic expressions for $T_{hff}^V$ are given as follows: 
\begin{align}
T_{hff}^Z&=\frac{\kappa_V}{256\pi^3m_h^3}\frac{g_Z^6m_Z^2}{16\pi^2(s-m_Z^2)} \Big\{c_W^4(v_f+a_f)^2[C_{12}(0,u,m_h^2,m_W,0,m_W) \notag\\
& - (C_0+C_{11})(t_Z^{},0,m_h^2;m_W,0,m_W) ]\notag\\
& + 4(v_f^4+6v_f^2a_f^2+a_f^4)\left[C_{12}(0,u,m_h^2;m_Z^{},0,m_Z^{}) - (C_0 + C_{11})(t_Z^{},0,m_h^2,m_Z^{},0,m_Z^{})\right] \Big\}\notag\\
&\times \left[s + \frac{(m_h^2-s-u)(u-m_Z^2)}{m_Z^2} \right], \label{eq:thff1}
 \end{align}
 \begin{align}
 T_{hff}^W& = \frac{\kappa_V}{256\pi^3m_h^3}\frac{g^6m_W^2}{32\pi^2(s-m_W^2)m_h^2}\Bigg\{C_{12}(0,u,m_h^2;m_W^{},0,m_W^{})-(C_0+C_{11})(t_W,0,m_h^2;m_W^{},0,m_W^{}) \notag\\ 
& + \frac{2}{c_W^4}\left[(v_f+a_f)^2C_{12}(0,u,m_h^2;m_Z^{},0,m_Z^{}) - (v_{f^\prime}+a_{f^\prime})^2(C_0+C_{11})(t_W,0,m_h^2;m_Z^{},0,m_Z^{})\right] \Bigg\}\notag\\
&\times \left[s + \frac{(m_h^2-s-u)(u-m_W^2)}{m_W^2} \right], 
 \end{align}
where $t_V = m_h^2 + m_V^2 -s -u $.

For the calculation of box diagrams, we define the Passarino-Veltman's $D$ functions: 
\begin{align}
&\frac{i}{16\pi^2}[D_0,D^\mu,D^{\mu\nu}](p_1^2,p_2^2,p_3^2,(p_1+p_2+p_3)^2,(p_1+p_2)^2,(p_2+p_3)^2;m_1,m_2,m_3,m_4) \notag\\
& = \int \frac{d^4k}{(2\pi)^4}\frac{[1,k^\mu, k^\mu k^\nu]}{N_1N_2N_3N_4}, 
\end{align}
where $N_1 = k^2 - m_1^2$, 
$N_2 = (k+p_1)^2 - m_2^2$, 
$N_3 = (k+p_1+p_2)^2 - m_3^2$, 
$N_4 = (k+p_1+p_2+p_3)^2 - m_4^2$.
We note that in our calculation up to the second rank tensors $D^{\mu\nu}$ appear, and these functions are UV finite. 
The first and second rank tensor functions are decomposed into the following scalar coefficients: 
\begin{align}
D^\mu &= p_1^\mu D_{11} + p_2^\mu D_{12} + p_3^\mu D_{13}, \\
D^{\mu\nu} &= p_1^\mu p_1^\nu D_{21} + p_2^\mu p_2^\nu D_{22} + p_3^\mu p_3^\nu D_{23}
 +(p_1^\mu p_2^\nu + p_2^\mu p_1^\nu) D_{24}\notag\\
& +(p_1^\mu p_3^\nu + p_3^\mu p_1^\nu) D_{25}
 +(p_2^\mu p_3^\nu + p_3^\mu p_2^\nu) D_{26}
+g^{\mu\nu}D_{27}. 
\end{align}
In the following, we shortly express the $D$ functions by 
$D_{0,i,ij}(p_1^2,p_2^2,p_3^2,(p_1+p_2+p_3)^2,(p_1+p_2)^2,(p_2+p_3)^2;a,b,c,d) \equiv D_{0,i,ij}(p_1^2,p_2^2,p_3^2,(p_1+p_2+p_3)^2,(p_1+p_2)^2,(p_2+p_3)^2;m_a,m_b,m_c,m_d)$. 

The contribution from box diagrams $B_V$ can be expressed as 
\begin{align}
B_V &= \frac{\kappa_V^{}}{256\pi^3 m_h^3}\frac{1}{16\pi^2}\frac{c_V^{} g_V^2}{(s-m_V^2)m_V} 
\Big[(m_V^2-t)(tu - m_h^2 m_V^2)(\Gamma_{Vff}^{V,{\rm tree}} {\cal B}_V^1 + \Gamma_{Vff}^{A,{\rm tree}} {\cal \bar{B}}_V^1) \notag\\
& +  (m_V^2-u)(tu-m_h^2 m_V^2)(\Gamma_{Vff}^{V,{\rm tree}} {\cal B}_V^2 + \Gamma_{Vff}^{A,{\rm tree}} {\cal \bar{B}}_V^2) 
 +2 (tu - m_h^2 m_V^2 + 2sm_V^2)(\Gamma_{Vff}^{V,{\rm tree}} {\cal B}_V^\gamma + \Gamma_{Vff}^{A,{\rm tree}} {\cal \bar{B}}_V^\gamma)\Big], 
\end{align}
where $c_V^{} = 1 (\sqrt{2})$ for $V = Z(W)$. 
Each factor ${\cal B}_V^i$ and ${\cal \bar{B}}_V^i$ ($i = 1,2,\gamma$) for $V =Z$ is given as follows:
\begin{align}
{\cal B}_{Z}^1&= g^4m_W\Big[-2c_W^{}I_f(D_0 + D_{11} + D_{13} + D_{25})(0,0,m_Z^2,m_h^2,s,u;W,f',W,W)\notag\\
& \hspace{18mm}-c_W^{}I_f(D_{13}-D_{12}+2D_{26})(0,0,m_Z^2,m_h^2,s,t;W,f',W,W) \notag\\
&\hspace{18mm}+\frac{s_W^{2}}{c_W^{}}I_f(D_{13} - D_{12})(0,0,m_Z^2,m_h^2,s,t;W,f',W,W) \notag\\
&\hspace{18mm}+\frac{v_{f'} + a_{f'}}{c_W} D_{26}(0,m_Z^2,0,m_h^2,u,t;W,f',f',W) \notag\\
&\hspace{18mm} +\frac{4}{c_W^5}(v_f^3 + 3v_fa_f^2)D_{26}(0,m_Z^2,0,m_h^2,u,t;Z,f,f,Z)\Big], \\
{\cal B}_{Z}^2&= g^4m_W\Big[-c_W^{}I_f (D_{13}-D_{12}+2D_{26})(0,0,m_Z^2,m_h^2,s,u;W,f',W,W)\notag\\
&\hspace{18mm}-2c_W^{}I_f(D_0 + D_{11} + D_{13} + D_{25})(0,0,m_Z^2,m_h^2,s,t;W,f',W,W)\notag\\
&\hspace{18mm} +\frac{s_W^{2}}{c_W^{}}I_f (D_{13} - D_{12})(0,0,m_Z^2,m_h^2,s,u;W,f',W,W)\notag\\
&\hspace{18mm}+\frac{v_{f'} + a_{f'}}{c_W}(D_0 + D_{11} + D_{12} + D_{24})(0,m_Z^2,0,m_h^2,u,t;W,f',f',W)\notag\\
&\hspace{18mm}+\frac{4}{c_W^5}(v_f^3 + 3v_fa_f^2)  (D_0 + D_{11} + D_{12} + D_{24})(0,m_Z^2,0,m_h^2,u,t;Z,f,f,Z)\Big], \\
%
{\cal B}_{Z}^\gamma & = g^4m_W\Big\{-\frac{I_f}{2}c_W^{}[-2C_0(s,m_Z^2,m_h^2;m_W,m_W,m_W)
+(t-s-m_Z^2)(D_0+D_{11})   \notag\\
&\hspace{18mm}+2(s+t-m_h^2)D_{12} + (s+t-m_h^2-m_Z^2)D_{13} -4D_{27} ](0,0,m_Z^2,m_h^2,s,u;W,f',W,W) \notag\\
&\hspace{18mm}+ \frac{I_f}{4}\frac{s_W^{2}}{c_W^{}}[C_0(s,m_Z^2,m_h^2;m_W,m_W,m_W)+2(s+t-m_Z^2)(D_0+D_{11})   \notag\\
&\hspace{18mm}- 2(s+t-m_h^2-m_Z^2)D_{13}](0,0,m_Z^2,m_h^2,s,u;W,f',W,W) + (t \leftrightarrow u)\Big\} \notag\\
& +g^4m_W\Big\{
\frac{v_{f'} + a_{f'}}{2c_W}[C_0(u,0,m_h^2;m_W,0,m_W)+(m_h^2-s-t)(D_0+D_{11}) + m_Z^2D_{12} \notag\\
&\hspace{18mm} -2D_{27}](0,m_Z^2,0,m_h^2,u,t;W,f',f',W) \notag\\
&+\frac{2}{c_W^5}(v_f^3 + 3v_fa_f^2)[C_0(u,0,m_h^2;m_Z,0,m_Z)+ (m_h^2-s-t)(D_0+D_{11}) + m_Z^2D_{12} \notag\\
&\hspace{18mm} -2D_{27}](0,m_Z^2,0,m_h^2,u,t;Z,f,f,Z) \Big\}, \\
{\cal \bar{B}}_Z^i &= {\cal B}_{Z}^i\big|_{v_f \leftrightarrow a_f}~~(i = 1,2,\gamma), 
\end{align}
Those for $V =W$ are given by 
\begin{align}
{\cal B}_{W}^1&= -\sqrt{2}g^4m_W\Big[2I_{f'}(v_{f'}+a_{f'}) (D_0 + D_{11} + D_{13} + D_{25})(0,0,m_W^2,m_h^2,s,u;W,f',Z,W)\notag\\
&\hspace{18mm}+2s_W^2 I_{f'}Q_{f'} (D_0 + D_{11} + D_{13} + D_{25})(0,0,m_W^2,m_h^2,s,u;W,f',\gamma,W)\notag\\
&\hspace{18mm}+I_f(v_f+a_f)(D_{13}-D_{12}+2D_{26})(0,0,m_W^2,m_h^2,s,t;W,f,Z,W)\notag\\
&\hspace{18mm}+s_W^2 I_{f}Q_{f}(D_{13}-D_{12}+2D_{26})(0,0,m_W^2,m_h^2,s,t;W,f,\gamma,W)\notag\\
&\hspace{18mm}+2I_f\frac{v_f+a_f}{c_W^2}(D_0 + D_{11} + D_{13} + D_{25})(0,0,m_W^2,m_h^2,s,u;Z,f,W,Z)\notag\\
&\hspace{18mm}+I_{f'}\frac{v_{f'}+a_{f'}}{c_W^2}(D_{13}-D_{12}+2D_{26})(0,0,m_W^2,m_h^2,s,t;Z,f',W,Z)\notag\\
&\hspace{18mm}+\frac{s_W^2}{c_W^2}I_{f}(v_{f} + a_{f})(D_{13} - D_{12})(0,0,m_W^2,m_h^2,s,t;W,f,Z,W)\notag\\
&\hspace{18mm}- s_W^2I_fQ_f(D_{13} - D_{12})(0,0,m_W^2,m_h^2,s,t;W,f,\gamma,W)\notag\\
&\hspace{18mm}-\frac{ (v_f + a_f)(v_{f'} + a_{f'})}{c_W^4} D_{26}(0,m_W^2,0,m_h^2,u,t;Z,f',f,Z) \Big],
\end{align}

\begin{align}
{\cal B}_{W}^2&= -\sqrt{2}g^4m_W\Big[I_{f'}(v_{f'}+a_{f'})(D_{13}-D_{12}+2D_{26})(0,0,m_W^2,m_h^2,s,u;W,f',Z,W)\notag\\
&\hspace{18mm}+s_W^2 I_{f'}Q_{f'}(D_{13}-D_{12}+2D_{26})(0,0,m_W^2,m_h^2,s,u;W,f',\gamma,W)\notag\\
&\hspace{18mm}+2I_{f}(v_{f}+a_{f}) (D_0 + D_{11} + D_{13} + D_{25})(0,0,m_W^2,m_h^2,s,t;W,f,Z,W)\notag\\
&\hspace{18mm}+2s_W^2 I_fQ_f (D_0 + D_{11} + D_{13} + D_{25})(0,0,m_W^2,m_h^2,s,t;W,f,\gamma,W)\notag\\
&\hspace{18mm}+I_f\frac{v_f+a_f}{c_W^2}(D_{13}-D_{12}+2D_{26})(0,0,m_W^2,m_h^2,s,u;Z,f,W,Z)\notag\\
&\hspace{18mm}+2I_{f'}\frac{v_{f'} + a_{f'}}{c_W^2}  (D_0 + D_{11} + D_{13} + D_{25})(0,0,m_W^2,m_h^2,s,t;Z,f',W,Z)\notag\\
&\hspace{18mm}+\frac{s_W^2}{c_W^2}I_{f'}(v_{f'} + a_{f'}) (D_{13} - D_{12})(0,0,m_W^2,m_h^2,s,u;W,f',Z,W)\notag\\
&\hspace{18mm}-s_W^2I_{f'}Q_{f'}(D_{13} - D_{12})(0,0,m_W^2,m_h^2,s,u;W,f',\gamma,W)\notag\\
&\hspace{18mm}-\frac{(v_f + a_f)(v_{f'} + a_{f'})}{c_W^4}  (D_0 + D_{11} + D_{12} + D_{24})(0,m_W^2,0,m_h^2,u,t;Z,f',f,Z)\Big],
\end{align}

\begin{align}
{\cal B}_{W}^\gamma &=
 -\frac{g^4m_W}{\sqrt{2}}\Big\{
    I_{f'}(v_{f'}+a_{f'})[ -2C_0(s,m_W^2,m_h^2;m_W,m_Z,m_W)+(t-s-m_W^2)(D_0+D_{11})  \notag\\
 &\hspace{18mm} +2(s+t-m_h^2)D_{12} + (s+t-m_h^2-m_W^2)D_{13}-4D_{27}](0,0,m_W^2,m_h^2,s,u;W,f',Z,W)\notag\\
 &\hspace{18mm}+s_W^2I_{f'}Q_{f'} [-2C_0(s,m_W^2,m_h^2;m_W,0,m_W) + (t-s-m_W^2)(D_0+D_{11})  \notag\\
 &\hspace{18mm} +2(s+t-m_h^2)D_{12}+ (s+t-m_h^2-m_W^2)D_{13}-4D_{27} ](0,0,m_W^2,m_h^2,s,u;W,f',\gamma,W)\notag\\
&\hspace{18mm}+ I_f\frac{v_f+a_f}{c_W^2} [-2C_0(s,m_W^2,m_h^2;m_W,m_Z,m_W) + (t-s-m_W^2)(D_0+D_{11}) \notag\\
&\hspace{18mm} +2(s+t-m_h^2)D_{12} + (s+t-m_h^2-m_W^2)D_{13}-4D_{27} ](0,0,m_W^2,m_h^2,s,u;Z,f,W,Z) \notag\\
 &\hspace{18mm} +\frac{s_W^2}{2c_W^2}I_{f'}(v_{f'} + a_{f'})[C_0(s,m_W^2,m_h^2;m_W,m_Z,m_W)+2(s+t-m_W^2)(D_0+D_{11}) \notag\\
 &\hspace{18mm} - 2(s+t-m_h^2-m_W^2)D_{13} ](0,0,m_W^2,m_h^2,s,u;W,f',Z,W)\notag\\
&\hspace{18mm} -\frac{s_W^2}{2}I_{f'}Q_{f'}[C_0(s,m_W^2,m_h^2;m_W,0,m_W) + 2(s+t-m_W^2)(D_0+D_{11}) \notag\\
&\hspace{18mm} - 2(s+t-m_h^2-m_W^2)D_{13} ](0,0,m_W^2,m_h^2,s,u;W,f',\gamma,W)\notag\\
&\hspace{18mm}  + (t \leftrightarrow u,~f \leftrightarrow f')\Big\} \notag\\
& +g^4m_W\frac{(v_f + a_f)(v_{f'} + a_{f'})}{\sqrt{2}c_W^4} [C_0(u,0,m_h^2;m_Z,0,m_Z) + (m_h^2-s-t)(D_0+D_{11}) \notag\\
&\hspace{18mm} + m_W^2D_{12} -2D_{27}](0,m_W^2,0,m_h^2,u,t;Z,f',f,Z), \\
{\cal \bar{B}}_W^i    &  = {\cal B}_W^i~~(i = 1,2,\gamma). 
\end{align}

\section{Real photon emissions in $h \to WW^* \to Wf\bar{f}'$  \label{sec:real_photon}}

\begin{figure}[t]
\centering
\includegraphics[scale=1.0]{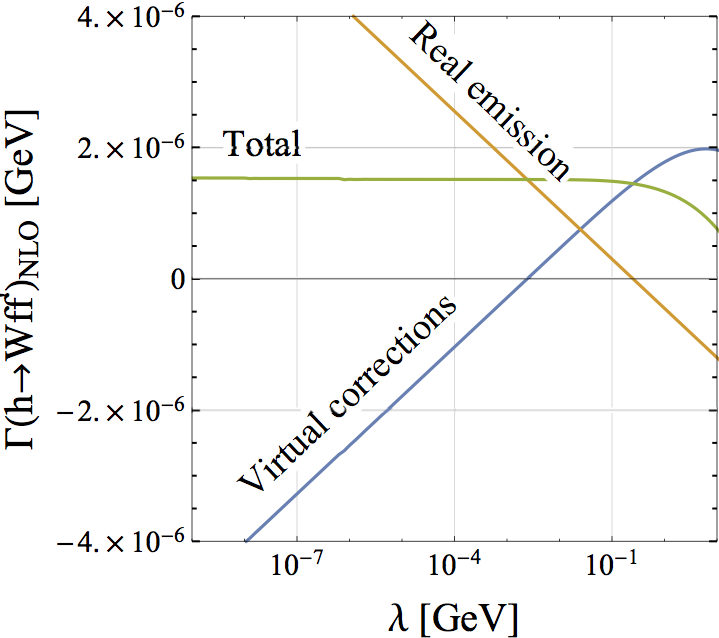}\\
\vspace{+0.8cm} 
\caption{Numerical check of cancellation of the soft divergence in the decay rate of $h \to Wf\bar{f}$ at NLO. 
The horizontal axis is the photon mass $\lambda$ { as a regulator }{to avoid the divergence}. 
The blue and orange curves show the contribution from the virtual corrections and real photon emissions, respectively. }
\label{FIG:lam_dep_Wffp}
\end{figure}

The contribution from the real photon emission in the $h \to WW^* \to Wf\bar{f}'$ process 
can be separately written by the soft-photon and hard-photon emission parts as 
\begin{align}
\Gamma({h\to Wff^\prime \gamma}) = \frac{1}{2m_h}\int_S|\mathcal{M}|^2d\Phi_4+\frac{1}{2m_h}\int_H|\mathcal{M}|^2d\Phi_4, 
\end{align}
where $\Phi_4$ is the four body phase space function. 
The first integral denoted as $\int_S$ is performed up to the cutoff of the photon energy $\Delta E$, while the second integral is done from $\Delta E$ to the maximal value of the photon energy. 
The $\Delta E$ dependence in the NLO decay rate, of course, disappears after summing up the soft and hard photon parts. 

The soft-photon part is calculated using the eikonal approximation by which the amplitude can be expressed by the product of 
the Born amplitude and the soft-photon factor. 
{Then, we can} separately perform the integration with respect to the 3-body phase space and the photon phase space. 
Therefore, the soft-photon part is expressed as 
\begin{align}
\frac{1}{2m_h}\int_S|\mathcal{M}|^2d\Phi_4  =  \int d \Gamma_0(h \to Wf\bar{f}')\delta_{W}^{\rm soft}, 
\end{align}
where 
\begin{align}
\notag
\delta_W^{\rm soft}&=-\frac{\alpha_{\rm em}}{2\pi}\Bigg[
Q_f^2\Big\{\Big(\log\frac{m_f^2}{s}+1\Big)\log\frac{4\Delta E}{\lambda^2}+\frac{1}{2}\Big(\log\frac{m_f^2}{4E_f^2}\Big)^2+\log\frac{m_f^2}{4E_f^2}+\frac{\pi^2}{3}\Big\} \\ \notag
&+Q_{f^\prime}^2\Big\{\Big(\log\frac{m_{f^\prime}^2}{s}+1\Big)\log\frac{4\Delta E}{\lambda^2}+\frac{1}{2}\Big(\log\frac{m_{f^\prime}^2}{4E_{f^\prime}^2}\Big)^2+\log\frac{m_{f^\prime}^2}{4E_{f^\prime}^2}+\frac{\pi^2}{3}\Big\} \\ \notag
&+\Big\{\Big(\log\frac{m_{W}^2}{s}+1\Big)\log\frac{4\Delta E}{\lambda^2}+\frac{1}{2}\Big(\log\frac{E_W-|\vec{p}_W|}{E_W+|\vec{p}_W|}\Big)^2+\frac{E_W}{|\vec{p}_W|}\log\frac{m_{W}^2}{(E_W+|\vec{p}_W|)^2}\Big\} \\ \notag
&+4Q_fI_f \Big\{\frac{1}{2}\log\frac{s^2}{(m_W^2-t)^2}\log\frac{4\Delta E}{\lambda^2}+{\rm Li_2}\left(1-\frac{2E_f(E_W-|\vec{p}_W|)}{t-m_W^2}\right) \\ \notag
&+{\rm Li_2}\left(1-\frac{2E_f(E_W+|\vec{p}_W|)}{t-m_W^2}\right)
\Big\} \\ \notag
&-4Q_{f^\prime}I_f\Big\{\frac{1}{2}\log\frac{s^2}{(m_W^2-u)^2}\log\frac{4\Delta E}{\lambda^2}+{\rm Li_2}\left(1-\frac{2E_{f^\prime}(E_W-|\vec{p}_W|)}{u-m_W^2}\right) \\ 
&+{\rm Li_2}\left(1-\frac{2E_{f^\prime}(E_W+|\vec{p}_W|)}{u-m_W^2}\right)
\Big\}\Bigg], 
\end{align}
with $t = m_h^2 + m_V^2 -s -u $, $E_W = m_h(1-x_s+x_W)/2$ and $|\vec{p}_W| = m_h\lambda^{1/2}(x_s,x_W)/2$. 
This expression agrees with~\cite{Ciccolini:2003jy}.
{Here, we introduced} the photon mass $\lambda$ to regularize the IR divergence and the fermion masses $m_f$
to regularize the collinear singularities.
{We numerically evaluate the hard-photon part by} using {\tt Madgraph5\_aMC@NLO}~\cite{Alwall:2014hca}.

In order to check the cancellation of the IR divergence, we show the partial decay rate at NLO in Fig.~\ref{FIG:lam_dep_Wffp}
as a function of the photon mass $\lambda$ {being a regulator} {to avoid the divergence}, where we take $\Delta E=$ 1 GeV.
We clearly see that the sum of the virtual corrections and the real emissions (denoted as ``Total'') does not depend on $\lambda$.

\end{appendix}

\bibliography{references}

\end{document}